%% file: main.tex
\documentclass[sigplan,screen,authorversion]{acmart}
\setcopyright{acmcopyright}
\acmPrice{}
\acmDOI{10.1145/3446804.3446854}
\acmYear{2021}
\copyrightyear{2021}
\acmSubmissionID{cc21main-p90-p}
\acmISBN{978-1-4503-8325-7/21/03}
\acmConference[CC '21]{Proceedings of the 30th ACM SIGPLAN International Conference on Compiler Construction}{March 2--3, 2021}{Virtual, USA}
\acmBooktitle{Proceedings of the 30th ACM SIGPLAN International Conference on Compiler Construction (CC '21), March 2--3, 2021, Virtual, USA}

\usepackage[utf8]{inputenc}
\usepackage[T1]{fontenc}
\usepackage{microtype}
\usepackage{booktabs}   %
\usepackage{subcaption} %
\usepackage{listings}
\usepackage[capitalise]{cleveref}
\usepackage[nomargin,inline,draft]{fixme}
\usepackage{amsmath}
\usepackage{cmll}
\usepackage{colortbl}
\usepackage{stmaryrd}
\usepackage[inline]{enumitem}
\newtoggle{fullversion}
\toggletrue{fullversion}

\fxsetup{theme=color,mode=multiuser}
\FXRegisterAuthor{FZ}{aFZ}{FZ}
\FXRegisterAuthor{NY}{aNY}{NY}
\FXRegisterAuthor{FF}{aFF}{FF}
\FXRegisterAuthor{AM}{aAM}{AM}
\newcommand{\FZ}[1]{\FZnote{#1}}

\crefname{section}{\S}{\SS}

\usepackage{bussproofs}
\usepackage{mathtools}
\usepackage{xspace}
\usepackage{mathpartir}
\usepackage{multirow}
\usepackage{tikz}
\usetikzlibrary{arrows}
\usetikzlibrary{positioning}
\usetikzlibrary{calc}
\input{macros.tex}

\begin{document}

\title{Communication-Safe Web Programming in TypeScript
with Routed Multiparty Session Types}
\author{Anson Miu}
\orcid{0000-0002-1666-6881}
\affiliation{
   \institution{Imperial College London and Bloomberg}            %
  \country{United Kingdom}                    %
}

\author{Francisco Ferreira}
\orcid{0000-0001-8494-7696}             %
\affiliation{
  \institution{Imperial College London}            %
  \country{United Kingdom}                    %
}

\author{Nobuko Yoshida}
\orcid{0000-0002-3925-8557}             %
\affiliation{
  \institution{Imperial College London}            %
  \country{United Kingdom}                    %
}

\author{Fangyi Zhou}
\orcid{0000-0002-8973-0821}             %
\affiliation{
  \institution{Imperial College London}            %
  \country{United Kingdom}                    %
}
\input{abstract}

\maketitle
\input{intro.tex}

\input{overview.tex}
\input{implementation}
\input{theory.tex}

\input{evaluation}

\input{related.tex}
\input{acks.tex}

\bibliography{references}

\fullversion{
\appendix
\input{appendix}

}{}
\end{document}

%% file: macros.tex
\newcommand{\code}[1]{\texttt{\upshape #1}}
\newcommand{\benchmark}[1]{\textbf{\code{#1}}}

\definecolor{ColourGlobal}{rgb}{.816,.125,.565} %
\definecolor{ColourLocal}{rgb}{0,0,.5} %
\definecolor{ColourLabel}{rgb}{.5,0,.5} %
\definecolor{ColourParticipant}{rgb}{0,.5,.5} %
\definecolor{Blue}{rgb}{0,0,1} %
\definecolor{ColourShade}{rgb}{0.94, 1.0, 0.96} %
\definecolor{Yellow}{rgb}{1,1,0}

\newcommand{\withcolour}[2]{{\color{#1} {#2}}}

\definecolor{keywordcolour}{rgb}{0.5,0,0.35}
\newcommand{\keyword}[1]{{\color{keywordcolour}\code{#1}}}
\definecolor{greencomments}{rgb}{0,0.5,0}
\lstdefinelanguage{Scribble}{%
  basicstyle=\footnotesize\ttfamily,
  stringstyle=\color{Blue},
  showstringspaces=false,
  keywords={nested,new,calls,and,as,at,by,catches,choice,continue,do,from,global,import,instantiates,interruptible,local,module,or,par,protocol,rec,role,sig,throws,to,type,with,int,aux},
  morestring=[b]",
  morestring=[b]',
  morecomment=[l][\color{greencomments}]{//},
}
\lstdefinelanguage{TypeScript}{%
  basicstyle=\footnotesize\ttfamily,
  stringstyle=\color{Blue},
  showstringspaces=false,
  keywords={
typeof, new, true, false, catch, function, return, null, catch, switch, var, if, in, while, do, else, case, break, class, export, boolean, throw, implements, import, this, const, let, extends, async, await, namespace, type, abstract, any, number, string, instanceof, typeof, declare, never,enum,constructor,public,private,void,super,undefined, as,protected,interface,keyof,infer,is,try,default,static,readonly
  },
  morestring=[b]",
  morestring=[b]',
  morestring=[b]`,
  morecomment=[l][\color{greencomments}]{//},
  morecomment=[s]{/*}{*/},
}
\newcommand{\ppt}[1]{\withcolour{ColourParticipant}{
  \textbf{\textsf{\upshape #1}}}
}

\newcommand{\lbl}[1]{\withcolour{ColourLabel}{#1}}

\newcommand{\gtype}[1]{\withcolour{ColourGlobal}{#1}}

\newcommand{\gtvar}[1]{\gtype{\mathbf{#1}}}
\newcommand{\gtrecur}[2]{\gtype{\mu\mathbf{#1}.#2}}
\newcommand{\gtend}{\gtype{\code{end}}}

\newcommand{\ltype}[1]{\withcolour{ColourLocal}{#1}}

\newcommand{\tvar}[1]{\ltype{\mathbf{#1}}}
\newcommand{\trecur}[2]{\ltype{\mu\mathbf{#1}.#2}}
\newcommand{\tend}{\ltype{\code{end}}}

\newcommand{\mrole}[1]{
\!\!\text{\ppt{#1}}\!\!
}
\newcommand{\mroles}[2]{
{\ppt{#1}\ppt{#2}}
}
\newcommand{\pt}[1]{
{\textnormal{\texttt{pt}}\left({#1}\right)}
}
\newcommand{\rulename}[1]{%
[\textsc{#1}]%
}

\newcommand{\projconf}[1]{
{\left\langle{\gtype{#1}}\right\rangle}
}
\newcommand{\subtype}{
\prec
}

\newcommand{\centroidop}{\circledast}
\newcommand{\centroid}[2]{
{\gtype{#1}} \centroidop \ppt{#2}
}
\newcommand{\trecvar}{\tvar{t}}
\newcommand{\trec}[1]{\trecur{t}{#1}}
\newcommand{\gcommone}[3]{
\gtype{\mrole{#1} \to \mrole{#2}: {#3} ~}
}
\newcommand{\gcomm}[4]{
\gtype{\mrole{#1} \to \mrole{#2} : \left\{ {#3} \right\}_{#4}}
}
\newcommand{\gtrans}[5]{
\gtype{\mrole{#1} \rightsquigarrow \mrole{#2}. ~ {\lbl{#3}} : \left\{ {#4} \right\}_{#5}}
}

\newcommand{\wf}[1]{
\textnormal{\texttt{wellFormed}}\left(#1\right)
}
\newcommand{\proj}[2]{
{\gtype{#1}} \!\mathrel{\projop}\! {\ppt{#2}}
}
\newcommand{\projop}{\upharpoonright}

\newcommand{\mergeop}{
\sqcap
}
\newcommand{\MERGEOP}{
{\mbox{\LARGE$\sqcap$}}
}
\newcommand{\tmerge}[2]{
{\ltype{#1}}\,{\mergeop}\,{\ltype{#2}}
}

\newcommand{\grouteone}[4]{
\gtype{\mrole{#1} -\mspace{-3mu}{\mrole{#3}}\!\shortrightarrow \mrole{#2}: ~ {#4} ~}
}
\newcommand{\groute}[5]{
\gtype{\mrole{#1} -\mspace{-3mu}{\mrole{#3}}\!\shortrightarrow \mrole{#2} : \left\{ {#4} \right\}_{#5}}
}
\newcommand{\gtransroute}[6]{
\gtype{\mrole{#1} \underset{\mrole{#3}}{\rightsquigarrow} \mrole{#2}. ~ \lbl{#4} : \left\{ {#5} \right\}_{#6}}
}

\newcommand{\wfnew}[2]{
\textnormal{\texttt{wellFormed}}^R\left(#1, \ppt{#2}\right)
}

\newcommand{\tbra}[3]{
\ltype{\mrole{#1} \with \left\{ #2 \right\}_{#3}}
}
\newcommand{\tsel}[3]{
\ltype{\mrole{#1} \oplus \left\{ {#2} \right\}_{#3}}
}
\newcommand{\tbraone}[2]{
\ltype{\mrole{#1} \with {#2}}
}
\newcommand{\tselone}[2]{
\ltype{\mrole{#1} \oplus {#2}}
}
\newcommand{\lty}[1]{
\ltype{T_{\mrole{#1}}}
}
\newcommand{\pinP}{
{\mrole{p} \in \mathcal{P}}
}
\newcommand{\qqinP}{
{\mroles{q}{q'} \in \mathcal{P}}
}

\newcommand{\router}[4]{
\ltype{\mrole{#1} \hookrightarrow \mrole{#2}: \left\{{#3}\right\}_{#4}}
}
\newcommand{\routerone}[4]{
\ltype{\mrole{#1} \hookrightarrow \mrole{#2}: {#3}~.~{#4}}
}
\newcommand{\routertrans}[5]{
\ltype{\mrole{#1} \looparrowright \mrole{#2}. ~ \lbl{#3} : \left\{{#4}\right\}_{#5}}
}

\newcommand{\tselproxy}[4]{
\ltype{\mrole{#1} \oplus \!\langle {\mrole{#2}} \rangle\left\{ {#3} \right\}_{#4}}
}
\newcommand{\tbraproxy}[4]{
\ltype{\mrole{#1} \with \!\langle {\mrole{#2}} \rangle\left\{ {#3} \right\}_{#4}}
}
\newcommand{\aout}[3]{
\ppt{#1}\ppt{#2}!{\lbl{#3}}
}
\newcommand{\ain}[3]{
\ppt{#1}\ppt{#2}?{\lbl{#3}}
}
\newcommand{\subj}[1]{
\text{subj}({#1})
}
\newcommand{\treduce}[3]{
\color{black}{{#1} \xrightarrow{\mathmakebox[1em]{#3}} {#2}}
}

\newcommand{\gtreduce}[3]{
\treduce{\gtype{#1}}{\gtype{#2}}{\lbl{#3}}
}
\newcommand{\ltreduce}[3]{
\treduce{\ltype{#1}}{\ltype{#2}}{\lbl{#3}}
}
\newcommand{\treducelong}[3]{
\color{black}{{#1} \xrightarrow{\mathmakebox[3.5em]{#3}} {#2}}
}
\newcommand{\treducelonger}[3]{
\color{black}{{#1} \xrightarrow{\mathmakebox[4.5em]{#3}} {#2}}
}
\newcommand{\gtreducelong}[3]{
\treducelong{\gtype{#1}}{\gtype{#2}}{\lbl{#3}}
}
\newcommand{\gtreducelonger}[3]{
\treducelonger{\gtype{#1}}{\gtype{#2}}{\lbl{#3}}
}
\newcommand{\ltreducelong}[3]{
\treducelong{\ltype{#1}}{\ltype{#2}}{\lbl{#3}}
}

\newcommand{\via}[2]{
{\texttt{via}{\langle\mrole{#1}\rangle}({#2})}
}

\newcommand{\enc}[2]{
  \color{black}{\mathrm{\llbracket} {#1},~\mrole{#2} \rrbracket}
}
\newcommand{\enclocal}[3]{
\llbracket {#1},~\mrole{#2},~\mrole{#3} \rrbracket
}

\newcommand{\newtheory}{\textsf{RouST}\xspace}
\newcommand{\codegen}{\textsf{STScript}\xspace}
\newcommand{\tprotocol}[1]{\textsf{#1}}
\newcommand{\tmsg}[1]{\code{#1}}
\newcommand{\trole}[1]{\textsf{#1}}

\newcommand{\myinferrule}[3][]{\ensuremath{\inferrule*[left={#1}, sep=1mm]{#2}{#3}}}
\newcommand*{\defeq}{\stackrel{\text{def}}{=}}

\newcommand{\TA}{\ltype{T_{\ppt A}}}
\newcommand{\TB}{\ltype{T_{\ppt B}}}

\newcommand{\reactjs}{\textsf{React.js}\xspace}
\newcommand{\nodejs}{\textsf{Node.js}\xspace}
\newcommand{\TS}{\textsf{TypeScript}\xspace}
\newcommand{\python}{\textsf{Python}\xspace}

\newcommand{\fullversion}[2]{\iftoggle{fullversion}{#1}{#2}}

%% file: abstract.tex
\begin{abstract}
Modern web programming
involves coordinating interactions
between browser clients and a server.
Typically,
the interactions
in web-based distributed systems
are
informally described, making
it hard to ensure
correctness, especially \emph{communication safety},
i.e.\ all endpoints progress %
without type errors or deadlocks, conforming to a specified protocol.

We present \codegen, a toolchain that
generates \TS APIs
for communication-safe web development
over WebSockets, and \newtheory,
a new session type theory
that supports multiparty communications with routing mechanisms.

\codegen provides
developers with \TS APIs generated
from a communication protocol specification
based on \newtheory.
The generated APIs
build upon \TS concurrency practices,
complement the event-driven style of programming
in full-stack web development,
and are compatible
with the \nodejs runtime for server-side endpoints
and the \reactjs framework
for browser-side endpoints.

\newtheory can express multiparty interactions routed
via an intermediate participant.
It supports
peer-to-peer communication
between browser-side endpoints by
routing communication via the server
in a way that avoids excessive serialisation.
\newtheory guarantees communication safety
for endpoint web applications
written using \codegen APIs.

We %
evaluate
the expressiveness of \codegen for modern
web programming
using several production-ready case studies deployed
as web applications.

\end{abstract}

%% file: intro.tex
\section{Introduction}\label{sec:intro}
Web technology advancements have changed the way people use computers.
Many services that required standalone applications, such as email,
chat, video conferences, or even games, are now provided in a
browser.
While the Hypertext Transfer Protocol (HTTP) is widely used
for serving web pages, its Request-Response model limits the communication
patterns --- the server may not send data to a client without the client first
making a request.

The \emph{WebSocket Protocol}~\cite{RFCWebSocket} addresses this
limitation by providing a bi-directional channel between the client
and the server, akin to a Unix socket.
Managing the correct usage of WebSockets introduces an additional concern in the
development process,
due to a lack of WebSocket testing tools, %
requiring an (often ad-hoc) specification of the communication protocol between
server and clients.

\input{figures/travel-agency-diagram.tex}

Consider the scenario in
\cref{fig:travel-agency-protocol-dia}, where an online travel agency operates a
``travelling with a friend'' scheme (ignoring the blue dashed arrows).
It starts when a traveller ($\ppt{B}$) suggests a trip destination to
their friend ($\ppt{A}$), who then queries the travel agency ($\ppt{S}$)
if the trip is available.
If so, the friends discuss among themselves whether to accept or reject the
quoted price.
If the trip was unavailable, the friends start again with a new
destination.

An implementation of the travel agency protocol may contain programming errors,
risking \emph{communication safety}.
For example, the following implementation
of the client-side endpoint for traveller $\ppt{A}$ sending a quote to
traveller $\ppt{B}$.%
\vspace{-1mm}
\begin{lstlisting}[tabsize=2,language=html]
<input type='number' id='quote' />£\label{line:number-input}£
<button id='submitQuote'>Send Quote to B</button>
<script>
document.getElementById('submitQuote')
	.addEventListener('click', () => {
		const quote = document.getElementById('quote').value;
		travellerB.send({ label: 'quote', quote });
		travellerB.onMessage( /* go to different screen */ );
		/* ...snip... */ }); </script>
\end{lstlisting}
\vspace{-1mm}
There are subtle errors that violate the communication protocol,
but these bugs are unfortunately left for the developer
to manually identify and test against:
\begin{description}[leftmargin=0cm]
\item[Communication Mismatch]
Whilst the input field mandates a \emph{numerical value}
(\cref{line:number-input}) for the
quote, the \code{value} from the input field is actually a \code{string}.
If $\ppt{B}$ expects a \code{number}
and performs arithmetic operations on the received payload from
$\ppt{A}$, the type mismatch may be left hidden due to
implicit type coercion and cause unintended errors.

\item[Channel Usage Violation]
As $\ppt{B}$ may take time to respond, $\ppt{A}$ can experience
a delay between sending the quote and receiving a response.
Notice that the button remains \emph{active} after sending
the quote --- $\ppt{A}$ could click on the button again, and send additional
quotes (thus reusing the communication channel), but $\ppt{B}$ may be unable to
deal with extra messages.

\item[Handling Session Cancellation]
An additional concern is how to handle browser disconnections,
as both travellers can freely close their browsers at
any stage of the protocol. Suppose $\ppt{S}$
temporarily reserves a seat on $\ppt{A}$'s query.
If $\ppt{A}$ closes their browser, the developer would need
to make sure that $\ppt{A}$ notifies $\ppt{S}$ prior to
disconnecting, and $\ppt{S}$ needs to implement recovery logic
(e.g.\ releasing the reserved seat) accordingly.
\end{description}

To prevent these errors and ensure deadlock-freedom, we propose to apply
\emph{session types}~\cite{ESOP98Session,JACM16MPST}
into practical interactive web programming.
The scenario described in \cref{fig:travel-agency-protocol-dia}
can be precisely described with a \emph{global type} using the typing discipline
of \emph{multiparty session types} (MPST)~\cite{JACM16MPST}.
Well-typed implementations conform to the given \emph{global protocol},
are guaranteed free from communication errors \emph{by construction}.

Whereas session type programming is well-studied~\cite{FTPL16BehavioralSurvey},
its application on web programming, in particular, interactive web
applications, remains relatively unexplored.
Integrating session types with web programming has been piloted by recent
work~\cite{ECOOP20MVU,PLACES19PureScript,SCP16Jolie}, yet none are
able to seamlessly implement the previous application scenario:
\Citet{ECOOP20MVU} uses \emph{binary} (2-party) session types;
and \citet{PLACES19PureScript} require each
non-server role to only communicate to the server,
hence preventing interactions between non-server roles
(cf.\ talking to a friend in the scenario).
The programming languages used in these works are, respectively,
Links~\cite{FMCO06Links} and PureScript~\cite{PureScript}, both not usually
considered mainstream
in the context of modern web programming.
The Jolie language~\cite{SCP16Jolie} focuses more on the server side, with
limited support for an interactive front end of web applications.

\textbf{This paper} presents a novel toolchain, \emph{Session TypeScript} (\codegen),
for implementing multiparty protocols safely in web programming.
\codegen integrates with \emph{modern} tools and practices,
utilising the popular programming language \TS, front end framework
\reactjs and back end runtime \nodejs.
Developers first specify a multiparty protocol and we generate
\emph{correct-by-construction} APIs for
developers to implement the protocol.
The generated APIs use WebSocket to establish communication between
participants, utilising its flexibility over the traditional HTTP model.
When developers use our generated APIs to correctly implement the protocol
endpoints, \codegen guarantees the freedom from communication errors,
including deadlocks, communication mismatches, channel usage violation or
cancellation errors.

Our toolchain is backed by a new session theory, a \emph{routed multiparty
session types theory} (\newtheory), to
endow servers with the capacity to \emph{route messages} between web clients.
The new theory addresses a practical limitation that WebSocket
connections still require clients to connect to a prescribed server,
constraining the ability for inter-client communication.
To overcome this, our API \emph{routes} inter-client messages through the
server, improving the expressiveness over previous work and enabling
developers to correctly implement multiparty protocols, as we show with blue
dashed arrows in \cref{fig:travel-agency-protocol-dia}.
In our travel agency scenario, the agency plays the server role:
it will establish WebSocket channels with each participant, and be
tasked with routing all the messages between the friends.
We formalise this routing mechanism as \newtheory and
prove deadlock-freedom of \newtheory and show a behaviour-preserving
encoding from the original MPST to \newtheory.
The formalism and results in \newtheory
directly guide a deadlock-free protocol implementation
in \nodejs via the router, preserving
communication structures of the original protocol written by a
developer.

Finally, we evaluate our toolchain (\codegen) by case studies.
We evaluate the expressiveness by implementing a number of web applications,
such as interactive multiplayer games (\tprotocol{Noughts and Crosses},
\tprotocol{Battleship}) and web services (\tprotocol{Travel Agency})
that require routed communication.

\paragraph{Contributions and Structure of the Paper}
  \cref{sec:overview} presents an overview of our toolchain
    \codegen, which generates APIs for communication-safe web applications in
    \TS from multiparty protocol descriptions.
  \cref{sec:implementation} motivates how the generated code
     executes the multiparty protocol descriptions, and present how
     \codegen prevents common errors in the context of web applications.
  \cref{sec:theory} presents \newtheory, multiparty
    session types (MPST) extended with \emph{routing}, and define a
    trace-preserving encoding of the original MPST into \newtheory.
  \cref{sec:eval}
    evaluates our toolchain \codegen via a case study of \tprotocol{Noughts
      and Crosses} and performance experiments.
  \cref{sec:related} gives related and future work.

\fullversion{Appendix}{The full version of the paper \FZ{arXiv link here}} includes omitted code,
definitions, performance benchmarks and detailed proofs.
The artifact accompanying this paper~\cite{artifact} is available via
DOI or at \url{https://github.com/STScript-2020/cc21-artifact},
containing the source code of \codegen, with implemented case studies and
performance benchmarks. \fullversion{See \cref{sec:artifact} for details about
the artifact.}{}

%% file: figures/travel-agency-diagram.tex
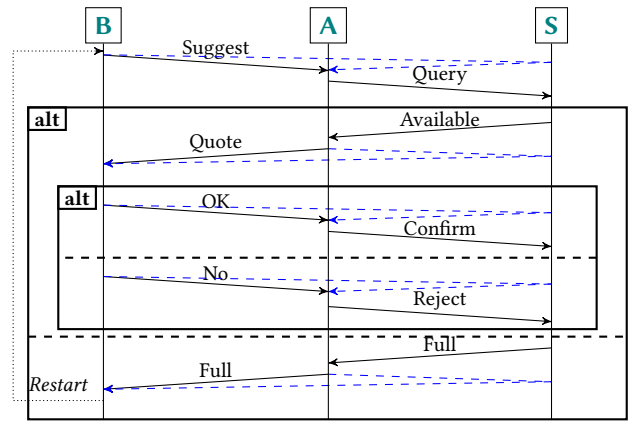
\begin{figure}[H]

\centering
\vspace{-3mm}
\begin{tikzpicture}[>=stealth']
  \node[draw] (A) {$\ppt A$};
  \node[draw, left = 2.5cm of A] (B) {$\ppt B$};
  \node[draw, right = 2.5cm of A] (S) {$\ppt S$};
  \node[below =5cm of A] (a) {};
  \node[below =5cm of B] (b) {};
  \node[below =5cm of S] (s) {};
  \draw[->] ($(B.south)!0.03!(b)$)
    -- node[above=-0.7mm,midway]{\footnotesize Suggest}
    ($(A.south)!0.07!(a)$);
  \draw[dashed,draw=blue] ($(B.south)!0.03!(b.north)$)
    --
    ($(S.south)!0.05!(s.north)$);
  \draw[dashed, ->, draw=blue] ($(S.south)!0.05!(s.north)$)
    --
    ($(A.south)!0.07!(a.north)$);
  \draw[->] ($(A.south)!0.10!(a.north)$)
    -- node[above=-0.7mm,midway]{\footnotesize Query}
    ($(S.south)!0.14!(s.north)$);
  \draw[->] ($(S.south)!0.21!(s.north)$)
    -- node[above=-0.7mm,midway]{\footnotesize Available}
    ($(A.south)!0.25!(a.north)$);
  \draw[->] ($(A.south)!0.28!(a.north)$)
    -- node[above=-0.7mm,midway]{\footnotesize Quote}
    ($(B.south)!0.32!(b.north)$);
  \draw[dashed, draw=blue] ($(A.south)!0.28!(a.north)$)
    --
    ($(S.south)!0.30!(s.north)$);
  \draw[dashed, draw=blue, ->] ($(S.south)!0.30!(s.north)$)
    --
    ($(B.south)!0.32!(b.north)$);
  \draw[->] ($(B.south)!0.43!(b.north)$)
    -- node[above=-0.7mm,midway]{\footnotesize OK}
    ($(A.south)!0.47!(a.north)$);
  \draw[dashed,draw=blue] ($(B.south)!0.43!(b.north)$)
    --
    ($(S.south)!0.45!(s.north)$);
  \draw[dashed, ->, draw=blue] ($(S.south)!0.45!(s.north)$)
    --
    ($(A.south)!0.47!(a.north)$);
  \draw[->] ($(A.south)!0.50!(a.north)$)
    -- node[above=-0.7mm,midway]{\footnotesize Confirm}
    ($(S.south)!0.54!(s.north)$);
  \draw[->] ($(B.south)!0.62!(b.north)$)
    -- node[above=-0.7mm,midway]{\footnotesize No}
    ($(A.south)!0.66!(a.north)$);
  \draw[dashed,draw=blue] ($(B.south)!0.62!(b.north)$)
    --
    ($(S.south)!0.64!(s.north)$);
  \draw[dashed, ->, draw=blue] ($(S.south)!0.64!(s.north)$)
    --
    ($(A.south)!0.66!(a.north)$);
  \draw[->] ($(A.south)!0.70!(a.north)$)
    -- node[above=-0.7mm,midway]{\footnotesize Reject}
    ($(S.south)!0.74!(s.north)$);
  \draw[->] ($(S.south)!0.81!(s.north)$)
    -- node[above=-0.7mm,midway]{\footnotesize Full}
    ($(A.south)!0.85!(a.north)$);
  \draw[->] ($(A.south)!0.88!(a.north)$)
    -- node[above=-0.7mm,midway]{\footnotesize Full}
    ($(B.south)!0.92!(b.north)$);
  \draw[dashed, draw=blue] ($(A.south)!0.88!(a.north)$)
    --
    ($(S.south)!0.90!(s.north)$);
  \draw[dashed, draw=blue, ->] ($(S.south)!0.90!(s.north)$)
    --
    ($(B.south)!0.92!(b.north)$);
  \draw (A) -- (a);
  \draw (B) -- (b);
  \draw (S) -- (s);
  \draw[thick]($(B.south)!0.17!(b.north)-(1,0)$)
    rectangle ($(B.south)!0.23!(b.north)-(0.5,0)$) node[pos=.5] {\footnotesize
      \bf alt};
  \draw[thick] ($(B.south)!0.17!(b.north)-(1,0)$)
    rectangle
    ($(S.south)!1.00!(s.north)+(1, 0)$);
  \draw[dashed, thick]
    ($(B.south)!0.78!(b.north)-(1,0)$)
    -- ($(S.south)!0.78!(s.north)+(1, 0)$);
  \draw[thick]($(B.south)!0.38!(b.north)-(0.6,0)$)
    rectangle ($(B.south)!0.44!(b.north)-(0.1,0)$) node[pos=.5] {\footnotesize
      \bf alt};
  \draw[thick] ($(B.south)!0.38!(b.north)-(0.6,0)$)
    rectangle
    ($(S.south)!0.76!(s.north)+(0.6, 0)$);
  \draw[thick, dashed]
    ($(S.south)!0.57!(s.north)+(0.6, 0)$)
    -- ($(B.south)!0.57!(b.north)-(0.6,0)$);
  \draw[densely dotted] ($(B.south)!0.95!(b.north)$) --
    node[above] {\footnotesize \textit{Restart}}
    ($(B.south)!0.95!(b.north)-(1.2, 0)$) --
    ($(B.south)!0.02!(b.north)-(1.2, 0)$);
  \draw[densely dotted, ->]
    ($(B.south)!0.02!(b.north)-(1.2, 0)$) --
    ($(B.south)!0.02!(b.north)$);
\end{tikzpicture}

\vspace{-4mm}
\caption{Travel Agency Protocol as a Sequence Diagram}\label{fig:travel-agency-protocol-dia}
\vspace{-3mm}

\end{figure}

%% file: overview.tex
\section{Overview}\label{sec:overview}
\begin{figure}[t]
    \includegraphics[width=0.9\columnwidth]{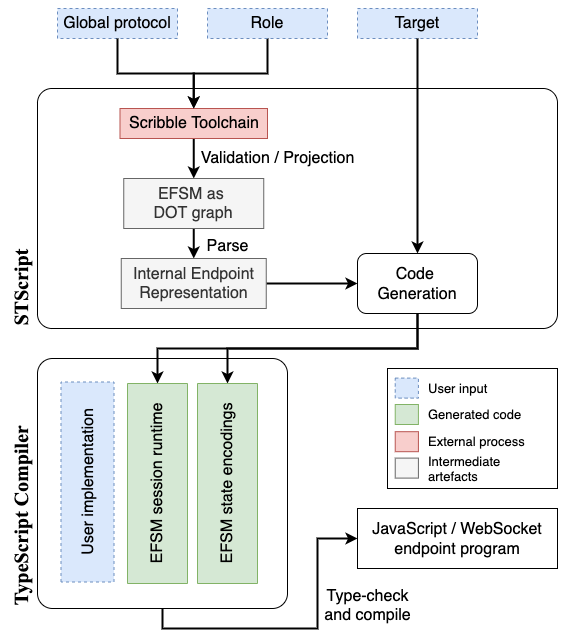}
    \vspace{-4mm}
    \caption{Overview of the toolchain \codegen}
    \label{fig:overview}
    \vspace{-5mm}
\end{figure}

In this section, we give an overview of our code
generation toolchain \codegen (\cref{fig:overview}),
demonstrate how to implement the travel agency scenario
(\cref{fig:travel-agency-protocol-dia})
as a \TS web application, and explain how \codegen prevents those
errors.

\paragraph{Multiparty Session Type Design Workflow}
Multiparty session types (MPST) \cite{JACM16MPST} use a top-down design methodology (\cref{fig:top-down}).
Developers begin with \emph{specifying} the global communication pattern of all
participants in a \emph{global type} or a \emph{global protocol}.
The protocol is described in the Scribble protocol description
language~\cite{ScribbleWebsite,TGC13Scribble,ScribbleBookChapter}.
We show the global protocol of the travel agency scenario (in \cref{sec:intro}) in \cref{fig:travel-agency-protocol}.

The Scribble language provides a user-friendly way to describe the global
protocol in terms of a sequence of message exchanges between roles.
A message is identified by its label
(e.g.\ \code{Suggest}, \code{Query}, etc), and carries payloads (e.g.\
\code{number}, \code{string}, etc).
The \keyword{choice} syntax (e.g.\ \cref{line:choice-1})
describes possible branches of the protocol -- in this case, the Server may
respond to the query either with \code{Available}, so the customer continues
booking, or with \code{Full}, so the customer retries by restarting the protocol
via the \keyword{do} syntax (\cref{line:do}).

In this scenario, we designate the roles $\ppt{A}$
and $\ppt{B}$ as \emph{client
  roles}, and role $\ppt{S}$ as a \emph{server role}.
Participating endpoints can obtain their local views of the communication
protocol, known as \emph{local types}, via \emph{projection} from the specified
global type (\cref{fig:top-down}).
The local type of an endpoint can be then used in the code generation process,
to generate APIs that are \emph{correct by
  construction}~\cite{FASE16EndpointAPI, PLACES19PureScript, OOPSLA20FStar}.

\input{figures/top-down.tex}
\input{figures/travel-agency.tex}

The code generation toolchain \codegen
(\cref{fig:overview}) follows the MPST design philosophy.
In \codegen, we take the global protocol as inputs, and generate
endpoint code for a given role as outputs, depending on the nature of the role.
We use the Scribble toolchain for
initial processing, and use an \emph{endpoint finite state machine} (EFSM)
based code generation technique targeting the \TS Language.

\paragraph{Targeting Web Programming}
The \TS~\cite{ECOOP14TypeScript} programming language is used for web
programming, with a static type system and a compiler to JavaScript.
\TS programs follow a similar syntax to JavaScript, but may contain type
annotations that are checked statically by the \TS type-checker.
After type-checking, the compiler converts \TS programs into
JavaScript programs, so they can be run in browsers and other hosts (e.g.\
\nodejs).

To implement a wide variety of communication patterns, we use the
\emph{WebSocket} protocol~\cite{RFCWebSocket}, enabling bi-directional
communication between the client and the server after connection.
This contrasts with the traditional request-response model of HTTP,
where the client needs to send a request and the server may only send a response after
receiving the request.
WebSockets require an endpoint to listen for
connections and the other endpoint connecting.
Moreover, clients, using the web application in a browser, may \emph{only}
start a connection to a WebSocket, and servers may \emph{only} listen for new
connections.
The design of WebSocket limits the ability for two clients to communicate
directly via a WebSocket (e.g.\ \cref{line:suggest} in
\cref{fig:travel-agency-protocol}).
\codegen uses the server to \emph{route} messages between client
roles, enabling communication between all participants via a star network
topology.

An important aspect of web programming is the interactivity of the user
interface (UI).
Viewed in a browser, the web application interacts with the user via UI events,
e.g.\ mouse clicks on buttons.
The handling of UI events may be implemented to send messages to the client
(e.g.\ when the ``Submit'' button on the form is clicked), which may lead to
practical problems.
For instance, would clicking ``Submit'' button twice create two bookings for
the customer?
We use the popular \emph{\reactjs} UI framework for generating client
endpoints, and generate APIs that prevent such errors from happening. %

\paragraph{Callback-Style API for Clients and Servers}
Our code generation toolchain \codegen produces \TS APIs in
a \emph{callback style}~\cite{OOPSLA20FStar} to
\emph{statically} guarantee channel linearity.
The input global protocol is analysed by the toolchain for well-formedness, and
an \emph{endpoint finite state machine} (EFSM) is produced for each endpoint.
We show the EFSM for role $\ppt{A}$ in \cref{fig:cfsm-a}.
The states in the EFSM represent local types (subject to reductions) and
transitions represent communication actions (Symbol $!$ stands for
sending actions, $?$ for receiving).

\input{figures/cfsm-a.tex}

In the callback API style, type signatures of callbacks are generated for
transitions in the EFSM.
Developers implement the callbacks to complete the program logic part of the
application, whilst a generated \emph{runtime} takes care of the communication
aspects.
For callbacks, sending actions correspond to callbacks prompting the payload
type as a \emph{return type}, so that the returned value can be sent by
the runtime.
Dually, receiving actions correspond to callbacks taking the payload type as an
\emph{argument}, so that the runtime invokes the callback with the received
value.

\paragraph{Implementing the Server Role}
In the travel agency protocol, as shown in \cref{fig:travel-agency-protocol},
we designate role $\ppt{S}$ as the server role.
The server role does not only interact with the two clients, but also
\emph{routes} messages for the two clients.
The routing will be handled automatically by the runtime, saving the need for
developers to specify manually.
As a result, the developer only handles the program logic regarding the
server, in this use case, namely providing quotes for holiday bookings and
handling booking confirmations.

\begin{lstlisting}[language=TypeScript]
import { Session, S } from "./TravelAgency/S";
const agencyProvider = (sessionID: string) => {
  const handleQuery = Session.Initial({
    Query: async (Next, dest) => { £\label{line:server-query}£
      // Provide quotes for holiday bookings
      const res = await checkAvailability(sessionID, dest);
      if (res.status === "available") {
        return Next.Available([res.quote], Next => ...);
      } else { return Next.Full([], handleQuery); } }, });
  return handleQuery; };
\end{lstlisting}

\begin{figure}[t]
    \includegraphics[width=\columnwidth]{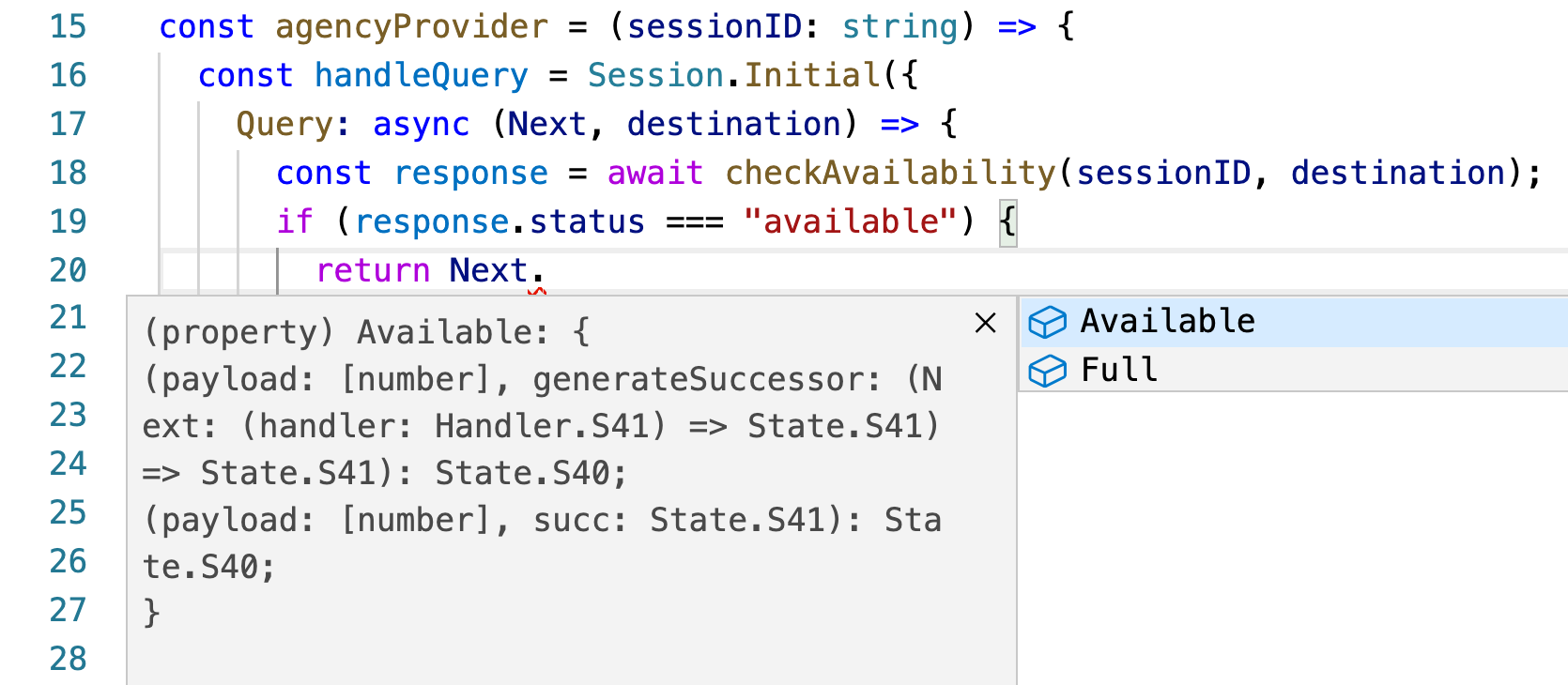}
    \vspace{-6mm}
	\caption{IDE Auto-Completion for Successor State}\label{fig:ide-autocomplete}
    \vspace{-6mm}
\end{figure}

All callbacks carry an extra parameter, \code{Next}, which acts as
a \emph{factory function} for constructing the successor state.
This empowers IDEs to provide auto-completion for developers.
For example, the factory function provided by the callback for
handling a \tmsg{Query} message (\cref{line:server-query})
prompts the permitted labels in the successor send state,
as illustrated in \cref{fig:ide-autocomplete}.

\paragraph{Implementing the Client Roles}
To implement client roles, merely implementing the callbacks for the program
logic is not sufficient --- unlike servers, web applications have interactive
user interfaces, additional to program logic.
As mentioned previously, our code generation toolchain targets \reactjs
for client roles.
For background, the smallest building blocks in \reactjs are \emph{components},
which can carry \emph{properties} (immutable upon construction) and
\emph{states} (mutable).
Components are \emph{rendered} into HTML elements, and they are re-rendered
when the component state mutates.

To bind the program logic with an interactive user interface, we provide
\emph{component factories} that allow the UI component to be interposed with
the current state of the EFSM.
Developers can provide the UI event handler to the component factory, and
obtain a component for rendering.
The generate code structure enforces that the state transition strictly follows
the EFSM, so programmer errors (such as the double ``submit'' problem) are
prevented by design.

\begin{lstlisting}[language=TypeScript]
render() {
  const OK = this.OK('onClick', () => [this.state.split]); £\label{line:component-factory}£
  const NO = this.No('onClick', () => []);
  return (...
    <NO><Button color='secondary'>No</Button></NO>
    <OK><Button color='primary'>OK</Button></OK> £\label{line:using-factory}£...); }
\end{lstlisting}

Using the send state component in the FSM
for the endpoint $\ppt{B}$ as an example,
\cref{line:component-factory} reads,
``generate a React component that sends the
\tmsg{OK} message with \code{this.state.split}
as payload on a click event''.
It is used on \cref{line:using-factory} as a wrapper
for a stylised \code{<Button>} component.
The runtime invokes the handler and
performs the state transition, which
prevents the double ``submit'' problem by design.

\paragraph{Guaranteeing Communication Safety}
Returning to the implementation in \cref{sec:intro},
we outline how \codegen prevents common errors
to enable type-safe web programming.

\begin{description}[leftmargin=0pt]
  \item[Communication Mismatch]
All generated callbacks are typed according to the
permitted payload data type
specified in the protocol, making it impossible
for traveller $\ppt{A}$ to send the quote as a string by accident.

\item[Channel Usage Violation]
The generated client-side runtime requires the developer
to provide different UI components for each EFSM state --
once traveller $\ppt{A}$ submits a quote, the runtime will
transition to, thus render the component of, a different
EFSM state. This guarantees that, whilst waiting for a response
from traveller $\ppt{B}$, it is impossible for traveller $\ppt{A}$
to submit another quote and violate channel linearity.

\item[Handling Session Cancellation]
If either traveller closes their browser before the protocol
runs to completion, the generated runtimes leverage the
events available on their WebSocket connections to notify
(via the server) other roles about the session cancellation.
The travel agency can implement the error handler
callback (generated by \codegen) to perform clean-up logic
in response to cancellations.
\end{description}

%% file: figures/top-down.tex
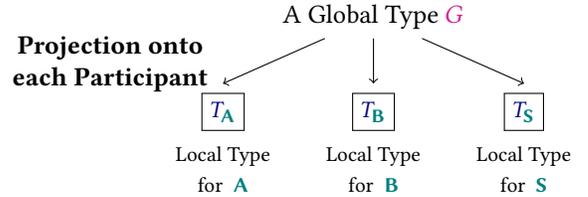
\begin{figure}[t]
  \begin{tikzpicture}
    \node (Gtext) {A Global Type $\gtype G$};
    \node[below=-1.5mm of Gtext, xshift=-35mm, align=center] (proj) {\bf
      Projection onto \\ \bf each
      Participant};
    \node[below=6mm of Gtext, xshift=-20mm, align=center] (LA) {\small
      \boxed{\TA} \\[1ex] \footnotesize Local Type \\ \footnotesize for \ppt A};
    \node[below=6mm of Gtext, align=center] (LB) {\small \boxed{\TB} \\[1ex]
      \footnotesize Local Type \\ \footnotesize for \ppt B};
    \node[below=6mm of Gtext, xshift=20mm, align=center] (LC) {\small
      \boxed{\ltype{T_{\ppt S}}} \\[1ex] \footnotesize Local Type \\
      \footnotesize for \ppt S};
    \draw[->] (Gtext) -- (LA.north);
    \draw[->] (Gtext) -- (LB.north);
    \draw[->] (Gtext) -- (LC.north);
  \end{tikzpicture}
  \vspace{-3mm}
  \caption{Top-down MPST Design Methodology}\label{fig:top-down}
  \vspace{-4mm}
\end{figure}

%% file: figures/travel-agency.tex
\begin{figure}[t]
\begin{lstlisting}[language=Scribble]
global protocol TravelAgency(role A, role B, role S)
{ Suggest(string) from B to A;//friend suggests place £\label{line:suggest}£
  Query(string) from A to S;
  choice at S £\label{line:choice-1}£
     { Available(number) from S to A;
       Quote(number) from A to B;//check price with friend
       choice at B £\label{line:choice-2}£
           { OK(number) from B to A;
             Confirm(credentials) from A to S; }
       or  { No() from B to A;
             Reject() from A to S; } }
  or { Full() from S to A; Full() from A to B;
       do TravelAgency(A, B, S); £\label{line:do}£} }
\end{lstlisting}

\vspace{-5mm}
\caption{Travel Agency Protocol in Scribble}\label{fig:travel-agency-protocol}
\vspace{-5mm}
\end{figure}

%% file: figures/cfsm-a.tex
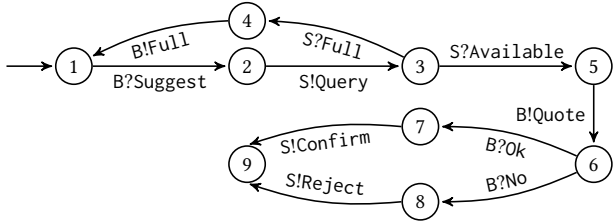
\begin{figure}
  \footnotesize
  \begin{tikzpicture}[->,>=stealth',shorten >=1pt,auto,node distance=1.8cm,
      semithick]
      \draw node[] (start) at (-1, 0) {};
      \draw node[circle, draw] (1) at (0, 0) {1};
      \draw node[circle, draw, right=of 1] (2) {2};
      \draw node[circle, draw, right=of 2] (3) {3};
      \draw node[circle, draw, above=0.1cm of 2] (4) {4};
      \draw node[circle, draw, right=of 3] (5) {5};
      \draw node[circle, draw, below=0.8cm of 5] (6) {6};
      \draw node[circle, draw, left=of 6, yshift=5mm] (7) {7};
      \draw node[circle, draw, left=of 6, yshift=-5mm] (8) {8};
      \draw node[circle, draw, left=of 7, yshift=-5mm] (9) {9};
      \path (start) edge node {} (1);
      \path (1) edge node[below] {\code{B}$?$\code{Suggest}} (2);
      \path (2) edge node[below] {\code{S}$!$\code{Query}} (3);
      \path (3) edge [bend right=12] node[below, sloped] {\code{S}$?$\code{Full}}
      (4.east);
      \path (4.west) edge [bend right=12] node[below, sloped] {\code{B}$!$\code{Full}} (1);
      \path (3) edge node {\code{S}$?$\code{Available}} (5);
      \path (5) edge node[left] {\code{B}$!$\code{Quote}} (6);
      \path (6) edge [bend right=12] node[below, sloped] {\code{B}$?$\code{Ok}}
      (7.east);
      \path (6) edge [bend left=12] node[sloped] {\code{B}$?$\code{No}}
      (8.east);
      \path (7.west) edge [bend right=12] node[below, sloped]
      {\code{S}$!$\code{Confirm}} (9.north);
      \path (8.west) edge [bend left=12] node[sloped] {\code{S}$!$\code{Reject}}
      (9.south);
  \end{tikzpicture}
  \vspace{-4mm}
  \caption{EFSM for \code{TravelAgency} role $\ppt{A}$}\label{fig:cfsm-a}
  \vspace{-5mm}
\end{figure}

%% file: implementation.tex
\section{Implementation}\label{sec:implementation}

In this section,
we explain how the generated code executes
the EFSM
for \nodejs and \reactjs targets.
We also present how \codegen APIs
handle errors in a dynamic web-based environment
(for complete code, see \fullversion{\cref{app:code}}{the full paper}).

\paragraph{Session Runtime}
The session runtime executes
the EFSM in a manner permitted by
the multiparty protocol description.
The runtime keeps track of the current state,
performs the required communication action (i.e. send or receive a message),
and transitions to the successor state.
The runtime provides \emph{seams} for the
developer to inject the callback implementations,
which define application-specific concerns for the EFSM,
such as what message payload to send (and dually,
how to process a received message).
This design conceals the WebSocket APIs from the developer
and entails that the developer cannot
trigger a send or receive action, so \codegen can
\emph{statically} guarantee protocol conformance.

\paragraph{Executing the  EFSM in \nodejs}
Each state of the EFSM is characterised by a (generated) \code{State} class
and a type describing the shape of the callback (supplied by the developer).
To allow the server to correctly manage concurrent sessions,
the developer can access a (generated) \emph{session ID}
when implementing the callbacks.
\codegen also generates \emph{IO interfaces} for each kind of EFSM state
-- send, receive, or terminal.
The generated \code{State} class implements the interface
corresponding to the type of communication action it performs.

\vspace{-2mm}
\begin{lstlisting}[language=TypeScript]
next(state: State.Type) {
  switch (state.type) {
  case 'Send': return state.performSend(
    this.next, this.cancel, this.send);
  case 'Receive': return state.prepareReceive(
    this.next, this.cancel, this.registerMessageHandler);
  case 'Terminal': return; }}
\end{lstlisting}
\vspace{-2mm}

The session runtime for \nodejs is a class that executes
the EFSM using a \emph{state transition function}
parameterised by the \code{State} class of the current EFSM state.
As the IO interfaces constitute a \emph{discriminated union},
the runtime can parse the type of the current EFSM state
and propagate the appropriate IO functions (for sending
or receiving) to the \code{State} class.
In turn, the \code{State} class invokes the callback supplied
by the developer to inject program logic into
the EFSM, perform the communication action (using \code{this.send}
or \code{this.registerMessageHandler}), and invoke the state
transition function (\code{this.next}) with the successor state.

Notably, the routed messages are completely absent because
the generated code transparently routes messages without
exposing any details.
As messages specify their intended recipient,
the runtime identifies messages not intended for the server
by inspecting the metadata, and forwards them to the
WebSocket connected to the intended recipient.

\paragraph{Executing the EFSM in \reactjs}
Each state in the EFSM is encoded as an
\emph{abstract} React component.
The developer implements the EFSM by extending
the abstract classes to provide their own
implementation -- namely, to build their
user interface.
Components for send states can access
\emph{component factories} to generate
React components that perform a send action
when a UI event (e.g. \code{onClick}, \code{onMouseOver}) is triggered.
Components for receive states must implement
abstract methods to handle all possible incoming messages.

The session runtime for \reactjs
is a React component, instantiated using
the developer's implementation of each EFSM state.
Channel communications are managed by the runtime,
so the developer's implementations cannot access
the WebSocket APIs, which prevents channel reuse
by construction.
The runtime renders the component of the current
EFSM state and binds the permitted communication action
through supplying component properties.

\paragraph{Error Handling}
An error handling mechanism is critical
for web applications.
Clients can disconnect from the session due to
network connectivity issues or simply by closing the browser.
Similarly, servers may also face connectivity issues.

Upon instantiating the session runtime,
\codegen requires developers to supply a
\emph{cancellation handler} to
handle \emph{local exceptions}
(e.g. errors thrown by application logic) and
\emph{global session cancellations}
(e.g. disconnection events by another endpoint).
The session runtime detects cancellation
by listening to the \emph{close event} on the WebSocket
connection, and invokes the cancellation
handler with appropriate arguments on a premature
close event.
We parameterise the cancellation handlers with
additional information (e.g. which role disconnected
from the session, the reason for the disconnection)
to let developers be more
specific in their error handling logic.

\paragraph{Cancellation Handlers for Servers}
Server endpoints define cancellation handlers
through a function, parameterised by the \emph{session ID},
the \emph{role} which initiated the cancellation, and
(optionally) the \emph{reason} for the cancellation ---
if the server-side logic throws an exception, the
handler can access the thrown error through
the \code{reason} parameter.

\vspace{-2mm}
\begin{lstlisting}[language=TypeScript]
const handleCancel = async (sessionID, role, reason) => {
  if (role === Role.Self) {
    console.error(`${sessionID}: internal server error`); }
  else { await tryRelease(sessionID); }}; £\label{line:svr-cancel}£
// Instantiate session runtime
new S(wss, handleCancel, agencyProvider);
\end{lstlisting}
\vspace{-2mm}

Using the \tprotocol{Travel Agency} scenario
introduced in \cref{sec:intro}, if the customer
prematurely closes their browser before responding to
a \tmsg{Quote}, the server can detect
this (\cref{line:svr-cancel}) and release the reservation
to preserve data integrity.

\paragraph{Cancellation Handlers for Clients}
Browser-side endpoints also define cancellation handlers
through a function parameterised in the same way as those
in \nodejs, but
must return a React component to be rendered by the
session runtime.
In the context of the \tprotocol{Travel Agency} scenario,
the customer can render a different UI depending on
whether the server disconnected or their friend closed
their web browser prematurely.
Browser endpoints can also respond to cancellations emitted by other client-side roles:
when a browser endpoint disconnects,
the server detects this and propagates the cancellation to the
other client-side roles.

%% file: theory.tex
\section{\newtheory: Routed Session Types}\label{sec:theory}
This section defines the syntax and semantics
of \newtheory and proves some important properties.
We show the sound and complete
trace correspondence between
a global type and a collection of endpoint types
projected from the global type (Theorem~\ref{th:traceeq}).
Using this result, we prove deadlock freedom (Theorem~\ref{th:deadlockfreedom}).
We then show that, in spite of the added routed communications,
\newtheory does not over-serialise communications
by proving \emph{communication preservations}
between the original MPST and \newtheory (Theorem~\ref{th:encwf}).
These three theorems ensure that \codegen endpoint programs
are communication-safe, always make progress, and
correctly conforms to the user-specified protocol.

\input{theory/syntax}

\input{theory/semantics}

\input{theory/relating}
\input{theory/encoding}

%% file: theory/syntax.tex
\subsection{Syntax of Routed Multiparty Session Types}
\label{subsec:syntax}
We define the syntax of \emph{global types} $\gtype{G}$ and \emph{local types}
(or \emph{endpoint types})
$\ltype{T}$ in \cref{def:syntax}.
Global types are also known as \emph{protocols}
and describe the communication behaviour between all
participating roles (participants), while
local types describe
the behaviour of a single participating role.
We shade additions to the original (or \emph{canonical})
multiparty session type (MPST)
\cite{POPL19MPST,DBLP:journals/mscs/CoppoDYP16,JACM16MPST,2013Automata}
in \colorbox{ColourShade}{this colour}.
\vspace{-1mm}
\begin{definition}[Global and Local Types]
The syntax of \emph{global} and \emph{local types} are defined below:
\label{def:syntax}
\vspace{-1mm}
\[
\small
\begin{array}{ll}
\hspace{-3mm}
\arraycolsep=1pt
\begin{array}{rcl}
  \gtype{G} & ::= & \gtend
     \ \mid \ \gtvar{t}
     \ \mid \ \gtrecur{t}{G}\\
    & \mid & \gcomm{p}{q}{\lbl{l_i}\!:\! G_i}{\lbl{i \in I}}\\
  \rowcolor{ColourShade}
  & \mid &
  \groute{p}{q}{s}{\lbl{l_i}\!:\! G_i}{\lbl{i \in I}}
\end{array} &
\hspace{-5mm}
\arraycolsep=1pt
\begin{array}{rcl>{\columncolor{ColourShade}}l}
  \ltype{T} & ::= & \tend
     \ \mid \ \trecvar
     \ \mid \ \trec{T}
    &\ \mid \ \router{p}{q}{\lbl{l_i}\!:\! T_i}{\lbl{i \in I}}\\
    & \mid & \tsel{p}{\lbl{l_i}\!:\! T_i}{\lbl{i \in I}}
    &\ \mid \ \tselproxy{p}{q}{\lbl{l_i}\!:\! T_i}{\lbl{i \in I}}\\
    & \mid & \tbra{p}{\lbl{l_i}\!:\! T_i}{\lbl{i \in I}}
    &\ \mid \ \tbraproxy{p}{q}{\lbl{l_i}\!:\! T_i}{\lbl{i \in I}}
\end{array}
\end{array}
\]
\end{definition}

\vspace{-2mm}
\paragraph{Global Types}
$\gcomm{p}{q}{\lbl{l_i}\!:\! G_i}{\lbl{i \in I}}$ describes
a \textbf{\emph{direct communication}} of
a message $l_i$ from a
role $\ppt p$ to $\ppt q$.
We require that $\ppt p \neq \ppt q$, that labels $\lbl{l_i}$ are pairwise
distinct, and that the index set $\lbl{I}$ is not empty.
The message in the communication can carry a label among a set of permitted
labels $\lbl{l_i}$ and some payload.
After a message with label $\lbl{l_i}$ is received by $\ppt q$,
the communication continues with
$\gtype{G_i}$, according to the chosen label.
For simplicity, we do not include payload types
(integers, strings, booleans, etc)
in the syntax.
We write $\gcommone{p}{q}{\lbl{l}:G}$ for single branches.
For recursion, we adopt an \emph{equi-recursive} view \cite[\S 21]{TAPL}, and
use $\gtrecur{t}{G}$ and $\gtvar{t}$ for a \textbf{\emph{recursive protocol}} and
a \textbf{\emph{type variable}}.
We require that recursive types are \emph{contractive
(guarded)}, i.e.\ the recursive type $\gtrecur{t}{G}$ progresses after the
substitution $\gtype{G[\gtrecur{t}{G}/\gtvar{t}]}$, prohibiting types such as
$\gtrecur{t}{\gtvar{t}}$.
We use $\gtend$ to mark the \textbf{\emph{termination}} of the protocol, and often omit the
final $\gtend$.

To support routed communication, we allow messages to be sent
through a \emph{router role}.
A \textbf{\emph{routed communication}}
$\groute{p}{q}{s}{\lbl{l_i}\!:\! G_i}{\lbl{i \in I}}$
describes a router role $\ppt s$ coordinating the communication
of a message from $\ppt p$ to $\ppt q$:
$\ppt q$ offers $\ppt p$ a choice in the index set $\lbl{I}$,
but $\ppt p$ sends the selected choice $\lbl{l_i}$ to the router
$\ppt s$ instead.
The router \emph{forwards} the selection from $\ppt p$ to $\ppt q$.
After $\ppt q$ receives $\ppt p$'s selection,
the communication continues with $\gtype{G_i}$.
$\ppt s$ ranges over the set of roles $\ppt p, \ppt q, \cdots$,
but we use $\ppt s$ by convention as the router is usually some server.
The syntax for routed communication shares the same properties
as direct communication, but we additionally require that
$\ppt p \neq \ppt q \neq \ppt s$.
We use $\pt{\gtype{G}}$ to denote the set of participants
in the global type $\gtype{G}$.

\begin{example}[Travel Agency]
  The travel agency protocol, as shown in \cref{fig:travel-agency-protocol},
  is described by the global type $\gtype{G_{\text{travel}}}$ in the
  original MPST, and $\gtype{G^R_{\text{travel}}}$ in \newtheory.
 \vspace{-1mm}
  \[
  \small
    \vspace{-1mm}
    \arraycolsep=2pt
    \begin{array}{l}
      \gtype{G_{\text{travel}}} =
      \gtrecur{t}{} \begin{array}[t]{@{}l@{}}
      \gcommone{B}{A}{\lbl{Suggest}} . \, \, \gcommone{A}{S}{\lbl{Query}} . \\
      \gcomm{S}{A}{
        \begin{array}{l}
          \lbl{Available}:\\
          \hspace{3mm}\begin{array}{l}
              \gcommone{A}{B}{\lbl{Quote}} . \, \,
              \gcommone{B}{A}{} \\
              \hspace{3mm}  \left\{\begin{array}{l}
                  \lbl{OK}: \gcommone{A}{S}{\lbl{Confirm}} \\
                  \lbl{No}: \gcommone{A}{S}{\lbl{Reject}}
                \end{array}\right\}
              \\
            \end{array} \\
          \lbl{Full}: \gcommone{A}{B}{\lbl{Full}}.\gtvar{t}
        \end{array}
      }{}
      \end{array} \\
      \gtype{G^R_{\text{travel}}} =
      \gtrecur{t}{} \begin{array}[t]{@{}l@{}}
      \grouteone{B}{A}{S}{\lbl{Suggest}} . \, \, \gcommone{A}{S}{\lbl{Query}} . \\
      \gcomm{S}{A}{
        \begin{array}{l}
          \lbl{Available}:\\
          \hspace{3mm}\begin{array}{l}
              \grouteone{A}{B}{S}{\lbl{Quote}} . \, \,
              \grouteone{B}{A}{S}{} \\
              \hspace{3mm}  \left\{\begin{array}{l}
                  \lbl{OK}: \gcommone{A}{S}{\lbl{Confirm}} \\
                  \lbl{No}: \gcommone{A}{S}{\lbl{Reject}}
                \end{array}\right\}
              \\
            \end{array} \\
          \lbl{Full}: \grouteone{A}{B}{S}{\lbl{Full}}.\gtvar{t}
        \end{array}
      }{}
      \end{array}
    \end{array}
  \]
\end{example}

\paragraph{Local Types}
We first describe the local types in the original MPST theory.
$\tbra{q}{\lbl{l_i}\!:\! T_i}{\lbl{i \in I}}$ stands for
\textbf{\emph{branching}}
and
$\tsel{q}{\lbl{l_i}\!:\! T_i}{\lbl{i \in I}}$ stands for
\textbf{\emph{selection}}.
From the perspective of $\ppt p$, branching (resp.\ selection) offers (resp.\
selects) a choice among an index set $\lbl{I}$ to (resp.\ from) $\ppt q$, and
communication continues with the corresponding $\ltype{T_i}$.
Local types $\trecur{t}{T}$, $\tvar{t}$ and $\tend$ have the same meaning
as their global type counterparts.

We add new syntax to express routed communication
from the perspective of each role involved.
The local type
$\tbraproxy{p}{s}{\lbl{l_i}\!:\! T_i}{\lbl{i \in I}}$
is a \textbf{\emph{routed branching}}:
the current role is offering a choice from an index set $\lbl{I}$
to $\ppt p$ (the intended sender),
but expects to receive
$\ppt p$'s choice via the router role $\ppt s$;
if the message received is labelled $\lbl{l_i}$,
$\ppt q$ will continue with local type $\ltype{T_i}$.
The local type
$\tselproxy{q}{s}{\lbl{l_i}\!:\! T_i}{\lbl{i \in I}}$
is a \textbf{\emph{routed selection}}:
the current role makes a selection from an index set $\lbl{I}$
to $\ppt q$ (the intended recipient),
but sends the selection to the router role $\ppt s$;
if the message sent is labelled $\lbl{l_i}$,
$\ppt p$ will continue with local type $\ltype{T_i}$.
The local type
$\router{p}{q}{\lbl{l_i}\!:\! T_i}{\lbl{i \in I}}$ is
a \textbf{\emph{routing communication}}.
The router role %
orchestrates the communication from
$\ppt p$ to $\ppt q$, and continues with local type $\ltype{T_i}$
depending on the label of the forwarded message.
We keep track of the router role to distinguish
between routing communications from normal selection and branching
interactions.

\input{figures/global-lts}
\paragraph{Endpoint Projection}
The local type $\ltype{T}$ of a participant $\ppt p$ in a global type $\gtype{G}$
is obtained by the \emph{endpoint projection} of $\gtype{G}$ onto
$\ppt p$, denoted by $G$ as $\proj{G}{p}$.

\vspace{-1mm}
\begin{definition}[Projection]
\label{def:projection}
The projection of $\gtype{G}$ onto $\ppt r$, written
$\proj{G}{r}$ is defined as: \\[1mm]
\hspace*{-4mm}
$
\small
\begin{array}{ll}
\begin{array}{l}
\proj{\left(\groute{p}{q}{s}{\lbl{l_i}\!:\! G_i}{\lbl{i \in I}}\right)}{r}\\
=
\begin{cases}
\tselproxy{q}{s}{\lbl{l_i}\!:\! \proj{G_i}{r}}{\lbl{i \in I}}
	& \text{if } \ppt r = \ppt p \\
\tbraproxy{p}{s}{\lbl{l_i}\!:\! \proj{G_i}{r}}{\lbl{i \in I}}
	& \text{if } \ppt r = \ppt q \\
\router{p}{q}{\lbl{l_i}\!:\! \proj{G_i}{r}}{\lbl{i \in I}}
	& \text{if } \ppt r = \ppt s \\
\MERGEOP_{\lbl{i \in I}}~\proj{G_i}{r}
	& \text{otherwise}\\
\end{cases}
\end{array}
\begin{array}{l}
\proj{(\gtrecur{t}{G})}{r}\\
= \
\begin{cases}
    \gtrecur{t}{(\proj{G}{r})} & \text{if }\proj{G}{r}\not = \tvar{t'}\\
    \gtend & \text{otherwise}\\
\end{cases}\\[1mm]
\begin{array}{rcl}
\proj{\gtend}{r}  & = &  \tend\\
\proj{\gtvar{t}}{r} & = &\ \tvar{t}
\end{array}
\end{array}
\end{array}
$\\
The projection
$\proj{(\gcomm{p}{q}{\lbl{l_i}\!:\! G_i)}{\lbl{i \in I}}}{r}$
is defined similar to
$\proj{\left(\groute{p}{q}{s}{\lbl{l_i}\!:\! G_i}{\lbl{i \in I}}\right)}{r}$
dropping $\ppt s$ (in the resulting local type) and the third case.
\end{definition}
\vspace{-1mm}
A \emph{merge operator} ($\mergeop$) is used
when projecting a communication
onto a non-participant. It checks that the projections of
all continuations must be ``compatible''
(see \fullversion{\cref{def:newmerge}}{the full paper}).
\vspace{-1mm}
\begin{example}[Merging Local Types]\label{ex:globaltypes}
  Two branching types from the same role with disjoint labels can merged into a
  type carrying both labels, e.g.\ $\tbraone{A}{\lbl{Hello}}.\tend \,
  \mergeop\, \tbraone{A}{\lbl{Bye}}.\tend = \tbra{A}{\lbl{Hello}: \tend;
    \lbl{Bye}: \tend}{}$.
  The same is not true for selections, $\tselone{A}{\lbl{Hello}}.\tend \,
  \mergeop\, \tselone{A}{\lbl{Bye}}.\tend$ is undefined.
  \vspace{-1mm}
  \[
  \small
  \vspace{-1mm}
    \begin{array}{l}
      \gtype{G_1} = \gcomm{A}{B}{
        \begin{array}{@{}l@{}}
          \lbl{Greet}:\gcommone{A}{C}{\lbl{Hello}}\,.\,\gtend \\
          \lbl{Farewell}:\gcommone{A}{C}{\lbl{Bye}}\,.\,\gtend
        \end{array}
      }{}\\
      \gtype{G_2} = \gcomm{A}{B}{
        \begin{array}{@{}l@{}}
          \lbl{Greet}:\gcommone{C}{A}{\lbl{Hello}}\,.\,\gtend \\
          \lbl{Farewell}:\gcommone{C}{A}{\lbl{Bye}}\,.\,\gtend
        \end{array}
      }{}
    \end{array}
  \]
  The global type $\gtype{G_1}$ can be projected to role $\ppt
  C$, but not $\gtype{G_2}$.
\end{example}

\paragraph{Well-formedness}
In the original theory, a global type $\gtype{G}$ is \emph{well-formed}
(or \emph{realisable}), denoted $\wf{\gtype{G}}$,
if the projection is defined for all its participants.
\vspace{-1mm}
\[
\vspace{-1mm}
\wf{\gtype{G}} \defeq
\forall \ppt p \in \pt{\gtype{G}}. ~ \proj{G}{p} \text{ exists}
\]
We assume that the global type $\gtype{G}$ is contractive (guarded).

In \newtheory, we say that a global type is
well-formed \emph{with respect to the role $\ppt s$
acting as the router}.
We define the characteristics that $\ppt s$
must display in $\gtype{G}$ to prove that it is a
router,
and formalise this as an \emph{inductive} relation,
$\centroid{G}{s}$ (\cref{def:centroid}), which reads \emph{$\ppt s$ is a
centroid in $\gtype{G}$}.
The intuition is that $\ppt s$ is at the centre
of all communication interactions.

\vspace{-1mm}
\begin{definition}[Centroid]
The relation $\centroid{G}{s}$ ($\ppt s$ is the
centroid of $\gtype{G}$) is defined by the
two axioms $\centroid{\gtend}{s}$ and $\centroid{\gtvar{t}}{s}$
and by the following rules:
{\small
\begin{mathpar}
\myinferrule%
{\centroid{G}{s}}
{\centroid{\gtrecur{t}{G}}{s}}
\myinferrule%
{
  \ppt{s} \in \left\{ \ppt{p}, \ppt{q} \right\}
  \
  \forall \lbl{i \in I}. ~ \centroid{G_i}{s}
}
{\centroid{\gcomm{p}{q}{l_i: G_i}{\lbl{i \in I}}}{s}}
\myinferrule%
{
  \mrole{r} = \mrole{s}
  \and
  \forall \lbl{i \in I}. ~ \centroid{G_i}{s}
}
{\centroid{\groute{p}{q}{r}{\lbl{l_i}\!:\! G_i}{\lbl{i \in I}}}{s}}
\vspace{-6mm}
\end{mathpar}
\vspace{-1mm}
}
\label{def:centroid}
\end{definition}
For direct communication, $\ppt{s}$ must
be a participant and a centroid of all continuations.
For routed communication, $\ppt{s}$ must
be the router and be a centroid of all continuations.
Now we define of well-formedness of a global type $\gtype G$ in \newtheory with
respect to the router $\ppt s$ (denoted $\wfnew{\gtype G}{s}$):
\vspace{-1mm}
\[
\vspace{-1mm}
\wfnew{\gtype{G}}{s} \defeq
(\forall \ppt p \in \pt{\gtype{G}}. ~ \proj{G}{p} \text{ exists})
\wedge
\centroid{G}{s}
\]

%% file: figures/global-lts.tex
\begin{figure*}[!t]
  \small
\begin{minipage}{0.48\textwidth}
\begin{prooftree}
\AxiomC{}
\RightLabel{\rulename{Gr1}}
\UnaryInfC{$
\treducelong
	{\gcomm{p}{q}{\lbl{l_i}\!:\! G_i}{i \in I}}
	{\gtrans{p}{q}{j}{\lbl{l_i}\!:\! G_i}{i \in I}}
	{\aout{p}{q}{j}}
$}
\end{prooftree}

\begin{prooftree}
\hspace{-8mm}
\AxiomC{}
\RightLabel{\rulename{Gr2}}
\UnaryInfC{$
\gtreducelong
	{\gtrans{p}{q}{j}{\lbl{l_i}\!:\! G_i}{\lbl{i \in I}}}
	{G_j}
	{\ain{p}{q}{j}}
        $}
\end{prooftree}

\begin{prooftree}
\AxiomC{$\forall \lbl{i \in I}. ~ \gtreduce{G_i}{G'_i}{l}$}
\AxiomC{$\subj{\lbl{l}} \notin \{\ppt{p}, \ppt{q}\}$}
\RightLabel{\rulename{Gr4}}
\BinaryInfC{$
\gtreducelong
	{\gcomm{p}{q}{\lbl{l_i}\!:\! G_i}{\lbl{i \in I}}}
	{\gcomm{p}{q}{\lbl{l_i}\!:\! G'_i}{\lbl{i \in I}}}
	{l}
$}
\end{prooftree}

\begin{prooftree}
\AxiomC{$\gtreduce{G_j}{G'_j}{l}$}
\AxiomC{$\subj{\lbl{l}} \neq \ppt{q}$}
\AxiomC{$\forall \lbl{i \in I \setminus \{ j \}} . ~ \gtype{G'_i} = \gtype{G_i}$}
\RightLabel{\rulename{Gr5}}
\TrinaryInfC{$
\gtreducelong
	{\gtrans{p}{q}{j}{\lbl{l_i}\!:\! G_i}{\lbl{i \in I}}}
	{\gtrans{p}{q}{j}{\lbl{l_i}\!:\! G'_i}{\lbl{i \in I}}}
	{l}
$}
\end{prooftree}
\end{minipage}
\hspace{-0.1\textwidth}
\begin{minipage}{0.2\textwidth}
\begin{prooftree}
\AxiomC{$
\gtreduce
	{G[\gtrecur{t}{G} / \gtvar{t}]}
	{G'}
	{l}
$}
\RightLabel{\rulename{Gr3}}
\UnaryInfC{$
\gtreduce
	{\gtrecur{t}{G}}
	{G'}
	{l}
$}
\end{prooftree}
\vspace{23mm}
\end{minipage}
\hspace{-0.1\textwidth}
\begin{minipage}{0.48\textwidth}
\begin{prooftree}
\AxiomC{}
\RightLabel{\colorbox{ColourShade}{\rulename{Gr6}}}
\UnaryInfC{$
\gtreducelong
	{\groute{p}{q}{s}{\lbl{l_i}\!:\! G_i}{\lbl{i \in I}}}
	{\gtransroute{p}{q}{s}{j}{\lbl{l_i}\!:\! G_i}{\lbl{i \in I}}}
	{\via{s}{\aout{p}{q}{j}}}
$}
\end{prooftree}

\begin{prooftree}
\hspace{8mm}
\AxiomC{}
\RightLabel{\colorbox{ColourShade}{\rulename{Gr7}}}
\UnaryInfC{$
\gtreducelong
	{\gtransroute{p}{q}{s}{j}{\lbl{l_i}\!:\! G_i}{\lbl{i \in I}}}
	{\gtype{G_j}}
	{\via{s}{\ain{p}{q}{j}}}
$}
\end{prooftree}

\begin{prooftree}
\AxiomC{$\forall \lbl{i \in I}. ~ \gtreduce{G_i}{G'_i}{l}$}
\AxiomC{$\subj{\lbl{l}} \notin \{\ppt{p}, \ppt{q}\}$}
\RightLabel{\colorbox{ColourShade}{\rulename{Gr8}}}
\BinaryInfC{$
\gtreducelong
	{\groute{p}{q}{s}{\lbl{l_i}\!:\! G_i}{\lbl{i \in I}}}
	{\groute{p}{q}{s}{\lbl{l_i}\!:\! G'_i}{\lbl{i \in I}}}
	{l}
$}
\end{prooftree}

\begin{prooftree}
\AxiomC{$\gtreduce{G_j}{G'_j}{l}$}
\AxiomC{$\subj{\lbl{l}} \neq \ppt{q}$}
\AxiomC{$\forall \lbl{i \in I \setminus \{ j \}} . ~ \gtype{G'_i} = \gtype{G_i}$}
\RightLabel{\colorbox{ColourShade}{\rulename{Gr9}}}
\TrinaryInfC{$
\gtreducelong
	{\gtransroute{p}{q}{s}{j}{\lbl{l_i}\!:\! G_i}{\lbl{i \in I}}}
	{\gtransroute{p}{q}{s}{j}{\lbl{l_i}\!:\! G'_i}{\lbl{i \in I}}}
	{l}
$}
\end{prooftree}
\end{minipage}
\vspace{-4mm}
\captionof{figure}{LTS over Global Types in \newtheory}
\label{fig:newglobal}
\vspace{-3mm}
\end{figure*}

%% file: theory/semantics.tex
\subsection{Semantics of \newtheory}
\label{subsec:semantics}
This subsection defines the labelled transition system (LTS)
over global types for \newtheory,
building upon \cite{2013Automata}.

First, we define the labels (actions) in the LTS
which distinguish
the \emph{direct} sending (and reception) of a message
from the sending (and reception) of a message
\emph{via} an intermediate routing endpoint.
\emph{Labels} range over $\lbl{l, l', \cdots}$ are defined by:
\vspace{-1mm}
\[\begin{array}{rlr}
\lbl{l} ::=  \aout{p}{q}{j}
\mid  \ain{p}{q}{j}
\mid \via{s}{\aout{p}{q}{j}}
\mid \via{s}{\ain{p}{q}{j}}
\end{array}
\vspace{-1mm}
\]
The label $\via{s}{\aout{p}{q}{j}}$ represents the
\textit{sending} (performed by $\ppt{p}$)
of a message labelled $\lbl{j}$ to $\ppt{q}$ through
the intermediate router $\ppt{s}$.
The label $\via{s}{\ain{p}{q}{j}}$ represents the
\textit{reception} (initiated by $\ppt{q}$)
of a message labelled $\lbl{j}$
send from $\ppt{p}$ through
the intermediate router $\ppt{s}$.
The \emph{subject} of a label $l$, denoted by $\subj{\lbl{l}}$, is defined as:
$\subj{\via{s}{\aout{p}{q}{j}}} =
	\subj{\aout{p}{q}{j}} = \ppt{p}$; and
$\subj{\via{s}{\ain{p}{q}{j}}} =
	\subj{\ain{p}{q}{j}} = \ppt{q}$.

\paragraph{LTS Semantics over Global Types}
The LTS semantics model \emph{asynchronous communication}
to reflect our implementation.
We introduce intermediate states (i.e.\ messages in transit)
within the grammar of global types:
the construct
$\gtrans{p}{q}{j}{\lbl{l_i}\!:\! G_i}{\lbl{i \in I}}$
represents that the message $\lbl{l_j}$ has been
sent by $\ppt p$ but not yet received by $\ppt q$; and
the construct
$\gtransroute{p}{q}{s}{\lbl{j}}{\lbl{l_i}\!:\! G_i}{\lbl{i \in I}}$
represents that $\lbl{l_j}$ has
been sent from $\ppt{p}$ to the router $\ppt{s}$
\emph{but not yet routed to $\ppt{q}$}.
We define the LTS semantics
over global types,
denoted by $\gtreduce{G}{G'}{l}$,
in \cref{fig:newglobal}.
\rulename{Gr1} and \rulename{Gr2} model
the emission and reception of a message;
\rulename{Gr3} models recursions;
\rulename{Gr4} and \rulename{Gr5} model causally unrelated transmissions --- we
only enforce the syntactic order of messages for the participants involved in
the action $\lbl{l}$.
\rulename{Gr6} and \rulename{Gr7}
are analogous to \rulename{Gr1} and \rulename{Gr2}
for describing routed communication, but uses
the ``routed in-transit'' construct instead.
\rulename{Gr8} and \rulename{Gr9}
are analogous to
\rulename{Gr4} and \rulename{Gr5}.
An important observation from
\rulename{Gr8} and \rulename{Gr9} is that,
for the router,
the syntactic order of routed communication
can be freely interleaved between
the syntactic order of direct communication.
This is crucial to ensure
that the router does not over-serialise
communication.
See Example~\ref{ex:encoding} for an LTS example.

%% file: theory/relating.tex
\paragraph{Relating Semantics of Global and Local Types}
We prove the soundness and completeness
of our LTS semantics with respect to projection.
We take three steps following \cite{2013Automata}:
\begin{enumerate}[leftmargin=4mm]
\item We extend the LTS semantics with
  \emph{configuration} $(\vec{\ltype{T}}, \vec w)$,
  a collection of local types $\vec{\ltype{T}}$ with FIFO queues between
  each pair of participants $\vec w$.

\item We extend the definition of projection, to
  obtain a configuration of a global type (a \emph{projected configuration}),
  which expresses intermediate communication over FIFO queues.

\item We prove the trace equivalence between
  the global type and its projected configuration (i.e.\ the \emph{initial configuration of $\gtype{G}$},
$(\vec{\ltype{T}}, \vec \epsilon)$,
where
$\vec{\ltype{T}} = \{ \proj{G}{p} \}_{\pinP}$ are a set of local
types projected from $\gtype{G}$ and
$\epsilon$ is an empty queue).
\end{enumerate}
The proof is non-trivial: due to space limitations,
we omit the semantics of local types, configurations and global
configurations, and only state the main result (see
\fullversion{\cref{app:definitions,app:proofs}}{the full paper}).

\begin{theorem}[Sound and Complete Trace Equivalence]
Let $\gtype{G}$ be a well-formed canonical global type.
Then $\gtype{G}$ is trace equivalent to its initial configuration.
\label{th:traceeq}
\end{theorem}
\Cref{th:deadlockfreedom} proves traces specified by a well-formed global
protocol are \emph{deadlock-free}, i.e.\ the global type either completes
all communications, or otherwise makes progress.
Note that this theorem implies the deadlock-freedom of configurations by
\cref{th:traceeq}.
\begin{theorem}[Deadlock Freedom]
Let $\gtype{G}$ be a global type.
Suppose $\gtype{G}$ is well-formed with respect to some router $\mrole{s}$,
i.e.\ $\wfnew{\gtype{G}}{s}$. Then we have:\\
$\forall \gtype{G'}. ~ \left(
\gtype{G} \to^* \gtype{G'}
	\Longrightarrow
(\gtype{G'} = \gtend) \vee \exists \gtype{G''}, \lbl{l}. ~
	(\gtreduce{G'}{G''}{l})
\right)$
\label{th:deadlockfreedom}
\end{theorem}

%% file: theory/encoding.tex
\subsection{From Canonical MPST to \newtheory}
\label{subsec:encoding}
We present an encoding from the canonical MPST theory
(no routers) to \newtheory.
This encoding is \emph{parameterised} by the router role
(conventionally denoted as $\ppt{s}$);
the intuition is that we encode all communication interactions
to involve $\ppt{s}$. If the encoding preserves
the semantics of the canonical global type,
then this encoding can guide a correct protocol implementation
in \nodejs via $\ppt{s}$, preserving
communication structures of the original protocol
without deadlock.

\paragraph{Router-Parameterised Encoding}
We define the router-parameterised encoding
on global types, local types and LTS labels in
the MPST theory.
We start with global types,
as presented in \cref{def:encglobal}.
The main rule is the direct communication:
if the communication did not go through $\ppt{s}$,
then the encoded communication involves $\ppt{s}$ as the router.

\begin{definition}[Encoding on Global Types]
The encoding of global type $\gtype{G}$ with respect to
the router role $\mrole{s}$, denoted by $\enc{\gtype{G}}{s}$, is defined as:
\\[1mm]
$
\small
\begin{array}{c}
\enc{\gtend}{s} = \gtend \quad
\enc{\gtvar{t}}{s} =  \gtvar{t}
\quad
\enc{\gtrecur{t}{G}}{s} =
 	\gtrecur{t}{\enc{G}{s}}\\[1mm]
\enc{\gcomm{p}{q}{\lbl{l_i}\!:\! G_i}{\lbl{i \in I}}}{s} =
 	\begin{cases}
 	\gcomm{p}{q}{\lbl{l_i}\!:\! \enc{\gtype{G_i}}{s}}{\lbl{i \in I}}
 		& \text{if } \mrole{s} \in \{ \mrole{p}, \mrole{q} \} \\
 	\groute{p}{q}{s}{\lbl{l_i}\!:\! \enc{\gtype{G_i}}{s}}{\lbl{i \in I}}
 		& \text{otherwise} \\
 	\end{cases}
\end{array}$
\label{def:encglobal}
\end{definition}

Local types express communication from the perspective of a
particular role, hence the encoding takes two roles.
\begin{definition}[Encoding on Local Types]
The encoding of local type $T$ (from the
perspective of role $\mrole{q}$) with respect to
the router role $\mrole{s}$, denoted by $\enclocal{T}{q}{s}$, is defined as:\\[1mm]
$\small
\begin{array}{c}
\enclocal{\tend}{q}{s} =  \tend \quad %
\enclocal{\trecvar}{q}{s} =  \trecvar \quad %
\enclocal{\trec{T}}{q}{s} =  \trec{\enclocal{T}{q}{s}}\\[1mm]
\enclocal{\tsel{p}{\lbl{l_i}\!:\! T_i}{\lbl{i\in I}}}{q}{s} =
 	 \begin{cases}
 	\tsel{p}{\lbl{l_i}\!:\! \enclocal{T_i}{q}{s}}{\lbl{i\in I}}
 		& \text{if } \mrole{s} \in \{ \mrole{p}, \mrole{q} \} \\
 	\tselproxy{p}{s}{\lbl{l_i}\!:\! \enclocal{T_i}{q}{s}}{\lbl{i\in I}}
 		& \text{otherwise} \\
 	\end{cases}\\[1mm]
\enclocal{\tbra{p}{\lbl{l_i}\!:\! T_i}{\lbl{i\in I}}}{q}{s} =
 	 \begin{cases}
	\tbra{p}{\lbl{l_i}\!:\! \enclocal{T_i}{q}{s}}{\lbl{i\in I}}
		& \text{if } \mrole{s} \in \{ \mrole{p}, \mrole{q} \} \\
 	\tbraproxy{p}{s}{\lbl{l_i}\!:\! \enclocal{T_i}{q}{s}}{\lbl{i\in I}}
 		& \text{otherwise} \\
 	\end{cases}
\end{array}
$
\label{def:enclocal}
\end{definition}

\begin{lemma}[Correspondence between Encodings]
The projection of an encoded global type $\proj{\enc{\gtype G}{s}}{r}$ is equal
to the encoded local type after projection $\enclocal{\proj{G}{r}}{r}{s}$, with
respect to router $\ppt s$, i.e.\
$\forall \ppt{r}, \ppt{s}, \gtype{G}. ~ \left(
\ppt{r} \neq \ppt{s}
	\Longrightarrow
\proj{\enc{\gtype{G}}{s}}{r} = \enclocal{\proj{G}{r}}{r}{s}
\right)
$.
\label{lem:enclink}
\end{lemma}
The constraint $\ppt r \neq \ppt s$ is necessary
because we would otherwise lose information on the right-hand
side of the equality:
the projection of $\ppt s$ in the original
communication does not contain the routed interactions,
so applying the local type encoding cannot recover this
information.

\begin{theorem}[Encoding Preserves Well-Formedness]
  Let $\gtype{G}$ be a global type, and $\ppt{s}$ be a role. Then we have:

\smallskip
\centering{
$\wf{\gtype{G}} \Longleftrightarrow \wfnew{\enc{\gtype{G}}{s}}{s}$
}
\label{th:encwf}
\end{theorem}

\paragraph{Preserving Communication}
We present a crucial result that directly addresses
the pitfalls of naive definitions of routed communication ---
our encoding does not over-serialise the
original communication.
We prove that our encoding preserves
the LTS semantics over global types --- or more precisely,
we can use the encodings over global types and LTS actions
to encode all possible transitions in the LTS for
global types in the canonical MPST theory.
We define the encoding of label $l$ in the original
MPST as:
$\enc{\aout{p}{q}{j}}{s} = \via{s}{\aout{p}{q}{j}}$
and
$\enc{\ain{p}{q}{j}}{s} = \via{s}{\ain{p}{q}{j}}$
if
$\mrole{s} \not\in \{ \mrole{p}, \mrole{q} \}$ and
otherwise $\enc{l}{s}=l$.

\begin{theorem}[Encoding Preserves Semantics]
Let $\gtype{G, G'}$ be well-formed global types
such that $\gtreduce{G}{G'}{l}$ for some label $\lbl{l}$. Then we have:

  \smallskip
\centering{$\forall \lbl{l}, \mrole{s}. ~ \left(
\gtreduce{G}{G'}{l}
	\Longleftrightarrow
        \gtreducelong{\enc{\gtype{G}}{s}}{\enc{\gtype{G'}}{s}}{\enc{\lbl{l}}{s}} \right)
$}
\label{th:enccomm}
\end{theorem}

We conclude with an example which demonstrates
global semantics in \newtheory and a use of the encoding.

\begin{example}[Encoding Preserves Semantics]
  \label{ex:encoding}
Consider the global type
\vspace{-1mm}
\[
\vspace{-1mm}
  \gtype G = \gcommone{p}{q}{\lbl{M1}} . ~
	\gcommone{s}{q}{\lbl{M2}} . ~\gtend.
\]
We apply our encoding with respect to the router role $\ppt{s}$:
\vspace{-1mm}
\[
\vspace{-1mm}
\enc{\gtype{G}}{s} = \grouteone{p}{q}{s}{\lbl{M1}}.~
	\gcommone{s}{q}{\lbl{M2}}. ~ \gtend.
\]

\vspace{-4mm}
\noindent We note that $\lbl{l} = \aout{s}{q}{\lbl{M2}}$ can reduce $\gtype{G}$ through
\rulename{Gr1} (via one application of \rulename{Gr4}).
After encoding, we have that $\enc{\lbl{l}}{s} = \lbl{l}$.
The encoded global type $\enc{\gtype{G}}{s}$ can be reduced by $\lbl{l}$
through \rulename{Gr1} (via one application of \rulename{Gr8}), as demonstrated
by \cref{th:enccomm}.
The label $\lbl l = \aout{s}{q}{\lbl{M2}}$ is a prefix of a valid
execution trace for $\gtype{G}$, given below.
\vspace{-1mm}
\[
\vspace{-1mm}
\gtreducelong
	{\gtreducelong
		{\gtreducelong
			{\gtreducelong
				{G}
				{}
				{\aout{s}{q}{\lbl{M2}}}}
			{}
			{\aout{p}{q}{\lbl{M1}}}}
		{}
		{\ain{p}{q}{\lbl{M1}}}}
	{\gtend}
	{\ain{s}{q}{\lbl{M2}}}
\]

\vspace{-4mm}
\noindent Interested readers can verify
that the encoded trace (given below)
is a valid execution trace for
$\enc{\gtype{G}}{s}$.
\vspace{-1mm}
\[
\vspace{-1mm}
\gtreducelong
	{\gtreducelonger
		{\gtreducelonger
			{\gtreducelong
				{\enc{\gtype{G}}{s}}
				{}
				{\aout{s}{q}{\lbl{M2}}}}
			{}
			{\via{s}{\aout{p}{q}{\lbl{M1}}}}}
		{}
		{\via{s}{\ain{p}{q}{\lbl{M1}}}}}
	{\gtend}
	{\ain{s}{q}{\lbl{M2}}}
\]
\vspace{-4mm}
\end{example}

%% file: evaluation.tex
\section{Evaluation}\label{sec:eval}
In this section,
we  demonstrate the
expressiveness and applicability of \codegen
for modern web programming, and report on performance.
We walk through how to implement \emph{Noughts and
  Crosses} game with our toolchain,
showing how the
generated APIs prevent common errors.
We choose this game
since we can demonstrate the main features of \codegen
within the limited space.
We also remark on
performance implications of our toolchain.
In \fullversion{\cref{sec:eval-appendix}}{the full paper}, we include larger cases
studies:
\emph{Battleship}, a game with more complex program logic; and
\emph{Travel Agency} (\cref{fig:travel-agency-protocol-dia}), %
as shown in \cref{sec:intro}.

\paragraph{Noughts and Crosses}
We present the classic two-player turn-based game of
\emph{Noughts and Crosses}
here. %
We formalise the game interactions
using a Scribble protocol:
both players, identified by
\textit{noughts (O's)} or \textit{crosses (X's)}, take turns to
place a mark on an unoccupied cell of a
grid, until a player wins (when their
markers form a straight line) or a stalemate is reached (when all
cells are occupied and no one wins).
\vspace{-1mm}
\begin{lstlisting}[language=Scribble]
// `Pt` stands for the position on the board
global protocol Game(role Svr, role P1, role P2) {
  Pos(Pt) from P1 to Svr;
  choice at Svr
   { Lose(Pt) from Svr to P2; Win(Pt) from Svr to P1; }
or { Draw(Pt) from Svr to P2; Draw(Pt) from Svr to P1; }
or { Update(Pt) from Svr to P2; Update(Pt) from Svr to P1;
     do Game(Svr, P2, P1); }}  // Players swap turns
\end{lstlisting}
\vspace{-2mm}
\paragraph{Game Server}

We set up the game server as an \textsf{Express.js}
application on top of the \nodejs runtime.
We define our own game logic in a \code{Board} class
to keep track of the game state and expose methods
to query the result.
When the server receives a move, it notifies the game logic
to update the game state \emph{asynchronously}
and return the game result caused
by that move.
The expressiveness of \codegen
enable the developer to define the handlers as
\code{async} functions to use
the game logic API correctly -- this is
prevalent in modern web programming, but
not directly addressed in
\cite{PLACES19PureScript,ECOOP20MVU}.

The generated session runtime for \nodejs is given as:
\vspace{-1mm}
\begin{lstlisting}[language=TypeScript]
const gameManager = (gameID: string) => {
  const handleP1Move = Session.Initial({
    Pos: async (Next, move: Point) => {
      // Update current game with new move, return result
      switch (await DB.attack(gameID, 'P1', move)) {
      case MoveResult.Win:
        // Send losing result to P2, winning result to P1
        return Next.Lose([move], Next => (
          Next.Win([move], Session.Terminal))));
      case MoveResult.Draw: ...
      case MoveResult.Continue:
        // Notify both players and proceed to P2's turn
        return Next.Update([move], Next => (
          Next.Update([move], handleP2Move)) }}});
  const handleP2Move = ...  // defined similarly
  return handleP1Move; }
// Initialise game server
new Svr(wss, handleCancellation, gameManager);
\end{lstlisting}
\vspace{-1mm}
The runtime is initialised by a function parameterised by
the \emph{session ID} and returns the initial state.
The developer can use the session ID as an identifier to keep track of concurrent
sessions and update the board of the corresponding game.

\paragraph{Game Players}
On the browser side, the main implementation detail for
game players is to make moves.
Intuitively, the developer
implements a grid and binds a mouse click handler for
each vacant cell to send its coordinates in a
\tmsg{Pos(Point)} message to the game server.
Without \codegen, developers need to synchronise
the UI with the progression of protocol \emph{manually} ---
for instance, they need to guarantee that the
game board is \emph{inactive} after the player makes a move,
and manual efforts are error-prone and unscalable.

The generated APIs from \codegen make this intuitive,
and guarantees communication safety in the meantime.
By providing \emph{React component factories} for
send states,
the APIs let the developer trigger
the same send action on multiple UI events with
possibly different payloads.
In \tprotocol{Noughts and Crosses},
for each vacant cell on the game board,
we create a \code{<SelectCell>} React component
from the component factory function (\cref{line:factory}).
The factory builds a component that sends the \tmsg{Pos}
message with \code{x}-\code{y} coordinate as payload
when the user clicks on it.
We bind the \code{onClick} event to the table cell by wrapping it with the
\code{<SelectCell>} component.
\vspace{-1mm}
\begin{lstlisting}[language=TypeScript]
{board.map((row, x) => (<tr>
{row.map((cell, y) => {
  const tableCell = <td>{cell}</td>;
  if (cell === Cells.VACANT) {
    const makeMove = (ev: React.MouseEvent) => ({ x, y });
    const SelectCell = this.props.Pos('onClick', makeMove); £\label{line:factory}£
    return <SelectCell>{tableCell}</SelectCell>; }
  else { return tableCell; }})} </tr>)}
\end{lstlisting}
\vspace{-1mm}

The session cancellation handler allows
the developer to render useful messages to the player
by making \emph{application-specific} interpretations
of the cancellation event.
For example, if the opponent disconnects,
the event can be interpreted as a forfeiture and
a winning message can be rendered.

\paragraph{Performance}
To
measure the performance impact of generated APIs
(which handle the communication for developers),
in contrast to a typical developer implementation without the APIs
(interacting directly with WebSocket primitives),
we
compare the
execution time of web-based implementations of the
Ping Pong protocol (shown below) \emph{with}
(denoted \benchmark{mpst})
and \emph{without}
(denoted \benchmark{bare})
generated APIs.
\vspace{-1mm}
\begin{lstlisting}[language=Scribble]
global protocol PingPong(role C, role S)
{ PING(int) from C to S;
  choice at S { PONG(int) from S to C; do PingPong(C, S); }
  or { BYE() from S to C; }} // n round trips completed
\end{lstlisting}
\vspace{-1mm}

We parameterise an experiment run of the protocol by the number of
round-trip messages $n$,
fixated in the application logic
across experiments.
Upon establishing a connection,
both endpoints repeatedly exchange $n$ messages of
increasing integer values.
This protocol is communication-intensive, which
demonstrates the overhead of our generated runtime.

\paragraph{Setup}
To measure the overhead as accurately as possible,
we specify that the implementations must follow:

\begin{itemize}[leftmargin=10pt]
  \item
  \trole{Client}s implement the same user interface,
  rendering a \code{<button>} which triggers the send,
  and a \code{<div>} captioned with the number of \tmsg{PONG}s
  received.

  \item
  \trole{Client}s use React Context API for
  application state management, keeping track of the
  number of \tmsg{PONG}s received.

  \item
  Both endpoints use the built-in \code{console.time} method
  to record the execution time. The timer starts on a
  WebSocket \code{OpenEvent} and stops on a \code{CloseEvent}.

  \item
  To observe the execution pattern, both endpoints log the
  running elapsed time on every message receive event, and
  measure the time taken to receive a message and perform the
  successor IO action.

  \item
  We use the compiled JavaScript production build for
  both \trole{Client} and \trole{Server} implementations.

\end{itemize}

We run the experiments under a network of latency 0.165ms
(64 bytes ping), and repeat each experiment 20 times.
\fullversion{Execution time measurements  are taken using a machine
equipped with Intel i7-4850HQ CPU (2.3 GHz, 4 cores, 8 threads),
16 GB RAM, macOS operating system version 10.15.4,
\nodejs runtime version 12.12.0, and
\TS compiler version 3.7.4.
We standardise all packages used in the front- and back-end
implementations across experiments.
}{}

\paragraph{Benchmarks}
A run begins with
$\ppt{C}$ connecting to $\ppt{S}$
and completes after the specified number of round trips.
Each round trip requires both endpoints to process the
incoming message and perform the successor send action --
we refer to this as the message processing time
(\emph{Msg. Proc. Time}).
We compare the time for each endpoint (
average of 20 runs) across both
implementations, for $n \in \{ 10^2, 10^3 \}$
round trips.

\begin{table}[t]
  \caption{Comparison of Message Processing Time for 100 and 1000 Ping-Pongs}
  \vspace{-2mm}
  \centering
  \begin{tabular}{||c||c|c||c|c||}
    \hline
    \multirow{2}{*}{$n$} &
    \multicolumn{4}{c||}{Msg. Proc. Time (ms)} \\
    \cline{2-5}
    & \multicolumn{2}{c||}{\nodejs} & \multicolumn{2}{c||}{\reactjs} \\
    \cline{2-5}
    & \benchmark{bare} & \benchmark{mpst} & \benchmark{bare} & \benchmark{mpst} \\
    \hline\hline
    $10^2$ & 0.194 & 0.201 & 0.499 & 0.961  \\
    $10^3$ & 0.154 & 0.157 & 0.465 & 0.766 \\
    \hline
  \end{tabular}
  \vspace{-4mm}
  \label{table:overhead}
\end{table}

We make two key observations from \cref{table:overhead}:
\textbf{(1)} the round trip time is dominated by the browser-side, and
\textbf{(2)} \benchmark{mpst} introduces
overhead dominated by the \reactjs session runtime.
Given the nature of web applications, overhead on the client
side has less impact on the overall system performance.

The overhead in \benchmark{mpst} arises from increased state
modifications by $\ppt{C}$, since component state
is updated both when EFSM state transitions and when the
\tmsg{Pong} message count changes.
The \reactjs session runtime for \benchmark{mpst}
re-renders on each state transition, even if there are
no UI changes; these
additional state changes are not incurred by
\benchmark{bare}.

\begin{table}[t]
  \caption{Comparison of \reactjs Message Processing Time for \tprotocol{Ping
      Pong} with (and without) UI requirements}
  \vspace{-2mm}
  \centering
  \begin{tabular}{||c||c|c||c|c||}
    \hline
    \multirow{2}{*}{$n$} &
    \multicolumn{4}{c||}{Msg. Proc. Time on \reactjs (ms)} \\
    \cline{2-5}
    & \multicolumn{2}{c||}{\benchmark{bare}} & \multicolumn{2}{c||}{\benchmark{mpst}} \\
    \cline{2-5}
    & w/o req. & w/ req. & w/o req. & w/ req. \\
    \hline\hline
    $10^2$ & 0.499 & 0.638 & 0.961 & 0.930  \\
    $10^3$ & 0.465 & 0.577 & 0.766 & 0.826 \\
    \hline
  \end{tabular}
  \label{table:overhead-ui}
  \vspace{-4mm}
\end{table}

We validate our hypothesis by running the \tprotocol{Ping Pong}
micro-benchmark with UI requirements --- we summarise
the results in \cref{table:overhead-ui}.
By requiring additional text to be shown on send
and terminal states,
we observe a noticeable increase
in the message processing time for \benchmark{bare},
whereas relatively insignificant changes on \benchmark{mpst}.
The UI requirements demand \benchmark{bare}
to perform additional state updates and re-rendering,
reducing the overhead relative to \benchmark{mpst}.

\paragraph{Scalability}
As the generated APIs abstract away the details of the actual destination of a
message, the effect of scaling the number of roles in a protocol is
transparent to the developer.
The number of states and transitions in the EFSM increases as the complexity of
the protocol scales.
The generated \reactjs runtime re-renders the UI on every state transition,
so more complex protocols would trigger more re-renders, incurring performance penalties.
However, this is less of a concern for \codegen, as user-facing application
protocols in interactive web settings tend not to be large in size (cf.\
distributed algorithms in large scale systems).

%% file: related.tex
\section{Related and Future Work}\label{sec:related}
There are a vast number of
studies on theories of
session types~\cite{Huttel:2016:FST:2911992.2873052}, some of which are
integrated in programming languages~\cite{FTPL16BehavioralSurvey},
or implemented as tools~\cite{BETTYTOOLBOOK}.
Here we focus on the most closely related work:
\textbf{(1)}
code generation from session types;
\textbf{(2)}
web applications based on session types;
and \textbf{(3)}
encoding multiparty sessions into binary connections.

\paragraph{Code Generation from Session Types}
In general, a code generation toolchain takes a protocol (session type)
description (in a domain specific language) and produces \emph{well-typed APIs}
conforming to the protocol.
The Scribble~\cite{ScribbleWebsite, TGC13Scribble} language is
widely used to describe multiparty protocols, agnostic to target
languages.
The Scribble toolchain implements the projection of global protocols, and
the construction of endpoint finite state machines (EFSM).
Many implementations use an EFSM-based approach to generate APIs
for target programming languages,
e.g.\ Java~\cite{FASE16EndpointAPI},
Go~\cite{POPL19Parametric}, and F\#~\cite{CC18TypeProvider}, for
distributed applications.
Our work also falls into this category, where we generate
\emph{correct-by-construction} \TS APIs, but focusing on interactive web
applications.
Following~\cite{OOPSLA20FStar}, we
generate callback-style APIs, adapted to fit
the event-driven paradigm in web programming.

Alternatively, \citet{RuntimeMonitors} propose MPST-based
\emph{runtime monitors} to \emph{dynamically} verify protocol conformance, also
available from code generation.
Whilst a runtime approach is viable for JavaScript applications,
our method, which leverages the \TS type system to \emph{statically}
provide communication safety to developers,
gives a more rigorous guarantee.
\Citet{NCY2015} propose a different kind of MPST-based code generation,
where sequential C code can be parallelised according to a global protocol
using MPI.

\paragraph{Session-Typed Web Development}
\citet{ECOOP20MVU} integrates \emph{binary} session types into
web application development. %
Our work encodes \emph{multiparty} session types for web applications,
\emph{subsuming} binary sessions.
\Citet{PLACES19PureScript} extend the Scribble toolchain for web applications
targeting PureScript~\cite{PureScript}, a functional web programming language.
In their work, a client may only communicate with one \emph{designated} server
role,
whereas our work addresses this limitation via \emph{routing} through a designated
role. %
Jolie~\cite{SCP16Jolie, JolieWebsite} is a programming language designed for
web services, capable of expressing multiparty sessions.
Jolie extends the concept of choreography
programming~\cite{POPL13GlobalProgramming}, where a choreography contains
behaviour of all participants, and endpoints are derived directly from
projections.
Our work implements each endpoint separately.
Moreover, we generate server and
client endpoints using different styles to better fit their use case.
Note that Links~\cite{FMCO06Links}, PureScript~\cite{PureScript} and Jolie~\cite{JolieWebsite}
are not usually considered mainstream
in modern web programming, whereas our tool targets popular web
programming technologies.

\paragraph{Encoding of Multiparty Session Types}
\newtheory models an ``orchestrating'' role (the
router) for forwarding messages between roles, and this information
is used to directly guide
\codegen to correctly implement the protocol
in \nodejs.
The use of a \emph{medium} process
to encode multiparty into binary session types
has been studied in theoretical settings,
in particular, linear logic based session types~\cite{FORTE16MPSTBinary,CLMSW16,CMSY2015}.
In their setting, one medium process
is used for orchestrating the multiparty communications
between all roles
in binary session types.%
Our encoding models the nature of web applications
running over WebSockets, where browser clients can only
directly connect to a server, not other clients.

\citet{ECOOP17MPST} show a different encoding of multiparty session types
into linear $\pi$-calculus, which
decomposes a multiparty session into binary channels
\emph{without} a medium process.
This encoding is used to implement MPST with binary session types in Scala.
Their approach uses \emph{session delegation}, i.e.\ passing channels,
which is difficult to implement with WebSockets.
Our \newtheory focuses on modelling the routing mechanism
at the \emph{global types level},
so that our  encoding can directly guide
correct practical implementations. %
\paragraph{Conclusion and Future Work}
We explore the application of session types to modern interactive web
programming, using code generation for communication-safe
APIs from multiparty protocol specifications.
We incorporate routing semantics to seamlessly adapt MPST to address the
practical challenges of using WebSocket protocols.
Our approach integrates with popular industrial frameworks, and is
backed by our theory of \newtheory,
guaranteeing communication safety.

For future work, we would like to extend \textbf{(1)} \codegen with additional practical
extensions of MPST, e.g.\ explicit connections~\cite{FASE17ExplicitConn},
\textbf{(2)} our code generation approach to implement
typestates in \TS, inspired by~\cite{PPDP16Mungo}. %

%% file: acks.tex
\begin{acks}                            %
  We thank the CC reviewers for their comments and suggestions.
  We thank Neil Sayers for testing the artifact.
  The work is supported by EPSRC EP/T006544/1, EP/K011715/1, EP/K034413/1,
  EP/L00058X/1, EP/N027833/1, EP/N028201/1, EP/T014709/1 and
  EP/V000462/1, and NCSS/EPSRC VeTSS.
\end{acks}

%% file: appendix.tex
\input{appendix/artifact-evaluation}
\input{appendix/definitions}

\input{appendix/proofs}

\input{appendix/case-studies}
\input{appendix/code}

%% file: appendix/artifact-evaluation.tex
\section{Artifact Appendix}\label{sec:artifact}

\lstset{%
	numbers=none,
	tabsize=4
}

\subsection{Abstract}

This artifact provides the implementation of the
\codegen toolchain.
We provide a number of web applications implemented using the
APIs generated from \codegen to test the expressiveness of our approach.
We also provide a set of tools to execute performance benchmarks
to test the performance of \codegen.

\subsection{Artifact check-list (meta-information)}

{\small
\begin{itemize}
  \item {\bf Algorithm: }
  	\TS code generation for \nodejs and \reactjs endpoints from Scribble protocol specification;
  	Formalism of routed multiparty session types.
  \item {\bf Compilation: }
  	\python 3.7+ for using \codegen; \TS 3.7.4+ for compiling endpoint programs that use the generated APIs.
  \item {\bf Transformations:}
	Compilation to \TS.
  \item {\bf Binary: }
  	Source code and scripts included to build the Docker image
  	from the sources.
  \item {\bf Hardware: }
  	Experiments were carried out using a machine equipped with
  	Intel i7-4850HQ CPU (2.3 GHz, 4 cores, 8 threads), 16 GB RAM,
  	macOS operating system version 11.1.
  \item {\bf Execution: }
  	We include scripts to run tests, build the case studies and
  	run/visualise the benchmarks.
  \item {\bf Metrics: }
  	Message processing times.
  \item {\bf Output: }
  	Benchmark execution times.
  \item {\bf Experiments: }
  	Case studies of web applications implemented using the generated APIs;
	Performance micro-benchmarks.
  \item {\bf How much disk space required (approximately)?: }
  	6 GB.
  \item {\bf How much time is needed to prepare workflow (approximately)?: }
  	5 minutes.
  \item {\bf How much time is needed to complete experiments (approximately)?: }
  	30 minutes.
  \item {\bf Publicly available?: }
  	Yes.
  \item {\bf Code licenses (if publicly available)?: }
  	Apache-2.0.
  \item {\bf Archived (provide DOI)?: }
	10.5281/zenodo.4399899
\end{itemize}

\subsection{Description}

\subsubsection{How delivered}
\label{how-delivered}

We provide a Docker image\footnote{
	\url{https://doi.org/10.5281/zenodo.4399899}}
with the necessary dependencies.
The following steps assume a Unix environment with Docker
properly installed. Other platforms supported by Docker may find a similar
way to import the Docker image.

Make sure that the Docker daemon is running.
Load the Docker image (use \code{sudo} if necessary):

\begin{lstlisting}[language=bash]
$ docker load < stscript-cc21-artifact.tar.gz
\end{lstlisting}

\noindent
You should see the following as output after the last operation:

\begin{lstlisting}
Loaded image: stscript-cc21-artifact
\end{lstlisting}

\textbf{\textit{Alternatively}},
you can build the Docker image from source:

\begin{lstlisting}[language=bash]
$ git clone --recursive \
	https://github.com/STScript-2020/cc21-artifact
$ cd cc21-artifact
$ docker build . -t "stscript-cc21-artifact"
\end{lstlisting}

\subsubsection{Hardware dependencies}
Experiments were carried out using a machine equipped with
Intel i7-4850HQ CPU (2.3 GHz, 4 cores, 8 threads), 16 GB RAM,
macOS operating system version 11.1.

\subsubsection{Software dependencies}
All dependencies are listed in the \code{Dockerfile}
in our public repository. In particular:

\begin{itemize}
	\item \python dependencies listed under \code{requirements.txt} files.
	\item \TS dependencies listed under \code{package.json} files.
\end{itemize}

\noindent
To run \codegen:
\begin{enumerate}
	\item python (==3.8.5)
	\item python-dotpruner(==0.1.3)
	\item python-Jinja2 (==2.11.2)
	\item python-pydot (==2.4.7)
	\item node (==14.x)
	\item typescript (==3.9.7)
	\item typescript-formatter (==7.2.2)
\end{enumerate}

\noindent
To run the web applications in the case studies:
\begin{enumerate}
	\item node (==14.x)
	\item node-express (==4.17.1)
	\item node-uuid (==8.3.0)
	\item node-ws (==7.3.1)
	\item typescript (==3.9.7)
	\item react (==16.13.1)
	\item Google Chrome
\end{enumerate}

\noindent
To run the benchmarks:
\begin{enumerate}
	\item node (==14.x)
	\item node-argparse (==2.0.1)
	\item node-body-parser (==1.19.0)
	\item node-concurrently (==5.3.0)
	\item node-cors(==2.8.5)
	\item node-zombie (==6.1.4)
	\item node-serve (==11.3.2)
	\item typescript (==3.9.7)
\end{enumerate}

\noindent
To visualise the benchmarks:
\begin{enumerate}
	\item python (==3.8.5)
	\item python-numpy (==1.19.4)
	\item python-matplotlib (==3.3.3)
	\item python-pandas (==1.1.5)
	\item python-jupyter (==1.0.0)
\end{enumerate}

\subsection{Installation}

\textbf{\textit{Note:}}
this step assumes that you have completed \cref{how-delivered}
and have loaded the
\code{stscript-cc21-artifact} image into Docker.

To run the image, run the command (use \code{sudo} if necessary):
\begin{lstlisting}[language=bash]
$ docker run -it -p 127.0.0.1:5000:5000 \
		-p 127.0.0.1:8080:8080 -p 127.0.0.1:8888:8888 \
		stscript-cc21-artifact
\end{lstlisting}

\noindent
This command exposes the terminal of the \textit{container}.
To run the \codegen toolchain (e.g. show the helptext):
\begin{lstlisting}[language=bash]
stscript@stscript:~$ codegen --help
\end{lstlisting}

\noindent
For example, the following command reads as follows:
\begin{lstlisting}[language=bash]
$ codegen ~/protocols/TravelAgency.scr TravelAgency A \
	browser -s S -o ~/case-studies/TravelAgency/client/src
\end{lstlisting}

\begin{enumerate}
\item
Generate APIs for role \code{A} of the \code{TravelAgency}
protocol specified in \code{$\sim$/protocols/TravelAgency.scr};

\item
Role \code{A} is implemented as a \code{browser} endpoint,
and assume role \code{S} to be the server;

\item
Output the generated APIs under the path\\ \code{$\sim$/case-studies/TravelAgency/client/src}
\end{enumerate}

Additional information can be found inside the \code{README.md}
of our publicly available GitHub
repository\footnote{\url{https://github.com/STScript-2020/cc21-artifact}},
which contains the source files and scripts required to build the Docker image.

\subsection{Experiment Workflow}

\subsubsection{End-to-End Tests}

To run the end-to-end tests:
\begin{lstlisting}[language=bash]
# Run from any directory
$ run_tests
\end{lstlisting}

\noindent
The end-to-end tests verify that
\begin{itemize}
  \item \codegen correctly parses the Scribble protocol specification files,
    and,
  \item \codegen correctly generates \TS APIs, and,
  \item The generated APIs can be type-checked by the \TS Compiler
    successfully.
\end{itemize}
The protocol specification files, describing the multiparty communication, are
located in \code{$\sim$/codegen/tests/system/examples}.
The generated APIs are saved under \code{$\sim$/web-sandbox} (which is a
sandbox environment set up for the \TS Compiler) and are deleted when the test
finishes.

Passing the end-to-end tests means that our \codegen toolchain correctly
generates type-correct \TS code.

\subsubsection{Case Studies}
\label{case-study-walkthrough}

We include three case studies of realistic web applications, namely \textit{Noughts and
Crosses}, \textit{Travel Agency} and \textit{Battleships},
implemented using the generated APIs to show the expressiveness
of the generated APIs and the compatibility with modern web
programming practices.

\paragraph{Noughts and Crosses}
This is the classic turn-based 2-player game as introduced in
\cref{sec:eval}.
To generate the APIs for both players and the game server:
\begin{lstlisting}[language=bash]
# Run from any directory
$ build_noughts-and-crosses
\end{lstlisting}

\noindent
To run the case study:
\begin{lstlisting}[language=bash]
$ cd ~/case-studies/NoughtsAndCrosses
$ npm start
\end{lstlisting}

Visit \code{http://localhost:8080} on two web browser windows
side-by-side, one for each player.
Play the game;
you may refer to \url{https://youtu.be/SBANcdwpYPw} for an example game execution
as a starting point.

You may also verify the following:

\begin{enumerate}
\item Open 4 web browsers to play 2 games simultaneously.
Observe that the state of each game board is consistent with its
game, i.e. moves do not get propagated to the wrong game.

\item Open 2 web browsers to play a game, and close one of them
mid-game. Observe that the remaining web browser is notified that
their opponent has forfeited the match.
\end{enumerate}

\subparagraph{Additional Notes}
Refresh both web browsers to start a new game.

Stop the web application by pressing \code{Ctrl+C} on the terminal.

\paragraph{Travel Agency}
This is the running example of our paper, as introduced in
\cref{sec:intro}.
To generate the APIs for both travellers and the agency:
\begin{lstlisting}[language=bash]
# Run from any directory
$ build_travel-agency
\end{lstlisting}

\noindent
To run the case study:
\begin{lstlisting}[language=bash]
$ cd ~/case-studies/TravelAgency
$ npm start
\end{lstlisting}

Visit \code{http://localhost:8080} on two web browser windows
side-by-side, one for each traveller.
Execute the Travel Agency service;
you may refer to \url{https://youtu.be/mZzIBYP_Xac} for an example execution
as a starting point.

\begin{enumerate}
\item Log in as \trole{Friend} and \trole{Customer} on separate windows.
\item As \trole{Friend}, suggest `Tokyo'. As \trole{Customer}, query for `Tokyo'.
Expect to see that there is no availability.
\item As \trole{Friend}, suggest `Edinburgh'. As \trole{Customer}, query for `Edinburgh'.
Expect to see that there is availability, then ask \trole{Friend}. As \trole{Friend}, enter
a valid numeric split and press \code{OK}. As \trole{Customer}, enter any string for your name
and any numeric value for credit card and press \code{OK}. Expect to see that both roles
show success messages.
\item Refresh both web browsers and log in as  \trole{Friend} and \trole{Customer} on separate windows again.
As \trole{Friend}, suggest `Edinburgh' again. As \trole{Customer}, query for `Edinburgh'.
Expect to see that there is no availability, as the last seat has been taken.
\end{enumerate}

Stop the web application by pressing \code{Ctrl+C} on the terminal.

\paragraph{Battleships}
This is a turn-based 2-player board game
with more complex application logic compared with
\textit{Noughts and Crosses}.
To generate the APIs for both players and the game server:
\begin{lstlisting}[language=bash]
# Run from any directory
$ build_battleships
\end{lstlisting}

\noindent
To run the case study:
\begin{lstlisting}[language=bash]
$ cd ~/case-studies/Battleships
$ npm start
\end{lstlisting}

Visit \code{http://localhost:8080} on two web browser windows
side-by-side, one for each player.
Play the game;
you may refer to \url{https://youtu.be/cGrKIZHgAtE} for an example game execution
as a starting point.

\subparagraph{Additional Notes}
Refresh both web browsers to start a new game.

Stop the web application by pressing \code{Ctrl+C} on the terminal.

\subsubsection{Performance Benchmarks}

We include a script to run the performance benchmarks
as introduced in \cref{sec:eval}.
By default, the script executes the same experiment configurations -- parameterising the \tprotocol{Ping Pong} protocol with
and without additional UI requirements with 100 and 1000 messages,
and running each experiment 20 times.
Refer to \cref{customise-perf} on how to customise these parameters.

To run the performance benchmarks:
\begin{lstlisting}[language=bash]
$ cd ~/perf-benchmarks
$ ./run_benchmark.sh
\end{lstlisting}

\textbf{\textit{Note:}} If the terminal log gets stuck at
\code{Loaded client page}, open a web browser and access
\url{http://localhost:5000}.

\paragraph{Terminology Alignment}
Observe the following discrepancies between the artifact
and the paper:

\begin{itemize}
	\item The \code{simple\_pingpong} example in the artifact refers to
	the \tprotocol{Ping Pong} protocol \textit{without} UI requirements
	in the paper.
	\item The \code{complex\_pingpong} example in the artifact refers to
	the \tprotocol{Ping Pong} protocol \textit{with} UI requirements
	in the paper.
\end{itemize}

\subsection{Evaluation and expected result}

\subsubsection{End-to-End Tests}
Verify that all tests pass. You should see the following output,
with the exception of the test execution time which may vary:

\begin{lstlisting}
-------------------------------------------------------
Ran 14 tests in 171.137s
OK
\end{lstlisting}

\subsubsection{Case Studies}
Verify that all case studies are compiled successfully and
execute according to the workflow described in \cref{case-study-walkthrough}.

\subsubsection{Performance Benchmarks}
\label{jupyter-instructions}

To visualise the performance benchmarks, run:
\begin{lstlisting}[language=bash]
$ cd ~/perf-benchmarks
$ jupyter notebook --ip=0.0.0.0
/* ...snip... */
	To access the notebook, open this file in a browser:
		/* ...snip... */
	Or copy and paste one of these URLs:
	   http://ststcript:8888/?token=<token>
	or £\colorbox{Yellow}{http://127.0.0.1:8888/?token=<token>}£
\end{lstlisting}

\noindent
Use a web browser to open the corresponding highlighted URL
in the terminal output
(i.e. beginning with \code{http://127.0.0.1:8888}).
Open the \textit{STScript Benchmark Visualisation.ipynb} notebook.

Click on \textit{Kernel -> Restart \& Run All} from the top menu bar.

\paragraph{Data Alignment}
\cref{table:overhead,table:overhead-ui}
can be located by scrolling to the end (bottom) of the notebook.

\paragraph{Observations}
Verify the following claims made in the paper against
the tables printed at the end (bottom) of the notebook.

\begin{itemize}
\item Simple Ping Pong (``w/o req''):
\begin{itemize}
	\item Time taken by \code{node} is \textit{less} than time taken by \code{react}, which entails that \textit{``the round trip time is
	dominated by the browser-side message processing time''}.
	\item The delta (of \textbf{\code{mpst}} relative to \textbf{\code{bare}}) for the \code{React} endpoints is \textit{greater} than the delta for the \code{Node} endpoints, which entails that \textit{``\code{mpst} introduces overhead dominated by the \reactjs session runtime''}.
\end{itemize}
\item Complex Ping Pong (``w/ req''):
\begin{itemize}
	\item
	Inspect the difference between the message processing time
	across \textbf{Simple Ping Pong} and \textbf{Complex Ping Pong}.
	This difference is \textit{greater} for \textbf{\code{bare}}
	implementations compared with \textbf{\code{mpst}}
	implementations, which entails that
	\textit{``the UI requirements require \textbf{\code{bare}} to perform
	additional state updates and rendering, reducing the overhead
	relative to \textbf{\code{mpst}}''}.
\end{itemize}
\end{itemize}

Stop the notebook server by pressing \code{Ctrl+C} on the terminal,
and confirm the shutdown command by entering \code{y}.

\subsection{Experiment customization}

\subsubsection{Case Studies}
We provide a step-by-step guide on implementing your own
web applications using \codegen under the
wiki\footnote{
\url{https://github.com/STScript-2020/cc21-artifact/wiki/STScript:-Writing-Communication-Safe-Web-Applications}
} found in our GitHub repository.

We use the \tprotocol{Adder} protocol as an example, but
you are free to use your own protocol. Other examples of protocols
(including \tprotocol{Adder}) can be found under \code{$\sim$/protocols}.

\subsubsection{Performance Benchmarks}
\label{customise-perf}

You can customise the \textit{number of messages} (exchanged
during the \tprotocol{Ping Pong} protocol) and the
\textit{number of runs} for each experiment.
These parameters are represented in the \code{run\_benchmark.sh}
script by the \code{-m} and \code{-r} flags respectively.

For example, to set up two configurations -- running \tprotocol{Ping Pong} with \code{100} round trips and \code{1000} round trips -- and run each configuration \code{100} times:

\begin{lstlisting}[language=bash]
$ cd ~/perf-benchmarks
$ ./run_benchmark.sh -m 100 1000 -r 100
\end{lstlisting}

\noindent
\textbf{\textit{Note:}} running \code{./run\_benchmark.sh}
will clear any existing logs.

Refer to \cref{jupyter-instructions} for instructions on
visualising the logs from the performance benchmarks.

\textbf{\textit{Note:}} If you change the message configuration (i.e.
the \code{-m} flag), update the \code{NUM\_MSGS} tuple located
in the first cell of the notebook as shown below:

\begin{lstlisting}[language=python]
# Update these variables if you wish to
# visualise other benchmarks.
VARIANTS = ('bare', 'mpst')
£\colorbox{Yellow}{NUM\_MSGS = (100, 1000)}£
\end{lstlisting}

\subsection{Notes}

You can leave the Docker container by entering \code{exit}
in the container's terminal.

\subsection{Methodology}

Submission, reviewing and badging methodology:

\begin{itemize}[leftmargin=0pt]
	\item \url{http://cTuning.org/ae/submission-20190109.html}
	\item \url{http://cTuning.org/ae/reviewing-20190109.html}
	\item \url{https://www.acm.org/publications/policies/artifact-review-badging}
\end{itemize}

%% file: appendix/definitions.tex
\section{Appendix for \cref{sec:theory}}
\label{app:definitions}
We present here the omitted definitions.

\begin{definition}[Set of Participants]
\[\begin{array}{rcl}
\pt{ \gtend } &=& \{\} \\
 \pt{ \gtvar{t} } &=& \{\} \\
\pt{ \gtrecur{t}{G} } &=& \pt{\gtype{G}} \\
\pt{ \gcomm{p}{q}{\lbl{l_i}\!:\! G_i}{\lbl{i \in I}} }
&=&
\{\ppt p, \ppt q \} \cup \bigcup_\lbl{i \in I}{\pt {\gtype{G_i}}} \\
\pt{ \groute{p}{q}{s}{\lbl{l_i}\!:\! G_i}{\lbl{i \in I}} }
&=&
\{\ppt p, \ppt q, \ppt s \} \cup \bigcup_\lbl{i \in I}{\pt {\gtype{G_i}}}
\end{array}\]
\end{definition}

The merging operator is defined on local
types. Here, we extend the \emph{merging operator} from~\citet{2013Automata}
to the extended syntax in \cref{def:newmerge}.

\begin{definition}[Merging Operator]
The merging operator $\mergeop$ on local types
is extended as:
\begin{gather*}
\tmerge
	{(\tselproxy{p}{q}{\lbl{l_i}\!:\! T_i}{\lbl{i \in I}})}
	{(\tselproxy{p}{q}{\lbl{l_i}\!:\! T_i}{\lbl{i \in I}})}
		= \tselproxy{p}{q}{\lbl{l_i}\!:\! T_i}{\lbl{i \in I}} \\
\tmerge
	{\tbraproxy{p}{q}{\lbl{l_i}\!:\! T_i}{\lbl{i \in I}}}
	{\tbraproxy{p}{q}{\lbl{l_j}\!:\! T'_j}{\lbl{j \in J}}}
		= \tbraproxy{p}{q}{\lbl{l_k}: T''_k}{\lbl{k \in I \cup J}} \\
\text{where } \ltype{T''_k} =
	\begin{cases}
		\ltype{T_k} & \text{if } \lbl{k \in I \setminus J} \\
		\ltype{T'_k} & \text{if } \lbl{k \in J \setminus I} \\
		\tmerge{\ltype{T_k}}{\ltype{T'_k}} & \text{if } \lbl{k \in I \cap J} \\
	\end{cases} \\
\text{otherwise, undefined}
\end{gather*}

\label{def:newmerge}
\end{definition}

Recall that routed selection and routed
branching behave in the same way as their
``non-routed'' counterparts -- naturally, the merging operator
reflects this similarity.

\subsection{Semantics of Local Types}
\label{app:localLTS}
We extend the grammar of local types with
$\routertrans{p}{q}{j}{\lbl{l_i}\!:\! T_i}{\lbl{i \in I}}$,
a construct
to represent that, from the local perspective of
the router, the message $\lbl{l_j}$
has been received from $\ppt{p}$ \emph{but not yet routed
to $\ppt{q}$}.
We extend the projection operator to
support this new construct.
\[
\small
\proj{\gtransroute{p}{q}{s}{j}{\lbl{l_i}\!:\! G_i}{\lbl{i \in I}}}{r} = \begin{cases}
\tbraproxy{p}{s}{\lbl{l_i}\!:\! \proj{G_i}{r}}{\lbl{i \in I}}
	& \text{if } \ppt{r} = \ppt{q} \\
\routertrans{p}{q}{j}{\lbl{l_i}\!:\! \proj{G_i}{r}}{\lbl{i \in I}}
	& \text{if } \ppt{r} = \ppt{p} \\
\proj{G_j}{r}
	& \text{otherwise}
\end{cases}
\]

\paragraph{LTS Semantics over Local Types}
We define the LTS semantics
over local types,
denoted by $\ltreduce{T}{T'}{l}$,
in \cref{fig:newlocal}.
We highlight and explain the new rules.

\begin{figure}[t]
\small
\begin{prooftree}
\AxiomC{}
\RightLabel{\rulename{Lr1}}
\UnaryInfC{$
\ltreducelong
	{\tsel{q}{\lbl{l_i}\!:\! T_i}{\lbl{i \in I}}}
	{\ltype{T_j}}
	{\aout{p}{q}{j}}
$}
\end{prooftree}

\begin{prooftree}
\AxiomC{}
\RightLabel{\rulename{Lr2}}
\UnaryInfC{$
\ltreducelong
	{\tbra{q}{\lbl{l_i}\!:\! T_i}{\lbl{i \in I}}}
	{T_j}
	{\ain{q}{p}{j}}
$}
\end{prooftree}

\begin{prooftree}
\AxiomC{$
\ltreduce
	{T[\trec{T} / \trecvar]}
	{T'}
	{l}
$}
\RightLabel{\rulename{Lr3}}
\UnaryInfC{$
\ltreduce
	{\trec{T}}
	{T'}
	{l}
$}
\end{prooftree}

\begin{prooftree}
\AxiomC{}
\RightLabel{\colorbox{ColourShade}{\rulename{Lr4}}}
\UnaryInfC{$
\ltreducelong
	{\tselproxy{q}{s}{\lbl{l_i}\!:\! T_i}{\lbl{i \in I}}}
	{T_j}
	{\via{s}{\aout{p}{q}{j}}}
$}
\end{prooftree}

\begin{prooftree}
\AxiomC{}
\RightLabel{\colorbox{ColourShade}{\rulename{Lr5}}}
\UnaryInfC{$
\ltreducelong
	{\tbraproxy{q}{s}{\lbl{l_i}\!:\! T_i}{\lbl{i \in I}}}
	{T_j}
	{\via{s}{\ain{q}{p}{j}}}
$}
\end{prooftree}

\begin{prooftree}
\AxiomC{}
\RightLabel{\colorbox{ColourShade}{\rulename{Lr6}}}
\UnaryInfC{$
\ltreducelong
	{\router{p}{q}{\lbl{l_i}\!:\! T_i}{\lbl{i \in I}}}
	{\routertrans{p}{q}{j}{\lbl{l_i}\!:\! T_i}{\lbl{i \in I}}}
	{\via{s}{\aout{p}{q}{j}}}
$}
\end{prooftree}

\begin{prooftree}
\AxiomC{}
\RightLabel{\colorbox{ColourShade}{\rulename{Lr7}}}
\UnaryInfC{$
\ltreducelong
	{\routertrans{p}{q}{j}{\lbl{l_i}\!:\! T_i}{\lbl{i \in I}}}
	{T_j}
	{\via{s}{\ain{p}{q}{j}}}
$}
\end{prooftree}

\begin{prooftree}
\AxiomC{$\forall \lbl{i \in I}. ~ \ltreduce{T_i}{T'_i}{l}$}
\AxiomC{$\subj{\lbl{l}} \notin \{\ppt{p}, \ppt{q}\}$}
\RightLabel{\colorbox{ColourShade}{\rulename{Lr8}}}
\BinaryInfC{$
\ltreducelong
	{\router{p}{q}{\lbl{l_i}\!:\! T_i}{\lbl{i \in I}}}
	{\router{p}{q}{\lbl{l_i}\!:\! T'_i}{\lbl{i \in I}}}
	{l}
$}
\end{prooftree}

\begin{prooftree}
\AxiomC{$\ltreduce{T_j}{T'_j}{l}$}
\AxiomC{$\subj{\lbl{l}} \neq \ppt{q}$}
\AxiomC{$\forall \lbl{i \in I \setminus \{ j \}}. ~ \ltype{T'_i} = \ltype{T_i}$}
\RightLabel{\colorbox{ColourShade}{\rulename{Lr9}}}
\TrinaryInfC{$
\ltreducelong
	{\routertrans{p}{q}{j}{\lbl{l_i}\!:\! T_i}{\lbl{i \in I}}}
	{\routertrans{p}{q}{j}{\lbl{l_i}\!:\! T'_i}{\lbl{i \in I}}}
	{l}
$}
\end{prooftree}

\begin{prooftree}
\AxiomC{$\lbl{l} = \lbl{\via{s}{\cdot}}$}
\AxiomC{$\subj{\lbl{l}} \neq \ppt{q}$}
\AxiomC{$\forall \lbl{i \in I}. ~ \ltreduce{T_i}{T'_i}{l}$}
\RightLabel{\colorbox{ColourShade}{\rulename{Lr10}}}
\TrinaryInfC{$
\ltreducelong
	{\tsel{q}{\lbl{l_i}\!:\! T_i}{\lbl{i \in I}}}
	{\tsel{q}{\lbl{l_i}\!:\! T'_i}{\lbl{i \in I}}}
	{l}
$}
\end{prooftree}

\begin{prooftree}
\AxiomC{$\lbl{l} = \lbl{\via{s}{\cdot}}$}
\AxiomC{$\subj{\lbl{l}} \neq \ppt{q}$}
\AxiomC{$\forall \lbl{i \in I}. ~ \ltreduce{T_i}{T'_i}{l}$}
\RightLabel{\colorbox{ColourShade}{\rulename{Lr11}}}
\TrinaryInfC{$
\ltreducelong
	{\tbra{q}{\lbl{l_i}\!:\! T_i}{\lbl{i \in I}}}
	{\tbra{q}{\lbl{l_i}\!:\! T'_i}{\lbl{i \in I}}}
	{l}
$}
\end{prooftree}
\captionof{figure}{LTS over Local Types in \newtheory}
\label{fig:newlocal}
\end{figure}

We walk through rules \rulename{Lr4} and \rulename{Lr5}
from the perspective of role $\ppt{p}$.

\begin{itemize}[wide, labelwidth=!, labelindent=0pt]

\item \rulename{Lr4} and \rulename{Lr5} are
analogue to \rulename{Lr1} and \rulename{Lr2}
for sending and receiving messages respectively.
The exception is that the new rules match
on the router role $\ppt{s}$ on the local type
and the routed label.

\end{itemize}

We walk through rules \rulename{Lr6}, \rulename{Lr7},
\rulename{Lr10} and \rulename{Lr11}
from the perspective of role $\ppt{s}$.

\begin{itemize}[wide, labelwidth=!, labelindent=0pt]

\item \rulename{Lr6} and \rulename{Lr7} are
analogue to \rulename{Gr1} and \rulename{Gr2}.
Intuitively, the router $\ppt{s}$ holds a
``global'' perspective on the interaction
between $\ppt{p}$ and $\ppt{q}$, which explains
the correspondence with the LTS semantics over global
types.

\item \rulename{Lr10} and \rulename{Lr11} allow
the router to perform routing actions before handling
their own direct communication.
The syntax $\lbl{l} = \lbl{\via{s}{\cdot}}$
means that the label $\lbl{l}$ is ``of the form'' of a routing
action, i.e. there exists some $\ppt{p}, \ppt{q}, \lbl{j}$
such that $\lbl{l} = \lbl{\via{s}{\aout{p}{q}{j}}}$ or
$\lbl{l} = \lbl{\via{s}{\ain{p}{q}{j}}}$.
The constraint of $\subj{\lbl{l}} \neq \ppt{q}$
prevents the violation of the syntactic order of messages
sent and received by $\ppt{q}$.

Curious readers can consider the examples
$\proj{G_1}{s}$ and $\proj{G_2}{s}$ below
to see why this constraint is needed.
\begin{align*}
\gtype{G_1} &=
	\gcommone{s}{r}{\lbl{M1}}.~
	\grouteone{r}{q}{s}{\lbl{M2}}.~ \gtend \\
\proj{G_1}{s} &=
	\tselone{r}{\lbl{M1}}. ~
	\routerone{r}{q}{\lbl{M2}}{\gtend} \\
\gtype{G_2} &=
	\gcommone{s}{r}{\lbl{M1}}.~
	\grouteone{p}{r}{s}{\lbl{M2}}.~ \gtend \\
\proj{G_2}{s} &=
	\tselone{r}{\lbl{M1}}. ~
	\routerone{p}{r}{\lbl{M2}}{\gtend}
\end{align*}
\end{itemize}

As for the remaining rules,
\rulename{Lr8} and \rulename{Lr9} are the local counterparts to
\rulename{Gr4} and \rulename{Gr5} because the router
holds a ``global'' perspective on the communication,
so transitions that do not violate the syntactic order
of messages between roles $\ppt{p}$ and $\ppt{q}$
are allowed.

\paragraph{LTS Semantics over Configurations}
Let $\mathcal{P}$ denote the set of participants in
the communication automaton.
Also let $\lty{p}$ denote the local type of a participant
$\pinP$.

A \emph{configuration} describes the
state of the communication automaton with respect to
each participant $\pinP$.
By definition of our LTS semantics, this includes
\emph{intermediate} states, so a configuration
would also need to express the state of messages
in transit.

We inherit the definition from \cite{2013Automata},
restated in \cref{def:newconfig}.

\begin{definition}[Configuration]
A configuration $s = (\vec{\ltype{T}}; ~ \vec w)$ of a system of
local types $\{ \lty{p} \}_{\pinP}$
is defined as a pair of:

\begin{itemize}[wide, labelwidth=!, labelindent=0pt]

\item $\vec{\ltype{T}} = (\lty{p})_{\pinP}$
is the collection of local types.
$\lty{p}$ describes the communication structure
from the local perspective of participant $\pinP$.

\item $\vec w = (w_{\mroles{p}{q}})_{\mrole{p} \neq \mrole{q} \in \mathcal{P}}$
is the collection of \emph{unbounded buffers}.
The unbounded buffer $w_{\ppt{p}\ppt{q}}$ represents a (FIFO)
queue of messages sent by $\ppt{p}$ but not yet
received by $\ppt{q}$.

\end{itemize}

\label{def:newconfig}
\end{definition}

\subparagraph{Remark:}
The \emph{subtyping} relation defined on
local types (see \cite{2013Automata})
can be extended to configurations:
\begin{prooftree}
\AxiomC{$\vec w = \vec w'$}
\AxiomC{$
\forall \ppt{p} \in \mathcal{P}. ~
\ltype{T}_\ppt{p} \subtype \ltype{T}'_\ppt{p}
$}
\BinaryInfC{$
(\vec{\ltype{T}}; \vec w) \subtype (\vec{\ltype{T'}}; \vec w')
$}
\end{prooftree}
We proceed to define the LTS over configurations in
\cref{def:newltsconfig}, highlighting the extensions
required for \newtheory.

\begin{definition}[LTS Semantics over Configurations]
The LTS semantics over configurations is defined by
the relation $\treduce{s_T}{s'_T}{l}$.

Let $s_T = (\vec{\ltype{T}}; ~ \vec w)$ and $s'_T = (\vec{\ltype{T'}}; ~ \vec w')$.
We define the specific transitions on $\vec T$ and $\vec w$
by case analysis on the label $\lbl{l}$.

\begin{itemize}[wide, labelwidth=!, labelindent=0pt]

\item $\lbl{l = \aout{p}{q}{j}}$

Then $\ltreduce{\ltype{T}_\ppt{p}}{\ltype{T}'_\ppt{p}}{l}~$
because $\ppt{p}$
initiates the action, so
$\ltype{\ltype{T}'_\ppt{p'} = \ltype{T}_{\ppt{p'}}}$
for all $\ppt{p'} \neq \ppt{p}$.

The message $\lbl{j}$ is in transit from $\ppt{p}$ to $\ppt{q}$,
so $w'_{\mroles{p}{q}} = w_{\mroles{p}{q}} \cdot \lbl{j}$
($\lbl{j}$ is appended to the queue of in-transit messages
sent from $\ppt{p}$ to $\ppt{q}$),
and unrelated buffers $w'_{\mroles{p'}{q'}} = w_{\mroles{p}{q}}$
are untouched for all $\mroles{p'}{q'} \neq \mroles{p}{q}$.

\item $\lbl{l = \ain{p}{q}{j}}$

Then $\ltreduce{\ltype{T}_\ppt{q}}{\ltype{T}'_\ppt{q}}{l}~$
because $\ppt{q}$
initiates the action, so
$\ltype{\ltype{T}'_\ppt{p'} = \ltype{T}_{\ppt{p'}}}$
for all $\ppt{p'} \neq \ppt{q}$.

The message $\lbl{j}$ is no longer in transit
from $\ppt{p}$ to $\ppt{q}$ as it is received by $\ppt{q}$,
so $w_{\mroles{p}{q}} = \lbl{j} \cdot w'_{\mroles{p}{q}}$
($\lbl{j}$ is removed from the front of the queue of in-transit
messages sent from $\ppt{p}$ to $\ppt{q}$),
and unrelated buffers $w'_{\mroles{p'}{q'}} = w_{\mroles{p}{q}}$
are untouched for all $\mroles{p'}{q'} \neq \mroles{p}{q}$.

\item \colorbox{ColourShade}{$\lbl{l} = \via{s}{\aout{p}{q}{j}}$}

Then $\ltreduce{\ltype{T}_\ppt{p}}{\ltype{T}'_\ppt{p}}{l}~$
because $\ppt{p}$
initiates the action.
Because the send action is routed, we also need
$\ltreduce{\ltype{T}_\ppt{s}}{\ltype{T}'_\ppt{s}}{l}$.
This means
$\ltype{T}'_\ppt{p'} = \ltype{T}_{\ppt{p'}}$
for all $\ppt{p'} \notin \{\ppt{p} , \ppt{s} \}$.

The message $\lbl{j}$ is in transit from $\ppt{p}$ to $\ppt{q}$,
so $w'_{\mroles{p}{q}} = w_{\mroles{p}{q}} \cdot \lbl{j}$
and unrelated buffers $w'_{\mroles{p'}{q'}} = w_{\mroles{p}{q}}$
are untouched for all $\mroles{p'}{q'} \neq \mroles{p}{q}$.

\item \colorbox{ColourShade}{$\lbl{l} = \via{s}{\ain{p}{q}{j}}$}

Then $\ltreduce{T_\ppt{q}}{T'_\ppt{q}}{l}~$
because $\ppt{q}$
initiates the action.
Because the receive action is routed, we also need
$\ltreduce{\ltype{T}_\ppt{s}}{\ltype{T}'_\ppt{s}}{l}$.
This means
$\ltype{T}'_\ppt{p'} = \ltype{T}_{\ppt{p'}}$
for all $\ppt{p'} \notin \{\ppt{q} , \ppt{s} \}$.

The message $\lbl{j}$ is no longer in transit
from $\ppt{p}$ to $\ppt{q}$ as it is received by $\ppt{q}$,
so $w_{\mroles{p}{q}} = \lbl{j} \cdot w'_{\mroles{p}{q}}$,
and unrelated buffers $w'_{\mroles{p'}{q'}} = w_{\mroles{p}{q}}$
are untouched for all $\mroles{p'}{q'} \neq \mroles{p}{q}$.

\end{itemize}
\label{def:newltsconfig}
\end{definition}

Routed actions are carried out by the router,
so it makes sense for the local type of the router
to also make a step.
The semantics of the message buffers for routed actions
are the same as their non-routed counterparts; the
only difference is that these message buffers are ``managed''
by the router.%

\paragraph{Projection for Configurations}
When considering the grammar of global types $\gtype{G}$
extended to include intermediate states,
we can obtain the \emph{projected configuration}
from a global type $\gtype{G}$ with participants $\mathcal{P}$:
\[
\projconf{G} =
\left(
	\{ \proj{G}{p} \}_{\pinP} ~ ; ~
	\projconf{G}_{\{ \epsilon \}_{\qqinP}}
\right)
\]
The collection of local types is obtained by
projecting $\gtype{G}$ onto each participant $\pinP$.
The contents of the buffers is defined as
$\projconf{G}_{\{ w_{\mroles{q}{q'}} \}_{\qqinP}}$.
$\epsilon$ denotes an empty buffer.
We inherit the definitions presented in \cite{2013Automata},
and introduce additional rules in
\cref{fig:buffer}.

\begin{figure}[t]
\begin{gather*}
\projconf{
\gtransroute{p}{p'}{s}{j}{\lbl{l_i}\!:\! G_i}{\lbl{i \in I}}
}_{\{ w_{\mroles{q}{q'}} \}_{\qqinP}}
	= \projconf{G_j}_{
	\{ w_{\mroles{q}{q'}} \}_{\qqinP}
	[w_{\mroles{p}{p'}} \mapsto w_{\mroles{p}{p'}} \cdot \lbl{j}]
	} \\
\projconf{
\groute{p}{p'}{s}{\lbl{l_i}\!:\! G_i}{\lbl{i \in I}}
}_{\{ w_{\mroles{q}{q'}} \}_{\qqinP}}
	= \projconf{G_i}_{
	\{ w_{\mroles{q}{q'}} \}_{\qqinP}
	} \text{ for }\lbl{i \in I}\\
\text{since}~\forall \lbl{i, j \in I}. ~
\projconf{G_i}_{
	\{ w_{\mroles{q}{q'}} \}_{\qqinP}
} = \projconf{G_j}_{
	\{ w_{\mroles{q}{q'}} \}_{\qqinP}
}
\end{gather*}
\captionof{figure}{Projection of Buffer Contents from Global Type in
\newtheory}
\label{fig:buffer}
\end{figure}

The semantics of the message buffers
for routed actions are the same as their
non-routed counterparts,
so the projected contents of the buffers
for routed communication are
the same as those under non-routed communication.

%% file: appendix/proofs.tex
\section{Appendix for Proofs in \cref{sec:theory}}
\label{app:proofs}
\subsection{Auxiliary Lemmas}
\begin{lemma}[Local LTS Preserves Merge]
Let $\ltype{T_1, T_2}$ be local types.
Suppose $\tmerge{T_1}{T_2}$ exists.
\[
\forall \lbl{l}, \ltype{T'_1}, \ltype{T'_2}. ~ \left(
(\ltreduce{T_1}{T'_1}{l}) \wedge (\ltreduce{T_2}{T'_2}{l})
	\Longrightarrow
(\tmerge{T'_1}{T'_2}) \text{ exists} \right)
\]
\label{lem:localltspreservemerge}
\end{lemma}
\vspace{-8mm}
\begin{proof}
By simultaneous induction on
$\tmerge{T_1}{T_2}$, $\ltreduce{T_1}{T'_1}{l}$,
and\\$\ltreduce{T_2}{T'_2}{l}$.
\end{proof}

\begin{lemma}[Projection and Participation]
\[
\forall \gtype{G}, \ppt{p}.~ \left(
\proj{G}{p} = \gtend \Longleftrightarrow \ppt{p} \notin \pt{\gtype{G}}
\right)
\]
\label{lem:projpt}
\end{lemma}
\vspace{-8mm}
\begin{proof}
Prove $(\Longrightarrow)$ by induction on the structure of $\gtype{G}$.
Prove $(\Longleftarrow)$ using the contrapositive (stated below)
by induction on the derivation of $\pt{\gtype{G}}$:
\[
\ppt{p} \in \pt{\gtype{G}} \Longrightarrow \proj{G}{p} \neq \gtend.
\]
\end{proof}

\begin{lemma}[Encoding Defines Centroid]
Given an encoding of global type $\gtype{G}$ with respect to the
router role $\ppt{s}$,
the router role is the centroid of the encoded communication:
\[
\centroid{\enc{\gtype{G}}{s}}{s}.
\]
\label{lem:enccenter}
\end{lemma}
\vspace{-8mm}
\begin{proof}
By induction on the structure of $\gtype{G}$.
\end{proof}

\begin{lemma}[Encoding and Substitution Permute]
Let $\gtype{G, G'}$ be global types,
and $\ppt{s}$ be a role.
\[
\enc{\gtype{G[G' / \gtvar{t}]}}{s} =
	\enc{\gtype{G}}{s}\left[ \enc{\gtype{G'}}{s} / \gtvar{t} \right]
\]
\label{lem:encsub}
\end{lemma}
\vspace{-8mm}
\begin{proof}
By induction on the structure of $\gtype{G}$.
\end{proof}

\begin{lemma}[Encoding Preserves Participants]
\[
\forall \gtype{G}, \ppt{s}. ~
\left( \pt{\gtype{G}} \subseteq \pt{\enc{\gtype{G}}{s}} \right)
\]
\label{lem:encpreservept}
\end{lemma}
\vspace{-8mm}
\begin{proof}
The following is logically equivalent:
\[
\forall \ppt{r}, \ppt{s}. ~ \left(
\ppt{r} \in \pt{\gtype{G}} \Longrightarrow \ppt{r} \in \pt{\enc{\gtype{G}}{s}}
\right)
\]
We prove this by induction on the structure of $\gtype{G}$.
\end{proof}

\begin{lemma}[Encoding Preserves Privacy]
The encoding on global types will not introduce
non-server roles that were not participants of the
original communication.
\[
\forall \ppt{r}, \ppt{s}, \gtype{G}. ~ \left(
\ppt{r} \neq \ppt{s} \wedge \ppt{r} \notin \pt{\gtype{G}}
\Longrightarrow \ppt{r} \notin \pt{\enc{\gtype{G}}{s}}
\right)
\]
\label{lem:encprivacy}
\end{lemma}
\vspace{-8mm}
\begin{proof}
The following is logically equivalent.
\[
\forall \ppt{r}, \ppt{s}, G. ~ \left(
\ppt{r} \neq \ppt{s} \wedge
\ppt{r} \in \pt{\enc{\gtype{G}}{s}}
\Longrightarrow  \ppt{r} \in \pt{\gtype{G}}
\right)
\]

We prove this by induction on the structure of $\gtype{G}$,
assuming that $\ppt{r} \neq \ppt{s}$ for arbitrary
roles $\ppt{r}$, $\ppt{s}$.
\end{proof}

\paragraph{Proof of \cref{lem:enclink}}
\begin{quote}
The projection of an encoded global type $\proj{\enc{\gtype G}{s}}{r}$ is equal
to the encoded local type after projection $\enclocal{\proj{G}{r}}{r}{s}$, with
respect to router $\ppt s$, i.e.
\vspace{0mm}
\[
\forall \ppt{r}, \ppt{s}, \gtype{G}. ~ \left(
\ppt{r} \neq \ppt{s}
\Longrightarrow
\proj{\enc{\gtype{G}}{s}}{r} = \enclocal{\proj{G}{r}}{r}{s}
\right)
\]

\end{quote}
\begin{proof}
By induction on the structure of $\gtype{G}$,
\cref{lem:encpreservept,lem:encprivacy}.
\end{proof}

\begin{lemma}[Local Type Encoding Preserves Equality of Projection]
  $\forall \gtype{G_1}, \gtype{G_2}, \ppt{r}, \ppt{s}.$
\[
\left(
(\proj{G_1}{r}) = (\proj{G_2}{r}) \wedge \ppt{r} \neq \ppt{s}
	\Longrightarrow
\proj{\enc{\gtype{G_1}}{s}}{r} = \proj{\enc{\gtype{G_2}}{s}}{r}
\right)
\]
\label{lem:encprojeq}
\end{lemma}
\vspace{-8mm}
\begin{proof} By consequence from \cref{lem:enclink}.

Take $\gtype{G_1}, \gtype{G_2}, \ppt{r}, \ppt{s}$ arbitrarily.
Assume $(\proj{G_1}{r}) = (\proj{G_2}{r})$ and $\ppt{r} \neq \ppt{s}$.

We need to show
$\proj{\enc{\gtype{G_1}}{s}}{r} = \proj{\enc{\gtype{G_2}}{s}}{r}$,
but by \cref{lem:enclink}, it is sufficient to show
\[
\enclocal{\proj{\gtype{G_1}}{r}}{r}{s}
= \enclocal{\proj{\gtype{G_2}}{r}}{r}{s}.
\]

The result follows by congruence from the assumption.

\end{proof}

\begin{lemma}[Encoding on Global Types Preserves Merge]
Take global types $\gtype{G_1}, \gtype{G_2}$
and roles $\ppt{r}, \ppt{s}$ such that $\ppt{r} \neq \ppt{s}$.
Suppose $\proj{G_1}{r}$ and $\proj{G_2}{r}$ exist.
\[
\tmerge{(\proj{G_1}{r})}{(\proj{G_2}{r})} \text{ exists}
	\Longrightarrow
\tmerge{(\proj{\enc{\gtype{G_1}}{s}}{r})}{(\proj{\enc{\gtype{G_2}}{s}}{r})}
	\text{ exists}
\]
\label{lem:encglobalpreservemerge}
\end{lemma}
\vspace{-8mm}
\begin{proof}
By induction on the structure of $\tmerge{T_1}{T_2}$,
\cref{lem:projpt,lem:encprojeq}.
\end{proof}

\begin{lemma}[Encoding Preserves Projection]
Let $\gtype{G}$ be a global type.
$\forall \ppt{r}, \ppt{s}$.
\[
\proj{G}{r} \text{ exists }
	\Longrightarrow
\proj{ \enc{\gtype{G}}{s} }{r} \text{ exists }
\]
\label{lem:encproj}
\end{lemma}
\vspace{-8mm}
\begin{proof}
By induction on the structure of $\gtype{G}$.

\begin{enumerate}[wide, labelwidth=!, labelindent=0pt]
\item $\gtype{G = \gtend}$, $\gtype{G = \gtvar{t}}$

As $\enc{\gtype{G}}{s} = G$, if $\proj{G}{r}$ exists,
so does $\proj{ \enc{\gtype{G}}{s} }{r}$.

\item $\gtype{G = \gtrecur{t}{G'}}$

By assumption, $\proj{\gtrecur{t}{G'}}{r}$ exists.
Note that $\proj{G'}{r}$ exists regardless of $\ppt{r}$'s
participation in $\gtype{G'}$.

By induction, $\proj{\enc{\gtype{G'}}{s}}{r}$ exists.

To show $\proj{\enc{\gtrecur{t}{G'}}{s}}{r}$ exists,
consider $\ppt{r}$ by case:

\begin{itemize}
\item $\ppt{r} \in \pt{\enc{\gtype{G'}}{s}}$:
\[
\proj{\enc{\gtrecur{t}{G'}}{s}}{r}
= \proj{\gtrecur{t}{\enc{\gtype{G'}}{s}}}{r}
= \gtrecur{t}{(\proj{\enc{\gtype{G'}}{s}}{r})}
\]
As $\proj{\enc{\gtype{G'}}{s}}{r}$ exists,
so does $\proj{\enc{\gtrecur{t}{G'}}{s}}{r}$.

\item $\ppt{r} \notin \pt{\enc{\gtype{G'}}{s}}$:
\[
\proj{\enc{\gtrecur{t}{G'}}{s}}{r}
= \proj{\gtrecur{t}{\enc{\gtype{G'}}{s}}}{r}
= \tend
\]
\end{itemize}

\item $\gtype{G = \gcomm{p}{q}{\lbl{l_i}\!:\! G_i}{\lbl{i \in I}}}$

To determine $\enc{\gtype{G}}{s}$, consider $\ppt{s}$ by case:

\begin{itemize}
\item $\ppt{s} \in \{ \ppt{p}, \ppt{q} \}$:

Then $\enc{\gtype{G}}{s} = \gcomm{p}{q}{\lbl{l_i}\!:\! \enc{\gtype{G_i}}{s}}{\lbl{i \in I}}$.

To show $\proj{\enc{\gtype{G}}{s}}{r}$ exists,
consider $\ppt{r}$ by case:

\begin{itemize}
\item $\ppt{r} = \ppt{p}$:
Then $\proj{G}{p} = \tsel{q}{\lbl{l_i}\!:\! \proj{G_i}{p}}{\lbl{i \in I}}$.

By induction, $\proj{\enc{\gtype{G_i}}{s}}{p}$ exists for $\lbl{i \in I}$.

$\proj{\enc{\gtype{G}}{s}}{p} = \tsel{q}{\lbl{l_i}\!:\! \proj{\enc{\gtype{G_i}}{s}}{p}}{\lbl{i \in I}}$.

As projections of the encoded continuations exist,
so does $\proj{\enc{\gtype{G}}{s}}{p}$.

\item $\ppt{r} = \ppt{q}$: Follows similarly from above.

\item $\ppt{r} \notin \{\ppt{p},\ppt{q}\}$:
Then $\proj{G}{r} = \underset{i \in I}{\MERGEOP}\proj{G_i}{r}$,
so the merge exists.

By induction, $\proj{\enc{\gtype{G_i}}{s}}{r}$ exists for $\lbl{i \in I}$.

$\proj{\enc{\gtype{G}}{s}}{r} =
	\underset{i \in I}{\MERGEOP}\proj{\enc{\gtype{G_i}}{s}}{r}$,
and this merge exists by \cref{lem:encglobalpreservemerge}.
\end{itemize}

\item $\ppt{s} \notin \{ \ppt{p}, \ppt{q} \}$:

Then $\enc{\gtype{G}}{s} = \groute{p}{q}{s}{\lbl{l_i}\!:\! \enc{\gtype{G_i}}{s}}{\lbl{i \in I}}$.

To show $\proj{\enc{\gtype{G}}{s}}{r}$ exists,
consider $\ppt{r}$ by case:
\begin{itemize}
\item $\ppt{r} = \ppt{p}$:
Then $\proj{G}{p} = \tsel{q}{\lbl{l_i}\!:\! \proj{G_i}{p}}{\lbl{i \in I}}$.

By induction, $\proj{\enc{\gtype{G_i}}{s}}{p}$ exists for $\lbl{i \in I}$.

$\proj{\enc{\gtype{G}}{s}}{p} =
	\tselproxy{q}{s}{\lbl{l_i}\!:\! \proj{\enc{\gtype{G_i}}{s}}{p}}{\lbl{i \in I}}$.

As projections of the encoded continuations exist,
so does $\proj{\enc{\gtype{G}}{s}}{p}$.

\item $\ppt{r} = \ppt{q}$: Follows similarly from above.
\item $\ppt{r} = \ppt{s}$:
Then $\proj{G}{s} = \underset{i \in I}{\MERGEOP}\proj{G_i}{s}$.

By induction, $\proj{\enc{\gtype{G_i}}{s}}{p}$ exists for $\lbl{i \in I}$.

$\proj{\enc{\gtype{G}}{s}}{s} =
	\router{p}{q}{\lbl{l_i}\!:\! \proj{\enc{\gtype{G_i}}{s}}{s}}{\lbl{i \in I}}$.

As projections of the encoded continuations exist,
so does $\proj{\enc{\gtype{G}}{s}}{s}$.

\item $\ppt{r} \notin \{\ppt{p},\ppt{q},\ppt{s}\}$:
Then $\proj{G}{r} = \underset{i \in I}{\MERGEOP}\proj{G_i}{r}$,
so the merge exists.

By induction, $\proj{\enc{\gtype{G_i}}{s}}{r}$ exists for $\lbl{i \in I}$.

$\proj{\enc{\gtype{G}}{s}}{r} =
	\underset{i \in I}{\MERGEOP}\proj{\enc{\gtype{G_i}}{s}}{r}$,
and this merge exists by \cref{lem:encglobalpreservemerge}.%
\end{itemize} %
\end{itemize} %
\end{enumerate}
\end{proof}

\vspace{-5mm}
\subsection{Proof of \cref{th:traceeq}}
\begin{quote}
Let $\gtype{G}$ be a global type with participants
$\mathcal{P} = \pt{\gtype{G}}$,
and let $\vec T = \{ \proj{G}{p} \}_{\pinP}$ be the local
types projected from $\gtype{G}$.
Then $\gtype{G} \approx (\vec T, \vec \epsilon)$.
\end{quote}
\vspace{-2mm}
\begin{proof}
Direct consequence of
\cref{lem:stepeq}.
\end{proof}

\begin{lemma}[Step Equivalence]
For all global types $\gtype{G}$ and configurations $s$,
if $\projconf{G} \subtype s$,
then $\gtreduce{G}{G'}{l} \Longleftrightarrow \treduce{s}{s'}{l}$
such that $\projconf{G'} \subtype s'$.
\label{lem:stepeq}
\end{lemma}
\vspace{-3mm}
\begin{proof}
By induction on the possible transitions in the LTSs
over global types (to prove $\Longrightarrow$,
i.e. \emph{soundness})
and configurations (to prove $\Longleftarrow$,
i.e. \emph{completeness}).

\paragraph{Notation conventions}
We use the following notation for decomposing configurations
and projected configurations.
\[\begin{array}{rcl}
s &= & \{ T_\ppt{q} \}_{\ppt{q} \in \mathcal{P}},~
	\{ w_{\mroles{q}{q'}} \}_{\qqinP} \\
s' &= & \{ T'_\ppt{q} \}_{\ppt{q} \in \mathcal{P}},~
	\{ w'_{\mroles{q}{q'}} \}_{\qqinP} \\
\projconf{G} &= & \{ \hat{T_\ppt{q}} \}_{\ppt{q} \in \mathcal{P}},~
	\{ \hat{w}_{\mroles{q}{q'}} \}_{\qqinP} \\
\projconf{G'} &= & \{ \hat{T'_\ppt{q}} \}_{\ppt{q} \in \mathcal{P}},~
	\{ \hat{w'}_{\mroles{q}{q'}} \}_{\qqinP} \\
\end{array}\]

\item \textbf{Soundness}

By rule induction on LTS semantics
over global types.

For each transition $\gtreduce{G}{G'}{l}$, we
take the configuration $s = \projconf{G}$,
derive $\gtreduce{G}{G'}{l}$ and $\treduce{s}{s'}{l}$
under the respective LTSs,
and show that $s' \subtype \projconf{G'}$.

The proofs for rules \rulename{Gr1-5} are
the same as in \cite{2013Automata}.
We focus on the new rules introduced for routing.

\begin{itemize}[wide, labelwidth=!, labelindent=0pt]

\item \rulename{Gr6},
  where $\gtype{G} = \groute{p}{p'}{s}{\lbl{l_i}\!:\! G_i}{\lbl{i \in I}}$,
  \hfill \hfill \linebreak
$\gtype{G'} = \gtransroute{p}{p'}{s}{j}{\lbl{l_i}\!:\! G_i}{\lbl{i \in I}}$
  with $\lbl{l} = \via{s}{\aout{p}{p'}{j}}$.

Then $s = \projconf{G}$ where
\begin{align*}
\ltype{T_\ppt{p}}
	&= \tselproxy{p'}{s}{\lbl{l_i}\!:\! \proj{G_i}{p}}{\lbl{i \in I}} \\
\ltype{T_\ppt{p'}}
	&= \tbraproxy{p}{s}{\lbl{l_i}\!:\! \proj{G_i}{p'}}{\lbl{i \in I}} \\
\ltype{T_\ppt{s}}
	&= \router{p}{p'}{\lbl{l_i}\!:\! \proj{G_i}{s}}{\lbl{i \in I}}\\
\ltype{T_\ppt{r}}
	&= \underset{i \in I}{\MERGEOP}~\proj{G_i}{r}
		\text{ for } \ppt{r} \notin \{ \ppt{p},\ppt{p'},\ppt{s} \}\\
\{ w_{\mroles{q}{q'}} \}_{\qqinP}
	&=  \projconf{G_i}_{\{\vec\epsilon\}}
		\text{ for some } \lbl{i \in I}
\end{align*}

\textbf{Global transition:}
We have
\begin{align*}
\ltype{\hat{T'_\ppt{p'}}} &= \tbraproxy{p}{s}{\lbl{l_i}\!:\! \proj{G_i}{p'}}{\lbl{i \in I}} \\
\ltype{\hat{T'_\ppt{s}}} &= \routertrans{p}{p'}{j}{\lbl{l_i}\!:\! \proj{G_i}{s}}{\lbl{i \in I}} \\
\ltype{\hat{T'_\ppt{r}}} &= \proj{G_j}{r}
	\text{ for } \ppt{r} \notin \{ \ppt{p'},\ppt{s} \}\\
\{ \hat{w'}_{\mroles{q}{q'}} \}_{\qqinP} &= \projconf{G_i}_
	{\{ \vec\epsilon \}[w_{\mroles{p}{p'}} \mapsto w_{\mroles{p}{p'}} \cdot \lbl{j}]} \text{ for some } \lbl{i \in I}
\end{align*}
So, $\hat{w'}_{\mroles{q}{q'}} = w_{\mroles{q}{q'}}$ for
$\mroles{q}{q'} \neq \mroles{p}{p'}$ and
$\hat{w'}_{\mroles{p}{p'}} = w_{\mroles{p}{p'}} \cdot \lbl{j}$.

\textbf{Configuration transition:}
Take $\ltype{T'_\ppt{r}} = \ltype{T_\ppt{r}}$
for $\ppt{r} \notin \{\ppt{p}, \ppt{s}\}$.

By \rulename{Lr4}, $\ltreduce{T_\ppt{p}}{T'_\ppt{p}}{l}$
where $\ltype{T'_\ppt{p}} = \proj{G_j}{p}$.

By \rulename{Lr6}, $\ltreduce{T_\ppt{s}}{T'_\ppt{s}}{l}$
where $\ltype{T'_\ppt{s}} =
\routertrans{p}{p'}{j}{\lbl{l_i}\!:\! \proj{G_i}{s}}{\lbl{i \in I}}$.

Also, ${w'}_{\mroles{q}{q'}} = w_{\mroles{q}{q'}}$ for
$\mroles{q}{q'} \neq \mroles{p}{p'}$ and
${w'}_{\mroles{p}{p'}} = w_{\mroles{p}{p'}} \cdot \lbl{j}$.

\textbf{Correspondence:}
We have
${w'}_{\mroles{q}{q'}} = \hat{w}_{\mroles{q}{q'}}$
for $\qqinP$ and
$\ltype{T'_\ppt{q}} = \ltype{\hat{T_\ppt{q}}}$
for $\ppt{q} \in \{ \ppt{p}, \ppt{p'}, \ppt{s} \}$.

For $\ppt{q} \notin \{ \ppt{p}, \ppt{p'}, \ppt{s} \}$,
we have
\[
\ltype{T'_\ppt{q}} = \underset{i \in I}{\MERGEOP}~\proj{G_i}{q}
	\subtype \proj{G_j}{q}
	= \ltype{\hat{T_\ppt{q}}}
\]

So $s' \subtype \projconf{G'}$.

\item \rulename{Gr7}
  where $\gtype{G} = \gtransroute{p}{p'}{s}{j}{\lbl{l_i}\!:\! G_i}{\lbl{i \in I}}$,
$\gtype{G'} = \gtype{G_j}$, and
$\lbl{l} = \via{s}{\ain{p}{p'}{j}}$

Then $s = \projconf{G}$ where
\begin{align*}
\ltype{T_\ppt{p'}}
	&= \tbraproxy{p}{s}{\lbl{l_i}\!:\! \proj{G_i}{p'}}{\lbl{i \in I}} \\
\ltype{T_\ppt{s}}
	&= \routertrans{p}{p'}{j}{\lbl{l_i}\!:\! \proj{G_i}{s}}{\lbl{i \in I}}\\
\ltype{T_\ppt{r}}
	&= \proj{G_j}{r}
		\text{ for } \ppt{r} \notin \{ \ppt{p'},\ppt{s} \}\\
\{ w_{\mroles{q}{q'}} \}_{\qqinP}
	&= \projconf{G_j}_
		{\{ \vec\epsilon \}[w_{\mroles{p}{p'}} \mapsto w_{\mroles{p}{p'}} \cdot \lbl{j}]}
\end{align*}

\textbf{Global transition:}
We have
\begin{align*}
\ltype{\hat{T'_\ppt{r}}} &=
	\proj{G_j}{r}
		\text{ for } \ppt{r} \in \mathcal{P} \\
\{ \hat{w'}_{\mroles{q}{q'}} \}_{\qqinP} &=
	\projconf{G_j}_
		{\{ \vec\epsilon \}}
\end{align*}
So,
$\hat{w'}_{\mroles{q}{q'}} = w_{\mroles{q}{q'}}$ for
$\mroles{q}{q'} \neq \mroles{p}{p'}$ and
$w_{\mroles{p}{p'}} = \lbl{j} \cdot \hat{w'}_{\mroles{p}{p'}}$.

\textbf{Configuration transition:}
Take $\ltype{T'_\ppt{r}} = \ltype{T_\ppt{r}}$
for $\ppt{r} \notin \{\ppt{p'}, \ppt{s}\}$.

By \rulename{Lr5}, $\ltreduce{T_\ppt{p}}{T'_\ppt{p}}{l}$
where $\ltype{T'_\ppt{p}} = \proj{G_j}{p}$.

By \rulename{Lr7}, $\ltreduce{T_\ppt{s}}{T'_\ppt{s}}{l}$
where $\ltype{T'_\ppt{s}} = \proj{G_j}{s}$.

Also, ${w'}_{\mroles{q}{q'}} = w_{\mroles{q}{q'}}$ for
$\mroles{q}{q'} \neq \mroles{p}{p'}$ and
$w_{\mroles{p}{p'}} = \lbl{j} \cdot {w'}_{\mroles{p}{p'}}$.

\textbf{Correspondence:}
We have
${w'}_{\mroles{q}{q'}} = \hat{w}_{\mroles{q}{q'}}$
for $\qqinP$ and
$\ltype{T'_\ppt{q}} = \ltype{\hat{T_\ppt{q}}}$
for $\ppt{q} \in \mathcal{P}$.

So, $s' = \projconf{G'}$.

\item \rulename{Gr8}
  where $\gtype{G} = \groute{p}{p'}{s}{\lbl{l_i}\!:\! G_i}{\lbl{i \in I}}$,

$\gtype{G'} = \groute{p}{p'}{s}{\lbl{l_i}\!:\! G'_i}{\lbl{i \in I}}$.

By hypothesis,
$\forall \lbl{i \in I}. ~ \gtreduce{G_i}{G'_i}{l}$
and $\subj{\lbl{l}} \notin \{ \ppt{p}, \ppt{p'} \}$.

By induction,
$\forall \lbl{i \in I}. ~ \treduce{\projconf{G_i}}{\projconf{G'_i}}{l}$.

To show that $\treduce{\projconf{G}}{\projconf{G'}}{l}$,
it is sufficient to show that\\
$\ltreduce{\proj{G}{q}}{\proj{G'}{q}}{l}$
for $\ppt{q} = \subj{\lbl{l}}$,
since the projections for \\
$\ppt{q'} \neq \subj{\lbl{l}}$ remain the same.

We know $\proj{G}{q} = \underset{i \in I}{\MERGEOP}\proj{G_i}{q}$
and $\proj{G'}{q} = \underset{i \in I}{\MERGEOP}\proj{G'_i}{q}$.

By induction,
$\ltreduce
	{\underset{i \in I}{\MERGEOP}\proj{G_i}{q}}
	{\underset{i \in I}{\MERGEOP}\proj{G'_i}{q}}
	{l}$,
so $\ltreduce{\proj{G}{q}}{\proj{G'}{q}}{l}$.

\item \rulename{Gr9}
  where $\gtype{G} = \gtransroute{p}{p'}{s}{j}{\lbl{l_i}\!:\! G_i}{\lbl{i \in I}}$, and

 $\gtype{G'} = \gtransroute{p}{p'}{s}{j}{\lbl{l_i}\!:\! G'_i}{\lbl{i \in I}}$.

By hypothesis,
$\ltreduce{G_j}{G'_j}{l}$,
$\ppt{p'} \neq \subj{\lbl{l}}$,
and $\forall \lbl{i \in I \setminus \{j\}}. ~ \gtype{G'_i = G_i}$.

By induction,
$\treduce{\projconf{G_j}}{\projconf{G'_j}}{l}$.

To show that $\treduce{\projconf{G}}{\projconf{G'}}{l}$,
it is sufficient to show that\\
$\ltreduce{\proj{G}{q}}{\proj{G'}{q}}{l}$
for $\ppt{q} = \subj{\lbl{l}}$,
since the projections for\\
$\ppt{q'} \neq \subj{\lbl{l}}$ remain the same.

We know $\proj{G}{q} = \proj{G_j}{q}$
and $\proj{G'}{q} = \proj{G'_j}{q}$.

By induction,
$\ltreduce
	{\proj{G_j}{q}}
	{\proj{G'_j}{q}}
	{l}$,
so $\ltreduce{\proj{G}{q}}{\proj{G'}{q}}{l}$.

\end{itemize}

\item \textbf{Completeness}

By considering the possible transitions in the LTS
over configurations, defined by
case analysis on the possible labels $\lbl{l}$.

For each transition $\treduce{s}{s'}{l}$,
we take the configuration $s$ from the reduction rule,
infer the structure of the global type $\gtype{G}$
such that $s = \projconf{G}$,
derive $\treduce{s}{s'}{l}$ and $\gtreduce{G}{G'}{l}$
under the respective LTSs,
and show that $s' \subtype \projconf{G'}$.

The proofs for $\lbl{l} = \aout{p}{q}{j}$
and $\lbl{l} = \ain{p}{q}{j}$
are the same as in \cref{subsec:semantics} of \cite{2013Automata}.
We focus on the new labels introduced for routing.
\begin{itemize}[wide, labelwidth=!, labelindent=0pt]

\item $\lbl{l} = \via{s}{\aout{p}{q}{j}}$:

Then $\ltype{T_\ppt{p}} =
\tselproxy{q}{s}{\lbl{l_i}\!:\! \proj{G_i}{p}}{\lbl{i \in I}}$.

Also, $\ltype{T_\ppt{s}}$ \emph{contains}
$\router{p}{q}{\lbl{l_i}\!:\! \proj{G_i}{s}}{\lbl{i \in I}}$ as subterm.
We denote this subterm $\ltype{\tilde{T_\ppt{s}}}$.

By definition of projection, $\gtype{G}$ has
$\groute{p}{q}{s}{\lbl{l_i}\!:\! G_i}{\lbl{i \in I}}$ as subterm.
We denote this subterm $\gtype{\tilde{G}}$.

Also by definition of projection, no action in $\gtype{G}$
will involve $\ppt{p}$ before $\gtype{\tilde{G}}$.

\textbf{Configuration transition:}

By \rulename{Lr4},
$\ltreduce{T_\ppt{p}}{T'_\ppt{p}}{l}$,
where $\ltype{T'_\ppt{p}} = \proj{G_j}{p}$.

By \rulename{Lr6},
$\ltreduce{\tilde{T}_\ppt{s}}{\tilde{T}'_\ppt{s}}{l}$,
where $\ltype{\tilde{T}'_\ppt{s}} =
	\routertrans{p}{q}{j}{\lbl{l_i}\!:\! \proj{G_i}{s}}{\lbl{i \in I}}$.

We get $\ltreduce{T_\ppt{s}}{T'_\ppt{s}}{l}$
by inversion lemma, as illustrated below.

\begin{prooftree}
\AxiomC{}
\RightLabel{\rulename{Lr6}}
\UnaryInfC{$\ltreduce{\tilde{T}_\ppt{s}}{\tilde{T}'_\ppt{s}}{l}$}
\UnaryInfC{$\vdots$}
\RightLabel{\rulename{Lr8,9,10,11} as needed}
\UnaryInfC{$\qquad\ltreducelong{T_\ppt{s}}{T'_\ppt{s}}{l}\qquad$}
\end{prooftree}

\textbf{Global transition:}

By \rulename{Gr6},
$\gtreduce{\tilde{G}}{\tilde{G}'}{l}$,
where $\gtype{\tilde{G}'} = \gtransroute{p}{q}{s}{j}{\lbl{l_i}\!:\! G_i}{\lbl{i \in I}}$.

We get $\gtreduce{G}{G'}{l}$ by inversion lemma,
as illustrated below.

\begin{prooftree}
\AxiomC{}
\RightLabel{\rulename{Gr6}}
\UnaryInfC{$\gtreduce{\tilde{G}}{\tilde{G}'}{l}$}
\UnaryInfC{$\vdots$}
\RightLabel{\rulename{Gr4,5,8,9} as needed}
\UnaryInfC{$\qquad\gtreducelong{G}{G'}{l}\qquad$}
\end{prooftree}

\textbf{Correspondence:}
Since the projections for
$\ppt{p'} \notin \{ \ppt{p}, \ppt{s} \}$
are unchanged,
it is sufficient to show that
$\ltype{T'_\ppt{p}} \subtype (\proj{\tilde{G'}}{p})$ and
$\ltype{\tilde{T}'_\ppt{s}} \subtype (\proj{\tilde{G'}}{s})$.
\begin{align*}
\proj{\tilde{G'}}{p}
	&= \proj{G_j}{p}
	= \ltype{T'_\ppt{p}} \\
\proj{\tilde{G'}}{s}
	&= \routertrans{p}{q}{j}{\lbl{l_i}\!:\! \proj{G_i}{s}}{\lbl{i \in I}}
	= \ltype{\tilde{T}'_\ppt{s}}
\end{align*}

\item $\lbl{l} = \via{s}{\ain{p}{q}{j}}$:

Then $\ltype{T_\ppt{q}} = \tbraproxy{p}{s}{\lbl{l_i}\!:\! \proj{G_i}{q}}{\lbl{i \in I}}$.

Also, $\ltype{T_\ppt{s}}$ \emph{contains}
$\routertrans{p}{q}{j}{\lbl{l_i}\!:\! \proj{G_i}{s}}{\lbl{i \in I}}$ as subterm.
We denote this subterm $\ltype{\tilde{T_\ppt{s}}}$.

By definition of projection, $\gtype{G}$ has
$\gtransroute{p}{q}{s}{j}{\lbl{l_i}\!:\! G_i}{\lbl{i \in I}}$ as subterm.
We denote this subterm $\gtype{\tilde{G}}$.

Also by definition of projection, no action in $\gtype{G}$
will involve $\ppt{q}$ before $\gtype{\tilde{G}}$.

\textbf{Configuration transition:}

By \rulename{Lr5},
$\ltreduce{T_\ppt{q}}{T'_\ppt{q}}{l}$,
where $\ltype{T'_\ppt{q}} = \proj{G_j}{q}$.

By \rulename{Lr7},
$\ltreduce{\tilde{T}_\ppt{s}}{\tilde{T}'_\ppt{s}}{l}$,
where $\ltype{\tilde{T}'_\ppt{s}} = \proj{G_j}{s}$.

We get $\ltreduce{T_\ppt{s}}{T'_\ppt{s}}{l}$
by inversion lemma, as illustrated below.

\begin{prooftree}
\AxiomC{}
\RightLabel{\rulename{Lr7}}
\UnaryInfC{$\ltreduce{\tilde{T}_\ppt{s}}{\tilde{T}'_\ppt{s}}{l}$}
\UnaryInfC{$\vdots$}
\RightLabel{\rulename{Lr8,9,10,11} as needed}
\UnaryInfC{$\qquad\ltreducelong{T_\ppt{s}}{T'_\ppt{s}}{l}\qquad$}
\end{prooftree}

\textbf{Global transition:}

By \rulename{Gr7},
$\gtreduce{\tilde{G}}{\tilde{G}'}{l}$,
where $\gtype{\tilde{G}'} = \gtype{G_j}$.

We get $\gtreduce{G}{G'}{l}$ by inversion lemma,
as illustrated below.

\begin{prooftree}
\AxiomC{}
\RightLabel{\rulename{Gr7}}
\UnaryInfC{$\gtreduce{\tilde{G}}{\tilde{G}'}{l}$}
\UnaryInfC{$\vdots$}
\RightLabel{\rulename{Gr4,5,8,9} as needed}
\UnaryInfC{$\qquad\gtreducelong{G}{G'}{l}\qquad$}
\end{prooftree}

\textbf{Correspondence:}
Since the projections for
$\ppt{p'} \notin \{ \ppt{q}, \ppt{s} \}$
are unchanged,
it is sufficient to show that
$\ltype{T'_\ppt{q}} \subtype (\proj{\tilde{G'}}{q})$ and
$\ltype{\tilde{T}'_\ppt{s}} \subtype (\proj{\tilde{G'}}{s})$.
\begin{align*}
\proj{\tilde{G'}}{q}
	&= \proj{G_j}{q}
	= \ltype{T'_\ppt{q}} \\
\proj{\tilde{G'}}{s}
	&= \proj{G_j}{s}
	= \ltype{\tilde{T}'_\ppt{s}}
\end{align*}

\end{itemize}

\end{proof}

\subsection{Proof of \cref{th:deadlockfreedom}}

\begin{quote}
Let $\gtype{G}$ be a global type.
Suppose $\gtype{G}$ is well-formed with respect to some router $\ppt{s}$,
i.e. $\wfnew{\gtype{G}}{s}$.
\[
\forall \gtype{G'}. ~ \left(
\gtype{G} \to^* \gtype{G'}
	\Longrightarrow
(\gtype{G'} = \gtend) \vee \exists \gtype{G''}, \lbl{l}. ~
	(\gtreduce{G'}{G''}{l})
\right)\]
\end{quote}
\begin{proof}
Direct consequence of
\cref{lem:preservewf,lem:progresswf}.
\end{proof}

\begin{lemma}[Preservation of Well-formedness]
Let $\gtype{G}$ be a global type.
Suppose $\gtype{G}$ is well-formed with respect to some router $\ppt{s}$,
i.e. $\wfnew{\gtype{G}}{s}$.
\[
\forall \gtype{G'}, \lbl{l}. ~
\left(\gtreduce{G}{G'}{l}
	\Longrightarrow
\wfnew{\gtype{G'}}{s}\right)
\]
\label{lem:preservewf}
\end{lemma}
\vspace{-8mm}
\begin{proof}
By rule induction on $\gtreduce{G}{G'}{l}$.

For each transition, we show the two conjuncts for
well-formedness, $\wfnew{\gtype{G'}}{s}$ :

\textbf{(1)} $\proj{G'}{r}$ exists for $\ppt{r}$
	such that $\proj{G}{r}$ exists;
and, \textbf{(2)} $\centroid{G'}{s}$.

\begin{itemize}[wide, labelwidth=!, labelindent=0pt]
\item \rulename{Gr1},
where
$\gtype{G} = \gcomm{p}{q}{\lbl{l_i}\!:\! G_i}{\lbl{i \in I}}$,

$\gtype{G'} = \gtrans{p}{q}{j}{\lbl{l_i}\!:\! G_i}{\lbl{i \in I}}$,

and $\lbl{l} = \aout{p}{q}{j}$.

\paragraph{(1)}
We know $\proj{G}{r}$ by assumption.
To show $\proj{G'}{r}$, consider $\ppt{r}$ by case:

\begin{itemize}
\item $\ppt{r} = \ppt{p}$:
Then $\proj{G}{p} = \tsel{q}{\lbl{l_i}\!:\! \proj{G_i}{p}}{\lbl{i \in I}}$,
so $\forall \lbl{i \in I}. ~ \proj{G_i}{p}$ exists.

$\proj{G'}{p} = \proj{G_j}{p}$, which exists as $\lbl{j \in I}$.

\item $\ppt{r} = \ppt{q}$:
Then $\proj{G'}{q} = \tbra{p}{\lbl{l_i}\!:\!
\proj{G_i}{q}}{\lbl{i \in I}} \proj{G}{q}$,
which exists.

\item $\ppt{r} \notin \{ \ppt{p}, \ppt{q} \}$:
Then $\proj{G}{r} = \underset{i \in I}{\MERGEOP}\proj{G_i}{r}$,
so $\forall \lbl{i \in I}. ~ \proj{G_i}{r}$ exists.

$\proj{G'}{r} = \proj{G_j}{r}$, which exists as $\lbl{j \in I}$.
\end{itemize}

\paragraph{(2)}
We know $\centroid{G}{s}$ by assumption.
We deduce $\centroid{G'}{s}$ by consequence.
\begin{align*}
\centroid{G}{s}
 &\Longrightarrow
	\ppt{s} \in \{ \ppt{p}, \ppt{q} \}
		\wedge
	\underset{i \in I}{\bigwedge}\centroid{G_i}{s} \\
&\Longrightarrow
	\ppt{s} \in \{ \ppt{p}, \ppt{q} \}
		\wedge
	\centroid{G_j}{s}\\
&\Longrightarrow \centroid{G'}{s}
\end{align*}

\item \rulename{Gr2},
where
$\gtype{G = \gtrans{p}{q}{j}{\lbl{l_i}\!:\! G_i}{\lbl{i \in I}}}$,
$\gtype{G' = G_j}$,
$\lbl{l} = \ain{p}{q}{j}$.

\paragraph{(1)}
We know $\proj{G}{r}$ by assumption.
To show $\proj{G'}{r}$, consider $\ppt{r}$ by case:

\begin{itemize}
\item $\ppt{r} = \ppt{p}$:
Then $\proj{G'}{p} = \proj{G_j}{p} = \proj{G}{p}$, which exists.

\item $\ppt{r} = \ppt{q}$:
Then $\proj{G}{q} = \tbra{p}{\lbl{l_i}\!:\! \proj{G_i}{q}}{\lbl{i \in I}}$,
so $\forall \lbl{i \in I}. ~ \proj{G_i}{q}$ exists.

$\proj{G'}{q} = \proj{G_j}{q}$, which exists as $\lbl{j \in I}$.

\item $\ppt{r} \notin \{ \ppt{p}, \ppt{q} \}$:
Then $\proj{G'}{r} = \proj{G_j}{r} = \proj{G}{r}$,
which exists.
\end{itemize}

\paragraph{(2)}
We know $\centroid{G}{s}$ by assumption.
We deduce $\centroid{G'}{s}$ by consequence.
\[
\centroid{G}{s}
	\Longrightarrow
		\ppt{s} \in \{ \ppt{p}, \ppt{q} \}
			\wedge
		\centroid{G_j}{s}
	\Longrightarrow
		\centroid{G'}{s}
\]
\item \rulename{Gr3},
where $\gtreduce{\gtrecur{t}{G}}{G'}{l}$.
By hypothesis, $\gtreduce{G[\gtrecur{t}{G} / \gtvar{t}]}{G'}{l}$.

We first show that $\wfnew{\gtype{G[\gtrecur{t}{G} / \gtvar{t}]}}{s}$.

\subparagraph{(1)}
$\proj{\gtrecur{t}{G}}{r}$ exists for some $\ppt{r}$.

Note that $\proj{G}{r}$ exists regardless of $\ppt{r}$'s
participation in $\gtype{G}$.

\begin{itemize}
\item
If $\ppt{r} \in \pt{\gtype{G}}$,
then $\proj{\gtrecur{t}{G}}{r} = \gtrecur{t}{\proj{G}{r}}$,
so $\proj{G}{r}$ exists.

\item
Otherwise, $\proj{G}{r} = \gtend$, which exists.
\end{itemize}

Projection is homomorphic under recursion, so
$\proj{G[\gtrecur{t}{G}/\gtvar{t}]}{r}$ exists.

\subparagraph{(2)}
By assumption, $\centroid{(\gtrecur{t}{G})}{s}$,
so $\centroid{G}{s}$.

The $\centroid{}{}$ relation is also homomorphic under recursion,
so we get $\centroid{G[\gtrecur{t}{G}/\gtvar{t}]}{s}$.

We conclude by induction to obtain $\wfnew{G'}{s}$.

\item \rulename{Gr4},
where
$\gtype{G} = \gcomm{p}{q}{\lbl{l_i}\!:\! G_i}{\lbl{i \in I}}$,

and
$\gtype{G'} = \gcomm{p}{q}{\lbl{l_i}\!:\! G'_i}{\lbl{i \in I}}$.

By hypothesis,
$\forall \lbl{i \in I}. ~ (\gtreduce{G_i}{G'_i}{l})$
and $\ppt{p} \neq \ppt{q} \neq \subj{l}$.

If $\proj{G}{r}$ exists, so does $\proj{G_i}{r}$ for $\lbl{i \in I}$.

By assumption, $\centroid{G}{s}$,
so $\ppt{s} \in \{ \ppt{p}, \ppt{q} \} \wedge
\underset{i \in I}{\bigwedge}\centroid{G_i}{s}$.

By induction, $\forall \lbl{i \in I}. ~
(\proj{G'_i}{r} \text{ exists } \wedge
\centroid{G'_i}{s})$.

\paragraph{(1)}
To show $\proj{G'}{r}$, consider $\ppt{r}$ by case:

\begin{itemize}
\item $\ppt{r} = \ppt{p}$:
Then $\proj{G'}{p} = \tsel{q}{\lbl{l_i}\!:\! \proj{G'}{p}}{\lbl{i \in I}}$.

\item $\ppt{r} = \ppt{q}$:
Then $\proj{G'}{q} = \tbra{p}{\lbl{l_i}\!:\! \proj{G'_i}{q}}{\lbl{i \in I}}$.

\item $\ppt{r} \notin \{ \ppt{p}, \ppt{q} \}$:
Then $\proj{G'}{r} = \underset{i \in I}{\MERGEOP} \proj{G'_i}{r}$.
We know that $\proj{G}{r} = \underset{i \in I}{\MERGEOP} \proj{G_i}{r}$
exists.
By \cref{lem:localltspreservemerge},
$\underset{i \in I}{\MERGEOP} \proj{G'_i}{r}$ exists too.
\end{itemize}

\paragraph{(2)}
We have $\ppt{s} \in \{ \ppt{p}, \ppt{q} \}$
from assumption and
$\underset{i \in I}{\bigwedge}\centroid{G'_i}{s}$ from induction,
so $\centroid{G'}{s}$.

\item \rulename{Gr5},
where
$\gtype{G} = \gtrans{p}{q}{j}{\lbl{l_i}\!:\! G_i}{\lbl{i \in I}}$,

and $\gtype{G'} = \gtrans{p}{q}{j}{\lbl{l_i}\!:\! G'_i}{\lbl{i \in I}}$.

By hypothesis,
$\gtreduce{G_j}{G'_j}{l}$,
$\forall \lbl{i \in I \setminus \{ j \}}. ~ \gtype{G'_i = G_i}$,
and $\ppt{q} \neq \subj{\lbl{l}}$.

If $\proj{G}{r}$ exists, so does $\proj{G_i}{r}$ for $\lbl{i \in I}$.

By assumption, $\centroid{G}{s}$,
so $\ppt{s} \in \{ \ppt{p}, \ppt{q} \} \wedge
\centroid{G_j}{s}$.

By induction on $\gtreduce{G_j}{G'_j}{l}$
and hypothesis\\
$\forall \lbl{i \in I \setminus \{ j \}}. ~ \gtype{G'_i = G_i}$,
we get
$\forall \lbl{i \in I}. ~
(\proj{G'_i}{r} \text{ exists } \wedge
\centroid{G'_i}{s})$.

\paragraph{(1)}
To show $\proj{G'}{r}$, consider $\ppt{r}$ by case:

\begin{itemize}
\item $\ppt{r} = \ppt{p}$:
Then $\proj{G'}{p} = \proj{G'_j}{p}$.

\item $\ppt{r} = \ppt{q}$:
Then $\proj{G'}{q} = \tbra{p}{\lbl{l_i}\!:\! \proj{G'_i}{q}}{\lbl{i \in I}}$.

\item $\ppt{r} \notin \{ \ppt{p}, \ppt{q} \}$:
Then $\proj{G'}{r} = \proj{G'_j}{r}$.

\end{itemize}

\paragraph{(2)}
We have $\ppt{s} \in \{ \ppt{p}, \ppt{q} \}$
from assumption and
$\centroid{G'_j}{s}$ from induction, so $\centroid{G'}{s}$.

\item \rulename{Gr6},
where
$\gtype{G} = \groute{p}{q}{t}{\lbl{l_i}\!:\! G_i}{\lbl{i \in I}}$,

$\gtype{G'} = \gtransroute{p}{q}{t}{j}{\lbl{l_i}\!:\! G_i}{\lbl{i \in I}}$,
and $\lbl{l} = \via{s}{\aout{p}{q}{j}}$.

By assumption, $\wfnew{\gtype{G}}{s}$, so $\ppt{t} = \ppt{s}$.

\paragraph{(1)}
We know $\proj{G}{r}$ by assumption.
To show $\proj{G'}{r}$, consider $\ppt{r}$ by case:

\begin{itemize}
\item $\ppt{r} = \ppt{p}$:
Then $\proj{G}{p} = \tselproxy{q}{s}{\lbl{l_i}\!:\! \proj{G_i}{p}}{\lbl{i \in I}}$,
so $\forall \lbl{i \in I}. ~ \proj{G_i}{p}$ exists.

$\proj{G'}{p} = \proj{G_j}{p}$, which exists as $\lbl{j \in I}$.

\item $\ppt{r} = \ppt{q}$:
Then $\proj{G'}{q} = \tbraproxy{p}{s}{\lbl{l_i}\!:\! \proj{G_i}{q}}{\lbl{i \in I}}
= \proj{G}{q}$,
which exists.

\item $\ppt{r} = \ppt{s}$:
Then $\proj{G}{s} = \router{p}{q}{\lbl{l_i}\!:\! \proj{G_i}{s}}{\lbl{i \in I}}$,
so $\forall \lbl{i \in I}. ~ \proj{G_i}{s}$ exists.

$\proj{G'}{s} = \routertrans{p}{q}{j}{\lbl{l_i}\!:\! \proj{G_i}{s}}{\lbl{i \in I}}$,
which exists.

\item $\ppt{r} \notin \{ \ppt{p}, \ppt{q}, \ppt{s} \}$:
Then $\proj{G}{r} = \underset{i \in I}{\MERGEOP}\proj{G_i}{r}$,
so $\forall \lbl{i \in I}. ~ \proj{G_i}{r}$ exists.

$\proj{G'}{r} = \proj{G_j}{r}$, which exists as $\lbl{j \in I}$.
\end{itemize}

\paragraph{(2)}
We know $\centroid{G}{s}$ by assumption.
We deduce $\centroid{G'}{s}$ by consequence.
\begin{align*}
\centroid{G}{s}
&\Longrightarrow
	\ppt{t} = \ppt{s}
		\wedge
	\underset{i \in I}{\bigwedge}\centroid{G_i}{s}\\
&\Longrightarrow
	\ppt{t} = \ppt{s}
		\wedge
	\centroid{G_j}{s}\\
&\Longrightarrow \centroid{G'}{s}
\end{align*}
\item \rulename{Gr7},
where
$\gtype{G} = \gtransroute{p}{q}{t}{j}{\lbl{l_i}\!:\! G_i}{\lbl{i \in I}}$,
$\gtype{G'} = \gtype{G_j}$ and
$\lbl{l} = \via{s}{\ain{p}{q}{j}}$.

\paragraph{(1)}
We know $\proj{G}{r}$ by assumption.
To show $\proj{G'}{r}$, consider $\ppt{r}$ by case:

By assumption, $\wfnew{\gtype{G}}{s}$, so $\ppt{t} = \ppt{s}$.

\begin{itemize}
\item $\ppt{r} = \ppt{p}$:
Then $\proj{G'}{p} = \proj{G_j}{p} = \proj{G}{p}$, which exists.

\item $\ppt{r} = \ppt{q}$:
Then $\proj{G}{q} = \tbraproxy{p}{s}{\lbl{l_i}\!:\! \proj{G_i}{q}}{\lbl{i \in I}}$,
so $\forall \lbl{i \in I}. ~ \proj{G_i}{q}$ exists.

$\proj{G'}{q} = \proj{G_j}{q}$, which exists as $\lbl{j \in I}$.

\item $\ppt{r} = \ppt{s}$:
Then $\proj{G}{s} = \routertrans{p}{q}{j}{\lbl{l_i}\!:\! \proj{G_i}{s}}{\lbl{i \in I}}$,
so $\forall \lbl{i \in I}. ~ \proj{G_i}{s}$ exists.

$\proj{G'}{s} = \proj{G_j}{s}$,
which exists as $\lbl{j \in I}$.

\item $\ppt{r} \notin \{ \ppt{p}, \ppt{q}, \ppt{s} \}$:
Then $\proj{G}{r} = \underset{i \in I}{\MERGEOP}\proj{G_i}{r}$,
so $\forall \lbl{i \in I}. ~ \proj{G_i}{r}$ exists.

$\proj{G'}{r} = \proj{G_j}{r}$,
which exists as $\lbl{j \in I}$.
\end{itemize}

\paragraph{(2)}
We know $\centroid{G}{s}$ by assumption.
We deduce $\centroid{G'}{s}$ by consequence.
\[
\centroid{G}{s}
	\Longrightarrow
		\ppt{s} \in \{ \ppt{p}, \ppt{q} \}
			\wedge
		\centroid{G_j}{s}
	\Longrightarrow
		\centroid{G'}{s}
\]
\item \rulename{Gr8},
where
$\gtype{G} = \groute{p}{q}{t}{\lbl{l_i}\!:\! G_i}{\lbl{i \in I}}$,

and
$\gtype{G'} = \groute{p}{q}{t}{\lbl{l_i}\!:\! G'_i}{\lbl{i \in I}}$.

By hypothesis,
$\forall \lbl{i \in I}. ~ (\gtreduce{G_i}{G'_i}{l})$
and $\ppt{p} \neq \ppt{q} \neq \subj{l}$.

If $\proj{G}{r}$ exists, so does $\proj{G_i}{r}$ for $\lbl{i \in I}$.

By assumption, $\centroid{G}{s}$,
so $\ppt{t} = \ppt{s} \wedge
\underset{i \in I}{\bigwedge}\centroid{G_i}{s}$.

By induction, $\forall \lbl{i \in I}. ~
(\proj{G'_i}{r} \text{ exists } \wedge
\centroid{G'_i}{s})$.

\paragraph{(1)}
To show $\proj{G'}{r}$, consider $\ppt{r}$ by case:

\begin{itemize}
\item $\ppt{r} = \ppt{p}$:
Then $\proj{G'}{p} = \tselproxy{q}{s}{\lbl{l_i}\!:\! \proj{G'}{p}}{\lbl{i \in I}}$.

\item $\ppt{r} = \ppt{q}$:
Then $\proj{G'}{q} = \tbraproxy{p}{s}{\lbl{l_i}\!:\! \proj{G'_i}{q}}{\lbl{i \in I}}$.

\item $\ppt{r} = \ppt{s}$:
Then $\proj{G'}{s} = \router{p}{q}{\lbl{l_i}\!:\! \proj{G'_i}{s}}{\lbl{i \in I}}$.

\item $\ppt{r} \notin \{ \ppt{p}, \ppt{q}, \ppt{s} \}$:
Then $\proj{G'}{r} = \underset{i \in I}{\MERGEOP} \proj{G'_i}{r}$.
We know that $\proj{G}{r} = \underset{i \in I}{\MERGEOP} \proj{G_i}{r}$
exists.
By \cref{lem:localltspreservemerge},
$\underset{i \in I}{\MERGEOP} \proj{G'_i}{r}$ exists too.

\end{itemize}

\paragraph{(2)}
We have $\ppt{t} = \ppt{s}$
from assumption and
$\underset{i \in I}{\bigwedge}\centroid{G'_i}{s}$ from induction,
so $\centroid{G'}{s}$.

\item \rulename{Gr9},
where
$\gtype{G} = \gtransroute{p}{q}{t}{j}{\lbl{l_i}\!:\! G_i}{\lbl{i \in I}}$,

and
$\gtype{G'} = \gtransroute{p}{q}{t}{j}{\lbl{l_i}\!:\! G'_i}{\lbl{i \in I}}$.

By hypothesis,
$\gtreduce{G_j}{G'_j}{l}$,

$\forall \lbl{i \in I \setminus \{ j \}}. ~ \gtype{G'_i = G_i}$,
and $\ppt{q} \neq \subj{\lbl{l}}$.

If $\proj{G}{r}$ exists, so does $\proj{G_i}{r}$ for $\lbl{i \in I}$.

By assumption, $\centroid{G}{s}$,
so $\ppt{t} = \ppt{s} \wedge
\centroid{G_j}{s}$.

By induction on $\gtreduce{G_j}{G'_j}{l}$
and hypothesis\\
$\forall \lbl{i \in I \setminus \{ j \}}. ~ \gtype{G'_i = G_i}$,
we get
$\forall \lbl{i \in I}. ~
(\proj{G'_i}{r} \text{ exists } \wedge
\centroid{G'_i}{s})$.

\paragraph{(1)}
To show $\proj{G'}{r}$, consider $\ppt{r}$ by case:

\begin{itemize}
\item $\ppt{r} = \ppt{p}$:
Then $\proj{G'}{p} = \proj{G'_j}{p}$.

\item $\ppt{r} = \ppt{q}$:
Then $\proj{G'}{q} = \tbraproxy{p}{s}{\lbl{l_i}\!:\! \proj{G'_i}{q}}{\lbl{i \in I}}$.

\item $\ppt{r} = \ppt{s}$:
Then $\proj{G'}{s} = \routertrans{p}{q}{j}{\lbl{l_i}\!:\! \proj{G'_i}{s}}{\lbl{i \in I}}$.

\item $\ppt{r} \notin \{ \ppt{p}, \ppt{q}, \ppt{s} \}$:
Then $\proj{G'}{r} = \proj{G'_j}{r}$.

\end{itemize}

\paragraph{(2)}
We have $\ppt{t} = \ppt{s}$
from assumption and
$\centroid{G'_j}{s}$ from induction, so $\centroid{G'}{s}$.

\end{itemize}
\end{proof}

\begin{lemma}[Progress for Well-formed Global Types]
Let $\gtype{G}$ be a global type.
Suppose $\gtype{G}$ is well-formed with respect to some router $\ppt{s}$,
i.e. $\wfnew{\gtype{G}}{s}$.
\[
(\gtype{G} = \gtend) \vee \exists \gtype{G'}, \lbl{l}. ~ (\gtreduce{G}{G'}{l})
\]
\label{lem:progresswf}
\end{lemma}
\vspace{-8mm}
\begin{proof}
The following is logically equivalent:
\[
(\gtype{G \neq \gtend})
	\Longrightarrow
\exists \gtype{G'}, \lbl{l}. ~ (\gtreduce{G}{G'}{l}).
\]
We prove this by induction on the structure of $\gtype{G}$.

We do not consider $\gtype{G = \gtend}$ by assumption.

We do not consider $\gtype{G = \gtvar{t}}$ as the type variable occurs
free.

\begin{enumerate}

\item $\gtype{G = \gtrecur{t}{G''}}$

$\gtvar{t}$ must occur in $\gtype{G}$,
so $\gtype{G[\gtrecur{t}{G} / \gtvar{t}]} \neq \gtend$.

By induction,
$\exists \gtype{G'}, \lbl{l}. ~
(\gtreduce{G[\gtrecur{t}{G} / \gtvar{t}]}{G'}{l})$.

Apply \rulename{Gr3} to get
$\exists \gtype{G'}, \lbl{l}. ~
(\gtreduce{\gtrecur{t}{G}}{G'}{l})$.

\item $\gtype{G = \gcomm{p}{q}{\lbl{l_i}\!:\! G_i}{\lbl{i \in I}}}$

Apply \rulename{Gr1} to get
$\gtreducelong
	{G}
	{\gtrans{p}{q}{j}{\lbl{l_i}\!:\! G_i}{\lbl{i \in I}}}
	{\aout{p}{q}{j}}$.

\item $\gtype{G = \groute{p}{q}{r}{\lbl{l_i}\!:\! G_i}{\lbl{i \in I}}}$

By assumption, $\wfnew{\gtype{G}}{s}$, so $\ppt{r} = \ppt{s}$.

Apply \rulename{Gr6} to get
$\gtreducelong
	{G}
	{\gtransroute{p}{q}{s}{j}{\lbl{l_i}\!:\! G_i}{\lbl{i \in I}}}
	{\via{s}{\aout{p}{q}{j}}}$.

\item $\gtype{G = \gtrans{p}{q}{j}{\lbl{l_i}\!:\! G_i}{\lbl{i \in I}}}$

Apply \rulename{Gr2} to get
$\gtreducelong
	{G}
	{G_j}
	{\ain{p}{q}{j}}$.

\item $\gtype{G = \gtransroute{p}{q}{r}{j}{\lbl{l_i}\!:\! G_i}{\lbl{i \in I}}}$

By assumption, $\wfnew{\gtype{G}}{s}$, so $\ppt{r} = \ppt{s}$.

Apply \rulename{Gr7} to get
$\gtreducelong
	{G}
	{G_j}
	{\via{s}{\ain{p}{q}{j}}}$.
\end{enumerate}

\end{proof}

\paragraph{Proof of \cref{th:encwf}}
\begin{quote}
Let $\gtype{G}$ be a global type, and $\ppt{s}$ be a role. Then we have:

\centering{
$\wf{\gtype{G}} \Longleftrightarrow \wfnew{\enc{\gtype{G}}{s}}{s}$
}
\end{quote}
\begin{proof}
  ($\Longrightarrow$) Direct consequence of
  \cref{lem:encproj,lem:enccenter}\\ %
  ($\Longleftarrow$) By definition.
\end{proof}

\subsection{Proof of \cref{th:enccomm}}
\begin{quote}
Let $\gtype{G, G'}$ be well-formed global types
such that $\gtreduce{G}{G'}{l}$ for some label $\lbl{l}$. Then we have:

\centering{$\forall \lbl{l}, \mrole{s}. ~ \left(
\gtreduce{G}{G'}{l}
	\Longleftrightarrow
        \gtreducelong{\enc{\gtype{G}}{s}}{\enc{\gtype{G'}}{s}}{\enc{\lbl{l}}{s}} \right)
$}
\end{quote}
\vspace{-5mm}
\begin{proof}
By rule induction on $\gtreduce{G}{G'}{l}$.
Take arbitrary router role $\ppt{s}$.

\begin{itemize}[wide, labelwidth=!, labelindent=0pt]

\item \rulename{Gr1},
  where $\gtype{G} = \gcomm{p}{q}{\lbl{l_i}\!:\! G_i}{\lbl{i \in I}}$,

$\gtype{G'} = \gtrans{p}{q}{j}{\lbl{l_i}\!:\! G_i}{\lbl{i \in I}}$,

and $\lbl{l} = \aout{p}{q}{j}$.

To show $\gtreducelong
{\enc{\gtype{G}}{s}}{\enc{\gtype{G'}}{s}}{\enc{\lbl{l}}{s}}$,
consider $\ppt{s}$ by case:

\begin{itemize}

\item $\ppt{s} \in \{ \ppt{p}, \ppt{q} \}$:
Then we have
\begin{align*}
\enc{\gtype{G}}{s} &= \gcomm{p}{q}{\lbl{l_i}\!:\! \enc{\gtype{G_i}}{s}}{\lbl{i \in I}} \\
\enc{\gtype{G'}}{s} &= \gtrans{p}{q}{j}{\lbl{l_i}\!:\! \enc{\gtype{G_i}}{s}}{\lbl{i \in I}} \\
\enc{\lbl{l}}{s} &= \aout{p}{q}{j}
\end{align*}
The encoded transition is possible using \rulename{Gr1}.

\item $\ppt{s} \notin \{ \ppt{p}, \ppt{q} \}$:
Then we have
\begin{align*}
\enc{\gtype{G}}{s} &= \groute{p}{q}{s}{\lbl{l_i}\!:\! \enc{\gtype{G_i}}{s}}{\lbl{i \in I}} \\
\enc{\gtype{G'}}{s} &= \gtransroute{p}{q}{s}{j}{\lbl{l_i}\!:\! \enc{\gtype{G_i}}{s}}{\lbl{i \in I}} \\
\enc{\lbl{l}}{s} &= \via{s}{\aout{p}{q}{j}}
\end{align*}
The encoded transition is possible using \rulename{Gr6}.

\end{itemize}

\item \rulename{Gr2},
where $\gtype{G = \gtrans{p}{q}{j}{\lbl{l_i}\!:\! G_i}{\lbl{i \in I}}},
\gtype{G' = G_j},
\lbl{l} = \ain{p}{q}{j}$.

We know $\enc{\gtype{G'}}{s} = \enc{\gtype{G_j}}{s}$

To show $\treducelong
{\enc{\gtype{G}}{s}}{\enc{\gtype{G'}}{s}}{\enc{\lbl{l}}{s}}$,
consider $\ppt{s}$ by case:

\begin{itemize}

\item $\ppt{s} \in \{ \ppt{p}, \ppt{q} \}$:
Then we have
\begin{align*}
\enc{\gtype{G}}{s} &= \gtrans{p}{q}{j}{\lbl{l_i}\!:\! \enc{\gtype{G_i}}{s}}{\lbl{i \in I}} \\
\enc{\lbl{l}}{s} &= \ain{p}{q}{j}
\end{align*}
The encoded transition is possible using \rulename{Gr2}.

\item $\ppt{s} \notin \{ \ppt{p}, \ppt{q} \}$:
Then we have
\begin{align*}
\enc{\gtype{G}}{s} &= \gtransroute{p}{q}{s}{j}{\lbl{l_i}\!:\! \enc{\gtype{G_i}}{s}}{\lbl{i \in I}} \\
\enc{\lbl{l}}{s} &= \via{s}{\ain{p}{q}{j}}
\end{align*}
The encoded transition is possible using \rulename{Gr7}.
\end{itemize}

\item \rulename{Gr3},
where $\gtype{G = \gtrecur{t}{G''}}$.

By hypothesis, $\gtreduce{G''[\gtrecur{t}{G''} / \gtvar{t}]}{G'}{l}$.

By induction, $
\gtreducelong
	{\enc{\gtype{G''[\gtrecur{t}{G''} / \trecvar]}}{s}}
	{\enc{\gtype{G'}}{s}}
	{\enc{\lbl{l}}{s}}
$.

By \cref{lem:encsub},
$\enc{\gtype{G''[\gtrecur{t}{G''} / \gtvar{t}]}}{s}
= \enc{\gtype{G''}}{s}\left[ \gtrecur{t}{\enc{\gtype{G''}}{s}} / \gtvar{t} \right]$.

We know $\enc{\gtype{G}}{s} = \enc{\gtrecur{t}{G''}}{s} = \gtrecur{t}{\enc{\gtype{G''}}{s}}$.

The encoded transition is possible using \rulename{Gr3} as shown:

\begin{prooftree}
\AxiomC{$
\gtreducelong
	{\enc{\gtype{G''}}{s}\left[ \gtrecur{t}{\enc{\gtype{G''}}{s}} / \gtvar{t}\right]}
	{\enc{\gtype{G'}}{s}}
	{\enc{\lbl{l}}{s}}
$}
\RightLabel{\rulename{Gr3}}
\UnaryInfC{$
\gtreducelong
	{\gtrecur{t}{\enc{\gtype{G''}}{s}}}
	{\enc{\gtype{G'}}{s}}
	{\enc{\lbl{l}}{s}}
$}
\end{prooftree}

\item \rulename{Gr4},
where $\gtype{G} = \gcomm{p}{q}{\lbl{l_i}\!:\! G_i}{\lbl{i \in I}}$,
$\gtype{G'} = \gcomm{p}{q}{\lbl{l_i}\!:\! G'_i}{\lbl{i \in I}}$.

By hypothesis,
$\forall \lbl{i \in I}. ~ \gtreduce{G_i}{G'_i}{l}$
and $\subj{\lbl{l}} \notin \{ \ppt{p}, \ppt{q} \}$.

By induction,
$\forall \lbl{i \in I}. ~ \left(
\gtreducelong{\enc{\gtype{G_i}}{s}}{\enc{\gtype{G'_i}}{s}}{\enc{\lbl{l}}{s}}
\right)$.

By definition of $\subj{\cdot}$,
$\subj{\enc{\lbl{l}}{s}} \notin \{ \ppt{p}, \ppt{q} \}$.

To show $\treducelong
{\enc{\gtype{G}}{s}}{\enc{\gtype{G'}}{s}}{\enc{\lbl{l}}{s}}$,
consider $\ppt{s}$ by case:

\begin{itemize}

\item $\ppt{s} \in \{ \ppt{p}, \ppt{q} \}$:
Then we have
\begin{align*}
\enc{\gtype{G}}{s} &= \gcomm{p}{q}{\lbl{l_i}\!:\! \enc{\gtype{G_i}}{s}}{\lbl{i \in I}} \\
\enc{\gtype{G'}}{s} &= \gcomm{p}{q}{\lbl{l_i}\!:\! \enc{\gtype{G'_i}}{s}}{\lbl{i \in I}}
\end{align*}
The encoded transition is possible using \rulename{Gr4}.

\item $\ppt{s} \notin \{ \ppt{p}, \ppt{q} \}$:
Then we have
\begin{align*}
\enc{\gtype{G}}{s} &= \groute{p}{q}{s}{\lbl{l_i}\!:\! \enc{\gtype{G_i}}{s}}{\lbl{i \in I}} \\
\enc{\gtype{G'}}{s} &= \groute{p}{q}{s}{\lbl{l_i}\!:\! \enc{\gtype{G'_i}}{s}}{\lbl{i \in I}}
\end{align*}
The encoded transition is possible using \rulename{Gr8}.

\end{itemize}

\item \rulename{Gr5},
  where $\gtype{G} = \gtrans{p}{q}{j}{\lbl{l_i}\!:\! G_i}{\lbl{i \in I}}$,

$\gtype{G'} = \gtrans{p}{q}{j}{\lbl{l_i}\!:\! G'_i}{\lbl{i \in I}}$.

By hypothesis,
$\gtreduce{G_j}{G'_j}{l}$,
$\ppt{p'} \neq \subj{\lbl{l}}$,\hfill \hfill \linebreak
and $\forall \lbl{i \in I \setminus \{j\}}. ~ \gtype{G'_i = G_i}$.

By induction,
$\gtreducelong{\enc{\gtype{G_j}}{s}}{\enc{\gtype{G'_j}}{s}}{\enc{\lbl{l}}{s}}$.

By definition of a set of subjects, we have:
$\subj{\enc{\lbl{l}}{s}} \neq \ppt{q}$.

To show $\treducelong
{\enc{\gtype{G}}{s}}{\enc{\gtype{G'}}{s}}{\enc{\lbl{l}}{s}}$,
consider $\ppt{s}$ by case:

\begin{itemize}

\item $\ppt{s} \in \{ \ppt{p}, \ppt{q} \}$:
Then we have
\begin{align*}
\enc{\gtype{G}}{s} &= \gtrans{p}{q}{j}{\lbl{l_i}\!:\! \enc{\gtype{G_i}}{s}}{\lbl{i \in I}} \\
\enc{\gtype{G'}}{s} &= \gtrans{p}{q}{j}
    {\lbl{l_i}\!:\! \enc{\gtype{G'_i}}{s}}{\lbl{i \in I}}
\end{align*}
The encoded transition is possible using \rulename{Gr5}.

\item $\ppt{s} \notin \{ \ppt{p}, \ppt{q} \}$:
Then we have
\begin{align*}
\enc{\gtype{G}}{s} &= \gtransroute{p}{q}{s}{j}{\lbl{l_i}\!:\! \enc{\gtype{G_i}}{s}}{\lbl{i \in I}} \\
\enc{\gtype{G'}}{s} &= \gtransroute{p}{q}{s}{j}{\lbl{l_i}\!:\! \enc{\gtype{G'_i}}{s}}{\lbl{i \in I}}
\end{align*}
The encoded transition is possible using \rulename{Gr9}.
\end{itemize}
\end{itemize}

($\Longleftarrow$) direction is similar by rule induction on
$$\gtreducelong{\enc{\gtype{G}}{s}}{\enc{\gtype{G'}}{s}}{\enc{\lbl{l}}{s}}.$$
\end{proof}

%% file: appendix/case-studies.tex
\section{Appendix for \cref{sec:eval}}\label{sec:eval-appendix}
We include
two more studies: \tprotocol{Battleships}
and \tprotocol{Travel Agency}.

\paragraph{Battleships}

We implement the \emph{Battleships} board game%
between two players, as used by \citet{PLACES19PureScript}.
The initialisation phase involves both players
placing rectangular battleships on a 2D grid.
The session proceeds to the game loop,
where players take turns to guess the location of their
opponent's ships, where the server responds with
a \tmsg{hit} or a \tmsg{miss}.
The game loop continues until all ships of one player
have been sunk.

\begin{lstlisting}[language=Scribble]
// Ship configuration
type <typescript> "Config" from "./Models" as Config;
// Coordinate on 2D grid
type <typescript> "Location" from "./Models" as Loc;
global protocol Battleships(role P1, role Svr, role P2) {
  Init(Config) from P1 to Svr;
  Init(Config) from P2 to Svr; do Game(P1, Svr, P2); }
aux global protocol Game(role Atk, role Svr, role Def) {
  Attack(Location) from Atk to Svr;
  choice at Svr
     { // Hit an opposing ship coordinate
       Hit(Loc) from Svr to Atk;
       Hit(Loc) from Svr to Def;
       do Game(Def, Svr, Atk); }
  or { Miss(Loc) from Svr to Atk;
       Miss(Loc) from Svr to Def;
       do Game(Def, Svr, Atk); }
  or { // Hit all coordinates of an opposing ship
       Sunk(Loc) from Svr to Atk;
       Sunk(Loc) from Svr to Def;
       do Game(Def, Svr, Atk); }
  or { // Sunk all opposing ships
       Winner(Loc) from Svr to Atk;
       Loser(Loc) from Svr to Def; }}
\end{lstlisting}

By interpreting the session ID (generated for $\ppt{Svr}$
by \codegen) as an unique game identifier,
the developer can keep track of
concurrent game sessions very easily.
As the generated session runtime for \nodejs needs to be
initialised by a \emph{function} that is parameterised by
the game ID and returns the initial state,
the game ID is bound inside the closure. This means that
the invocation of callbacks (as part of the game logic)
can access the game ID of the current game, so the
developer can update the application state of the
corresponding game.

\begin{lstlisting}[language=TypeScript]
const gameManager = (gameID: string) => {
  const handleP1 = Session.S176({
    Attack: async (Next, location) => {
      // Handle attack by P1 in current game
      const result = await
        DB.attack(gameID, GamePlayers.P1, location);
      ...}, });
  const handleP2 = ...	// defined similarly
  return Session.Initial({
    Init: (Next, p1Config) => Next({
      Init: (_, p2Config) => {
        // Initialise new game in database bound to 'gameID'
        DB.initialiseGame(gameID, p1Config, p2Config);
        return handleP1; }, }), }); };
// Initialise session runtime
new Svr(webSocketServer, cancellationHandler, gameManager);
\end{lstlisting}

Because the API for the session cancellation
handler also exposes the session ID of the cancelled session,
the $\ppt{Svr}$ can use the session ID to free up resources
allocated to the corresponding game session accordingly.
In fact, given that the API also exposes the role that
initiated the cancellation, the $\ppt{Svr}$ could identify
which player forfeited the game and update leaderboard
details to reflect the forfeit.

\begin{lstlisting}[language=TypeScript]
const cancellationHandler = async (
    id: string, role: Role.All, reason: any) => {
  console.log(`${id}: ${role} disconnected - ${reason}`);
  // Free up resources allocated by this game
  await DB.deleteGame(id); };
\end{lstlisting}

\paragraph{Travel Agency}

We implement the Travel Agency scenario
motivated in \cref{sec:intro}.
We specify the Scribble protocol
in \cref{fig:travel-agency-protocol}, and present
the \codegen APIs implemented by
the developer for both server and client endpoints
in \cref{sec:overview}.

This scenario involves routed communications
-- namely, between the customer and their friend.
In \cref{sec:implementation}, we discuss how the
routing mechanism is transparent to the \nodejs callbacks
implemented by the developer.
Here, we show that the routing mechanism is equally
transparent to the \reactjs components implemented
by the developer for browser-side endpoints.

\begin{lstlisting}[language=TypeScript]
export default class WaitSuggestion extends S11 {
  /* ...snip... */
  Suggest(place: string) {
    this.context.setSuggestion(place);
  }
  /* ...snip... */
}
\end{lstlisting}

\begin{lstlisting}[language=TypeScript]
export default class WaitResponse extends S14 {
  /* ...snip... */
  Available(quote: number) {
    this.context.setQuote(quote);
  }

  Full() {
    this.context.setErrorMessage(`
      No availability for ${this.context.suggestion}
    `);
    this.context.setSuggestion('');
  }
  /* ...snip... */
}
\end{lstlisting}

Focusing on the role $\ppt{A}$,
the \code{WaitSuggestion} component expects a message
from the other client role $\ppt{B}$, and
the \code{WaitResponse} component expects to receive
from the server role $\ppt{S}$.
Because the client roles $\ppt{A}$ and $\ppt{B}$
cannot directly communicate over a
WebSocket connection, the message to be received by
\code{WaitSuggestion} is in fact routed by $\ppt{S}$.
However, this is transparent to the developer
-- both components
handle incoming messages in the same manner.

Additionally, the APIs generated by \codegen
offer developers the flexibility to integrate
existing third-party libraries and frameworks
when designing their user interfaces.
The client endpoints are written using
the \emph{Material-UI} framework \citep{MaterialUI}
and leverage the \emph{React Context API}
to manage internal application state.

%% file: appendix/code.tex
\section{Code for \tprotocol{Travel Agency} scenario}\label{app:code}
This appendix contains
the implementation for the \tprotocol{Travel Agency}
scenario (specified in \cref{fig:travel-agency-protocol}).
We walk through key aspects of the APIs
generated by \codegen are used to implement
the role $\ppt{S}$ (for the server endpoint)
and role $\ppt{B}$ (for the client endpoint).
The full implementations can be found in the accompanying artifact
\citep{artifact}.

\subsection{Server Role $\ppt{S}$: Generated APIs}

\begin{lstlisting}[
language=TypeScript,tabsize=2,title=\code{namespace Message}
]
export interface S40_Available {
	label: "Available", payload: [number] };
export interface S40_Full { label: "Full", payload: [] };
export type S40 = | S40_Available | S40_Full;

export interface S38_Query {
	label: "Query", payload: [string] };
export type S38 = | S38_Query;

export interface S41_Confirm {
	label: "Confirm", payload: [Cred] };
export interface S41_Reject { label: "Reject", payload: [] };
export type S41 = | S41_Confirm | S41_Reject;
\end{lstlisting}

EFSM state transitions are characterised by the
possible \emph{messages} to be sent or received.
We generate an \emph{interface} for each message,
specifying types for the label (as a string literal)
and payload (as defined on the protocol).
Hence, each state is defined
as a \emph{union type} of the possible messages.

\begin{lstlisting}[
tabsize=2,language=TypeScript,title=\code{namespace Handler}
]
export type S40 = MaybePromise<
	| ["Available", Message.S40_Available['payload'], State.S41]
	| ["Full", Message.S40_Full['payload'], State.S38]>;

export interface S38 {
	"Query": (Next: typeof Factory.S40,
		...payload: Message.S38_Query['payload']
		) => MaybePromise<State.S40>, };

export interface S41 {
	"Confirm": (Next: typeof Factory.S39,
		...payload: Message.S41_Confirm['payload']
		) => MaybePromise<State.S39>,
	"Reject": (Next: typeof Factory.S39,
		...payload: Message.S41_Reject['payload']
		) => MaybePromise<State.S39>, };
\end{lstlisting}

Handlers define the callback-style APIs that
the developer needs to implement.
The handler for a \emph{send state} (i.e. \code{S40})
is a tuple of the message label (as a string literal),
payload, and the successor state.
The handler for a \emph{receive state} (i.e.
\code{S38}, \code{S41}) is an object literal
defining labelled callbacks.
Each callback is parameterised by
a factory function for the successor state
(discussed shortly) and the payload for this
particular message type, and is expected to return
the successor state.
The \code{MaybePromise<>} generic type allows
developers to write \emph{asynchronous} handlers
in their implementation.

\begin{lstlisting}[
tabsize=2,
language=TypeScript,
title=\code{namespace State}
]
interface ISend {
	readonly type: 'Send';
	performSend(next: StateTransitionHandler,
		cancel: Cancellation, send: SendStateHandler): void; };

interface IReceive {
	readonly type: 'Receive';
	prepareReceive(next: StateTransitionHandler,
		cancel: Cancellation,
		register: ReceiveStateHandler): void; };

interface ITerminal { readonly type: 'Terminal'; };

export type Type = ISend | IReceive | ITerminal;

export class S40 implements ISend {
	readonly type: 'Send' = 'Send';
	constructor(public handler: Handler.S40) { }

	performSend(next: StateTransitionHandler, £\label{line:perform-send}£
		cancel: Cancellation, send: SendStateHandler) {
		const thunk = (
			[label, payload, succ]: FromPromise<Handler.S40>) => {
			send(Role.Peers.A, label, payload);£\label{line:send-message}£
			return next(succ); };
		if (this.handler instanceof Promise) {
			this.handler.then(thunk).catch(cancel); }
		else { try { thunk(this.handler); }
			catch (error) { cancel(error); }}}};

export class S38 implements IReceive {
	readonly type: 'Receive' = 'Receive';
	constructor(public handler: Handler.S38) { }

	prepareReceive(next: StateTransitionHandler,
		cancel: Cancellation, register: ReceiveStateHandler) {
		const onReceive = (message: any) => {
			const parsed = JSON.parse(message) as Message.S38;
			switch (parsed.label) {
			case "Query": { try {
				const successor = this.handler[parsed.label](
					Factory.S40, ...parsed.payload);
					if (successor instanceof Promise) {
						successor.then(next).catch(cancel); }
					else { next(successor); }
				} catch (error) { cancel(error); }
				return; }}};
		register(Role.Peers.A, onReceive); }};
\end{lstlisting}

The handler for each EFSM state is used to
instantiate its \code{State} class instance,
which is used by the session runtime to trigger
the communication action.
For \emph{send states}, the runtime triggers the \code{performSend}
method (\cref{line:perform-send})
to perform the send using the label and payload
defined in the handler (\cref{line:send-message}).
For \emph{receive states}, the \code{State} class instance
\emph{registers} the handler to the runtime, so the handler
can be invoked when the message is received.
The \code{State} class will check whether the handler
is defined asynchronously (i.e. a \TS \code{Promise})
and invoke the handler accordingly.

\begin{lstlisting}[
language=TypeScript,tabsize=2,title=\code{namespace Factory}
]
type S40_Available =
	| [Message.S40_Available['payload'],
	   (Next: typeof S41) => State.S41]
	| [Message.S40_Available['payload'], State.S41];

function S40_Available(
	payload: Message.S40_Available['payload'],
	generateSucc: (Next: typeof S41) => State.S41): State.S40;
function S40_Available(
	payload: Message.S40_Available['payload'],
	succ: State.S41): State.S40;
function S40_Available(...args: S40_Available) {
	if (typeof args[1] === 'function') {
		const [payload, generateSucc] = args;
		const succ = generateSucc(S41);
		return new State.S40(["Available", payload, succ]); }
	else {
		const [payload, succ] = args;
		return new State.S40(["Available", payload, succ]); }}

type S40_Full =
	| [Message.S40_Full['payload'],
	   (Next: typeof S38) => State.S38]
	| [Message.S40_Full['payload'], State.S38];

// function S40_Full defined similarly

export const S40 = {
	Available: S40_Available, Full: S40_Full, };£\label{line:send-factory}£

export function S38(handler: Handler.S38) { £\label{line_recv-factory}£
	return new State.S38(handler); };
export function S41(handler: Handler.S41) {
	return new State.S41(handler); };

export const Initial = S38;

export const S39 = () => new State.S39();
export const Terminal = S39;
\end{lstlisting}

The \code{Factory} namespace exposes \emph{developer-friendly}
APIs for instantiating the \code{State} class instance.
The factory API for a \emph{send state}
is an object literal defining labelled callbacks for each
possible selection. Each callback is parameterised by the
message payload and successor state.
The factory API for a \emph{receive state} is an alias
for the constructor function of the corresponding \code{State} class.
\code{Initial} and \code{Terminal} aliases are also exported as
convenient references to the initial and terminal EFSM state
respectively.\\

\subsection{Server Role $\ppt{S}$: API Usage}

\begin{lstlisting}[
tabsize=2,
language=TypeScript
]
import express from "express";
import http from "http";
import WebSocket from "ws";

const app = express();
const server = http.createServer(app);
const wss = new WebSocket.Server({ server });

import { Session, S } from "./TravelAgency/S";

const agencyProvider = (sessionID: string) => {
	const handleQuery = Session.Initial({ £\label{line:impl-recv}£
		Query: async (Next, dest) => {
			const res = await checkAvailability(sessionID, dest);
			if (res.status === "available") {
				return Next.Available([res.quote], handleResponse); } £\label{line:impl-send}£
			else { return Next.Full([], handleQuery); }},});

	const handleResponse = Session.S41({
		Confirm: async (End, credentials) => {
			// Handle confirmation
			await confirmBooking(sessionID, credentials);
			return End(); },
		Reject: async (End) => {
			await release(sessionID);
			return End(); },});
	return handleQuery; };

new S(wss, async (sessionID, role, reason) => {£\label{line:new-svr}£
	if (role === Role.Self) {
		console.error(`${sessionID}: internal server error`); }
	else { await tryRelease(sessionID); }}, agencyProvider);
\end{lstlisting}

The developer instantiates the session (\cref{line:new-svr})
using the WebSocket server, cancellation handler,
and the EFSM implementation --- a function to be invoked
for every new session, parameterised by the session ID,
and returns the \code{State} class instance of the initial state.
\cref{line:impl-recv} implements the API generated for
a receive state, specifying how to handle
a \tmsg{Query} message.
\cref{line:impl-send} implements the API generated for
a send state --- send the \tmsg{Available}
message with the price and proceed to the
\code{handleResponse} continuation.\\

\subsection{Client Role $\ppt{B}$: Generated APIs}

\begin{lstlisting}[
language=TypeScript,tabsize=2,title=\code{src/TravelAgency/B/S27.tsx}
]
type Props = { factory: SendComponentFactoryFactory };
export default abstract class S27<ComponentState = {}>
	extends React.Component<Props, ComponentState> {
	protected Suggest: SendComponentFactory<[string]>;
	constructor(props: Props) {
		super(props);
		this.Suggest = props.factory<[string]>( £\label{line:send-fact-fact}£
			Roles.Peers.A, 'Suggest', ReceiveState.S29); }}
\end{lstlisting}

The generated React component for a \emph{send state}
receives a \code{factory} function from the session runtime
to generate \emph{component factories} for each permitted selection.
\cref{line:send-fact-fact} reads, ``generate a component factory
which sends a message (labelled \tmsg{Suggest} to
role $\ppt{A}$ with one \code{string}-typed payload) and
transitions to state \code{S29}''.
The component factory is defined as  a \emph{protected} property to allow
access by subclasses implemented by the developer.

\begin{lstlisting}[
language=TypeScript,tabsize=2,title=\code{src/TravelAgency/B/S29.tsx}
]
enum Labels { Quote = 'Quote', Full = 'Full' };

interface QuoteMessage {
	label: Labels.Quote, payload: [number] };
interface FullMessage {
	label: Labels.Full, payload: [] };
type Message = | QuoteMessage | FullMessage

type Props = {
	register: (role: Roles.Peers,
		handle: ReceiveHandler) => void };

export default abstract class S29<ComponentState = {}>
	extends React.Component<Props, ComponentState> {
	componentDidMount() {
		this.props.register(Roles.Peers.A,
			this.handle.bind(this)); }

	handle(data: any): MaybePromise<State> { £\label{line:msg-handler}£
		const message = JSON.parse(data) as Message;
		switch (message.label) {
		case Labels.Quote: {
			const thunk = () => SendState.S30;
			const continuation = this.Quote(...message.payload);
			if (continuation instanceof Promise) {
				return new Promise((resolve, reject) => {
					continuation.then(() => resolve(thunk()))
						.catch(reject); }); }
			else { return thunk(); }}
		case Labels.Full: {
			const thunk = () => SendState.S27;
			const continuation = this.Full(...message.payload);
			if (continuation instanceof Promise) {
				return new Promise((resolve, reject) => {
					continuation.then(() => resolve(thunk()))
						.catch(reject); }); }
			else { return thunk(); } }}}

	abstract Quote(payload1: number, ): MaybePromise<void>;£\label{line:abs-meth-1}£
	abstract Full(): MaybePromise<void>; }£\label{line:abs-meth-2}£
\end{lstlisting}

The generated React component for a \emph{receive state}
registers a message handler (\cref{line:msg-handler})
to the session runtime component,
to be invoked on a WebSocket \code{onmessage} event.
An abstract method is exposed for each possible branch
(\cref{line:abs-meth-1,line:abs-meth-2}),
requiring the developer to explicitly handle the received
message.
When invoked, the message handler parses
the WebSocket message and invokes the abstract method
(implemented by the developer) corresponding to the
label of the received message.
The message handler returns the successor state to the
runtime to advance the EFSM.
\\

\subsection{Client Role $\ppt{B}$: API Usage}

\begin{lstlisting}[
language=TypeScript,tabsize=2,title=\code{src/components/MakeSuggestion.tsx}
]
import { S27 } from "../../TravelAgency/B";
export default class MakeSuggestion extends S27 {
	static contextType = FriendState;
	declare context: React.ContextType<typeof FriendState>;

	render() {
		const options: DestinationOption[] = places.map(
			(name) => ({ name,
				buildClickComponent: () => (['Suggest',
					this.Suggest('onClick', () => { £\label{line:use-fact}£
						this.context.setDestination(name);
						this.context.setErrorMessage(undefined);
						return [name]; }),]),}};
		return (<div>
			<Typography variant='h3' gutterBottom>
				Offer Suggestion
			</Typography>
			<Container>
				{this.context.errorMessage !== undefined &&
				<Alert
					style={{ marginBottom: '1rem' }}
					severity='error'>
					{this.context.errorMessage}</Alert>}
				<Grid container spacing={3}>
					{options.map((option, key) => (
						<Grid item xs={6} sm={4} key={key}>
							<TravelCard content={option} />
						</Grid>))}
			</Grid></Container></div>); }};
\end{lstlisting}

The developer implements the send state \code{S27}
by extending from the corresponding abstract class,
and uses the component factory property to generate
UI components that are bound to the communication action.
\cref{line:use-fact} creates a React component
that sends the \tmsg{Suggest} message on a click event,
for each destination option in \code{places}.

\begin{lstlisting}[
language=TypeScript,tabsize=2,title=\code{WaitResponse.tsx}
]
import { S29 } from "../../TravelAgency/B";
export default class WaitResponse extends S29 {
	static contextType = FriendState;
	declare context: React.ContextType<typeof FriendState>;

	Full() { £\label{line:use-recv-full}£
		this.context.setErrorMessage(
			`No availability for ${this.context.destination}`);
		this.context.setDestination(undefined); }

	Quote(quote: number) { this.context.setQuote(quote); } £\label{line:use-recv-quote}£

	render() {
		return (<div>
			<div><Typography variant='h3' gutterBottom>
				Pending enquiry for {this.context.destination}
			</Typography></div>
			<div><CircularProgress /></div>
		</div>); }};
\end{lstlisting}

The developer implements the receive state \code{S29}
by extending from the corresponding abstract class,
and implements the required abstract methods to
define how to handle a \tmsg{Full} message
(\cref{line:use-recv-full}) or a \tmsg{Quote} message
(\cref{line:use-recv-quote}).

\begin{lstlisting}[
language=TypeScript,tabsize=2,title=\code{FriendView.tsx}
]
import { B } from "../../TravelAgency/B";

export default class FriendView extends React.Component {
	render() {
		const origin = process.env.REACT_APP_PROXY
			?? window.location.origin;
		const endpoint = origin.replace(/^http/, 'ws');
		return (<div><FriendState>
			<B £\label{line:new-react}£
				endpoint={endpoint} £\label{line:react-endpoint}£
				states={{ S27: MakeSuggestion, S28: Completion, £\label{line:react-mapping}£
					S29: WaitResponse, S30: MakeDecision }}
				waiting={<CircularProgress />} £\label{line:react-waiting}£
				connectFailed={<Alert severity='error'> £\label{line:react-connect-failed}£
					Connect Failed</Alert>}
				cancellation={(role, reason) => { £\label{line:react-cancel}£
					console.error(reason);
					return <Alert severity='error'>
						Session Cancelled by {role}: {reason}</Alert>; }}
			/></FriendState></div>); }};
\end{lstlisting}

The developer instantiates the session (\cref{line:new-react})
by supplying: the WebSocket endpoint (\cref{line:react-endpoint}),
a React component to render whilst waiting (\cref{line:react-waiting}),
a React component to render on a connection failure (\cref{line:react-connect-failed}),
a function to build a React component to respond to session cancellation
(\cref{line:react-cancel}), and an object literal (\cref{line:react-mapping})
mapping each EFSM state to the corresponding subclass implemented by the developer.